\documentclass[aps,prx,amsmath,amssymb,amsfonts,superscriptaddress,notitlepage,reprint,longbibliography,floatfix%,linenumbers%,linenumbers
]{revtex4-2}

\usepackage{graphicx}
\usepackage{dcolumn}
\usepackage{bm}
\usepackage{amssymb}
\usepackage{amsmath}
\usepackage[svgnames]{xcolor}
\usepackage[colorlinks=true,
            linkcolor=blue,
            urlcolor=blue,
            citecolor=blue]{hyperref}
\usepackage[english]{babel}
\usepackage{braket}
\usepackage{microtype}
\usepackage{bbm}
\usepackage{tikz}
\usepackage{siunitx}
\usepackage{booktabs}
\usepackage{physics}
\usepackage{csquotes}
\usepackage{multirow}
\usepackage{interval}
\usepackage{ulem}
\usepackage{booktabs}

\usepackage[capitalise]{cleveref}

\usepackage{float}
\makeatletter
\let\newfloat\newfloat@ltx
\makeatother
\usepackage{algorithm}
\usepackage{algpseudocode}
\usepackage[capitalise]{cleveref}

\definecolor{manu_color}{rgb}{1, 0, 0}
\definecolor{alex}{rgb}{0,0.58,0.81}
\definecolor{leo}{rgb}{0.87, 0.40, 0.11}

\newcommand{\alex}[1]{\textcolor{alex}{#1}}

\definecolor{patrice}{rgb}{1, 0.5, 0}
\definecolor{dv}{rgb}{1, 0, 1}
\definecolor{darkgreen}{rgb}{0, 0.6, 0}
\definecolor{purple}{rgb}{0.5, 0, 0.5}

\definecolor{leon_color}{rgb}{0.541, 0.671, 0.8}

\newcommand{\pth}{p_{\text{eq}}}
\newcommand{\preset}{p_{\text{eq},\text{reset}}}
\newcommand{\wpsrh}{W/$\sqrt{\text{Hz}}$ }

\begin{document}

\makeatother
\author{Alexandre S. May}
\altaffiliation[These authors contributed equally to this work.]{}
\affiliation{Quantronics group, Service de Physique de l'\'Etat Condens\'e (CNRS, UMR\ 3680),\\IRAMIS, CEA-Saclay, Universit\'e Paris-Saclay, 91191 Gif-sur-Yvette, France}
\affiliation{Alice\&Bob, 53 boulevard du Général Martial Valin, 75015 Paris, France}

\author{Leo Šutevski}
\altaffiliation[These authors contributed equally to this work.]{}
\affiliation{Quantronics group, Service de Physique de l'\'Etat Condens\'e (CNRS, UMR\ 3680),\\IRAMIS, CEA-Saclay, Universit\'e Paris-Saclay, 91191 Gif-sur-Yvette, France}

\author{Jeanne Solard}
\affiliation{Alice\&Bob, 53 boulevard du Général Martial Valin, 75015 Paris, France}

\author{Gil Cardoso}
\affiliation{Alice\&Bob, 53 boulevard du Général Martial Valin, 75015 Paris, France}

\author{L{\ifmmode\acute{e}\else\'{e}\fi}on Carde}
\affiliation{
Laboratoire de Physique de l’Ecole Normale Supérieure, Inria, ENS, Mines ParisTech, Université PSL, Sorbonne Université, Paris, France
}
\affiliation{Alice\&Bob, 53 boulevard du Général Martial Valin, 75015 Paris, France}

\author{Louis Pallegoix}
\affiliation{Quantronics group, Service de Physique de l'\'Etat Condens\'e  (CNRS, UMR\ 3680),\\IRAMIS, CEA-Saclay, Universit\'e Paris-Saclay, 91191 Gif-sur-Yvette, France}

\author{Raphael Lescanne}
\affiliation{Alice\&Bob, 53 boulevard du Général Martial Valin, 75015 Paris, France}

\author{Denis Vion}
\affiliation{Quantronics group, Service de Physique de l'\'Etat Condens\'e  (CNRS, UMR\ 3680),\\IRAMIS, CEA-Saclay, Universit\'e Paris-Saclay, 91191 Gif-sur-Yvette, France}
\author{James O’Sullivan}
\affiliation{Quantronics group, Service de Physique de l'\'Etat Condens\'e  (CNRS, UMR\ 3680),\\IRAMIS, CEA-Saclay, Universit\'e Paris-Saclay, 91191 Gif-sur-Yvette, France}
\author{Patrice Bertet}
\email{patrice.bertet@cea.fr}
\affiliation{Quantronics group, Service de Physique de l'\'Etat Condens\'e  (CNRS, UMR\ 3680),\\IRAMIS, CEA-Saclay, Universit\'e Paris-Saclay, 91191 Gif-sur-Yvette, France}
\author{Emmanuel Flurin}
\email{emmanuel.flurin@cea.fr}
\affiliation{Quantronics group, Service de Physique de l'\'Etat Condens\'e  (CNRS, UMR\ 3680),\\IRAMIS, CEA-Saclay, Universit\'e Paris-Saclay, 91191 Gif-sur-Yvette, France}

\title{Noise Mitigation in Microwave Photon Counting\\ by Cascaded Quantum Measurements}

\begin{abstract}
Single-photon detection in the microwave domain has recently emerged as a powerful tool for quantum sensing, offering a path beyond the standard quantum limit of linear amplification since their sensitivity is set by their dark-count rate rather than by the vacuum fluctuations of the detected field. Superconducting qubit–based single microwave photon detectors (SMPDs) encode the absorption of itinerant photons into discrete quantum states via engineered, non-destructive interactions. A previously demonstrated implementation based on four-wave mixing achieved high efficiency, but remained limited by residual qubit excitation and thermal noise at the detector input. Here, we propose and demonstrate a new SMPD that relies on two cascaded four-wave mixing processes. The information on photon detection is encoded on two spatially separated qubits, which mitigates noise due to uncorrelated qubit errors. Furthermore, increasing the detection frequency up to 11.7 GHz reduces input thermal noise. These combined advances lead to a record-low power sensitivity of $2.4 \cdot10^{-23}\mathrm{W}/\sqrt{\mathrm{Hz}}$ over a bandwidth of $230\ \mathrm{kHz}$.
\end{abstract}

\maketitle

\section{Introduction}
Single photon counting at optical frequencies has been a key enabling technology for more than 40 years. Its application range is wide, from fluorescence microscopy~\cite{orrit_single_1990, klar_fluorescence_2000, betzig_imaging_2006, bruschini_single-photon_2019} to measurement-based quantum computing~\cite{hadfield_single-photon_2009}. 
It is challenging to detect single photons at microwave frequencies due to the five orders of magnitude energy difference with optics, which prohibits room-temperature operation and requires instead millikelvin temperatures. Single Microwave Photon Detectors (SMPDs) are instrumental for the detection of incoherent microwave emitters, such as electronic spins in crystals~\cite{albertinale_detecting_2021, wang_single-electron_2023}, or dark matter candidates~\cite{lamoreaux_analysis_2013, dixit_searching_2021, braggio_quantum-enhanced_2024}. SMPDs may also be applied for primary thermometry~\cite{scigliuzzo_primary_2020} and for quantum illumination protocols~\cite{assouly_quantum_2023}. In quantum computing, SMPDs enable the implementation of measurement-based protocols~\cite{raussendorf_measurement-based_2003, bartolucci_fusion-based_2023, briegel_measurement-based_2009,narla_robust_2016}, the development of novel qubit readout schemes~\cite{opremcak_high-fidelity_2021}, and the generation of quantum states~\cite{besse_single-shot_2018}. Finally, the SMPD is a building block for a novel quantum computing platform based on nuclear spins in a crystal~\cite{osullivan_nuclearspinregister2024} and for Nuclear Magnetic Resonance (NMR) at the single atom level~\cite{travesedo_all-microwave_2024}.

Microwave photon counting has advanced along two distinct routes. On one hand, cavity‐based detectors have been developed to resolve single microwave photons confined within long-lived resonators, thereby enhancing their interaction with the measurement apparatus~\cite{schuster_2007, gleyzes_quantum_2007, dixit_searching_2021, dassonneville_number-resolved_2020, essig_multiplexed_2021, hutin_monitoring_2024}. On the other hand, alternative schemes target itinerant microwave photons impinging on the detector for which neither the arrival time nor the frequency is known beforehand. Our detector falls in the latter. In these setups, one finds qubit‐based detectors capable of single-photon resolution~\cite{chen_microwave_2011, inomata_single_2016, kono_quantum_2018, besse_single-shot_2018, opremcak_high-fidelity_2021}, as well as devices that push the sensitivity toward to the single-photon limit using either bolometric techniques~\cite{lee_graphene-based_2020, kokkoniemi_nanobolometer_2019, richards_bolometers_1994} or nonlinear Josephson junction circuits~\cite{leppakangas_multiplying_2018, pankratov_observation_2024, albert_microwave_2024, stanisavljevic_efficient_2024}.

\begin{figure}[t]
    \centering
    \includegraphics[width=8.4cm]{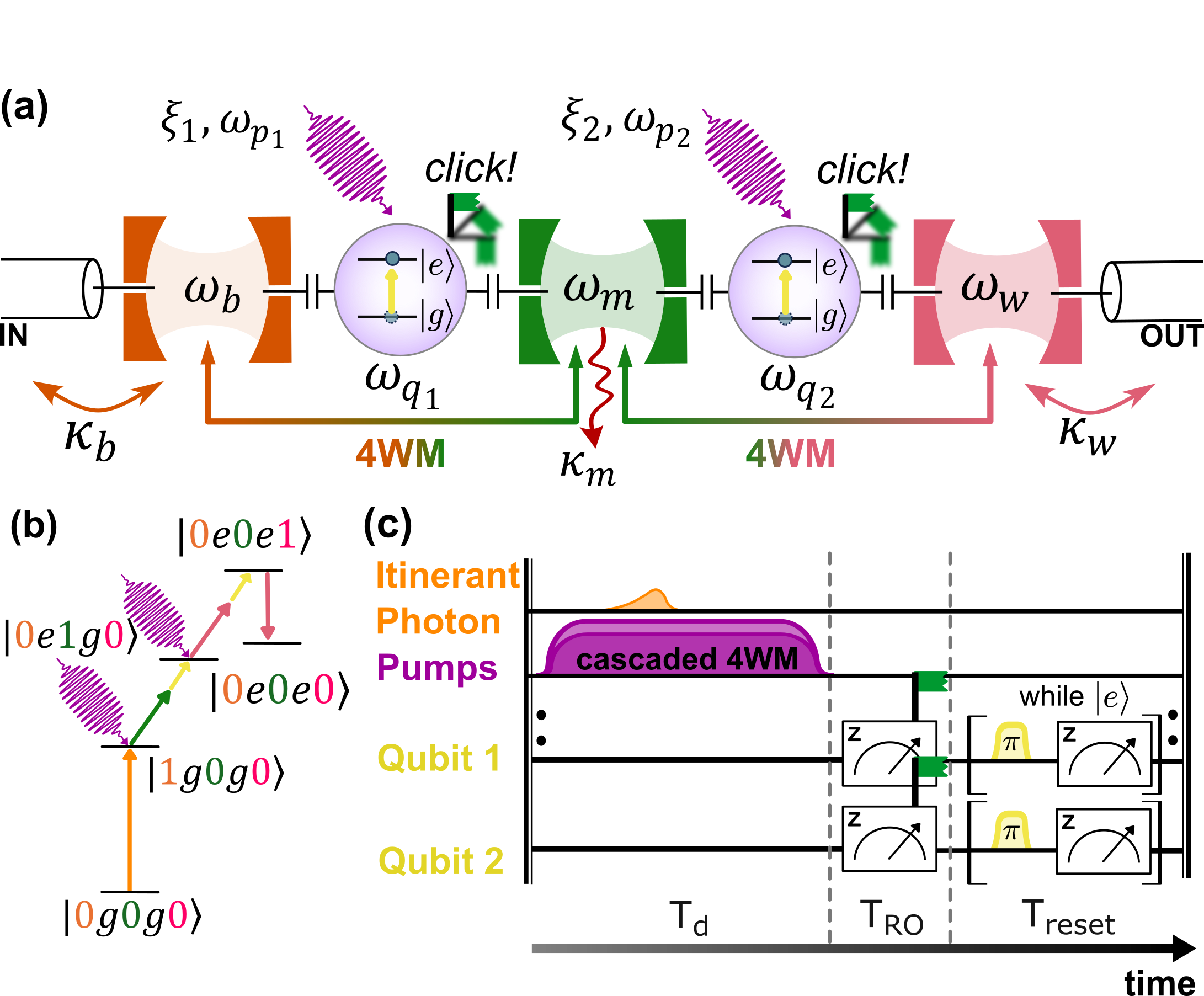}
    \caption{\textbf{cSMPD principle}. (a) Schematic description of the device. The input mode (buffer) at $\omega_b$ is parametrically coupled to the memory mode (frequency $\omega_m$) via qubit 1, pumped by a microwave tone (pump 1) of amplitude $\xi_1$ and frequency $\omega_{p1}$. The memory mode is itself parametrically coupled to a third mode (the waste) of frequency $\omega_w$ via qubit 2, pumped by a microwave tone (pump 2) of amplitude $\xi_2$ and frequency $\omega_{p2}$. The buffer (resp. waste) is coupled to the input  (resp. output) line at a rate $\kappa_b$ (resp. $\kappa_w$), and the memory has an internal loss rate $\kappa_m$. The two cascaded four-wave-mixing (4WM) processes result in the frequency conversion and transmission of an input photon into the output line. (b) Cascaded 4WM processes and states reached by the system during the detection step. (c) Operation cycle of the detector. The cycle includes the detection step of duration $T_d$, the qubit readout step (duration $T_\text{RO}$), and the conditional reset to ground state (duration $T_\text{reset}$).}
    \label{Fig1}
\end{figure}

In all applications, it is essential to improve the SMPD sensitivity. The latter depends on two figures of merit: the dark count rate $\alpha$, defined as the number of false positive detections per unit of time, and the operational efficiency $\eta$ defined as the ratio of true positive count rate over the incoming photon flux. Combining those metrics allows to define the power sensitivity

\begin{equation}\label{eq: sensitivity}
    S=\hbar\omega\dfrac{\sqrt{\alpha}}{\eta}
\end{equation}

\noindent
of the detector, as the noise equivalent power~\cite{richards_bolometers_1994} (NEP,  \cref{Appendix_NEP}) for an integration time of 1s. Note that our operational definition of dark counts includes both intrinsic detection errors, as well as undesired counts caused by thermal photons at the detector input. 

The finite efficiency of the device, on the other hand, is due to photon loss within the device, imperfect quantum dynamics, dead times or readout failures of the device.

Devices based on superconducting qubits and Ramsey interferometry protocols~\cite{kono_quantum_2018} as well as a lambda system dynamics~\cite{inomata_single_2016} have achieved respectively $\alpha=10^4-10^5$ s$^{-1}$ and an operational efficiency $\eta\sim0.5-0.7$ over a bandwidth of $10-20$ MHz. In the 7 GHz range, those properties are equivalent to sensitivities which are respectively $S=2-9\times 10^{-21}$ \wpsrh. These numbers can be compared with state-of-the-art devices based on a parametric four-wave mixing (4WM) process, which are also specifically designed to achieve lower bandwidth~\cite{balembois_cyclically2024, pallegoix_smpd2024}.
They achieve an operational efficiency of $\eta\sim0.75$ and a dark count rate $\alpha\sim40$ s$^{-1}$ over a bandwidth of a few $100$ \SI{}{\kilo\Hz} resulting in a sensitivity $S=3\times 10^{-23}$ \wpsrh, thus two orders of magnitude lower than the other detectors operating in the microwave range.

A single qubit 4WM-based SMPDs consists of two resonator modes of different frequency and a long-lived quantum bit, depicted as a flag, coupled via parametric pumping. When a propagating photon enters the detector in the first mode, its frequency is converted to the second one, simultaneously raising the flag. The information about the presence of the microwave photon can then be extracted by reading out the state of the flag. Spurious excitation of the "flag qubit" leads to detection errors and SMPD dark counts. 

In this work, we introduce a new SMPD design, called Cascaded SMPD (cSMPD). The cSMPD design introduces redundancy in the detection encoding, leveraging the non-demolition nature of the four-wave-mixing process~\cite{lescanne_irreversible_2020}.The principle is shown in \cref{Fig1}(a): two SMPD units are combined in the same device, using a resonator mode as an interface between the two units. An incoming photon should then result in the excitation of the two qubits (final state $\ket{e_1 e_2}$), by successive four-wave-mixing on each SMPD unit. All other outcomes ($\ket{g_1 g_2}$, $\ket{e_1 g_2}$, $\ket{g_1 e_2}$) are discarded. 

Because a large fraction of spurious qubit excitations are uncorrelated, this encoding is expected to reduce the number of false positives. It comes at the expense of a reduced efficiency compared to single-unit SMPDs, since the cSMPD is sensitive to the decay of each qubit. We note that this approach can be extended to arbitrarily long chains of cascaded SMPDs [see \cref{Appendix: N qubit CPD}]. Redundant encoding in photon detection has been demonstrated in the optical domain to enhance the signal-to-noise ratio (SNR)~\cite{distante_detecting_2021}, and similar concepts have been proposed in the microwave domain using DC-biased Josephson junctions~\cite{leppakangas_multiplying_2018}. 

This manuscript reports the operation of cSMPD devices, demonstrating reduced dark counts compared to single-qubit SMPDs. In \cref{section: 3-mode model main}, we discuss the cSMPD modelling, using an approximate three-coupled-mode model. In \cref{Device description}, we describe the device implementation and the experimental setup. \cref{section: cSMPD tuning} deals with cSMPD calibration, with a focus on how pump parameters are chosen for optimal device operation. In \cref{section: cSMPD detectors metrics exp.}, we measure the cSMPD efficiency and dark counts, and report a record power sensitivity of $2.4\cdot10^{-23}\mathrm{W}/\sqrt{\mathrm{Hz}}$ for $230\ \mathrm{kHz}$ detection bandwidth for one of the devices. Finally, in \cref{section: app thermometry}, we show that cSMPDs are useful to measure photon occupation as low as $5\cdot 10^{-6}$, and discuss their application to millikelvin microwave thermometry.

\section{Cascaded SMPD modeling}
\label{section: cascaded}

\subsection{Circuit description and Hamiltonian}
\label{section: cascaded hamiltonian}

The cSMPD schematics is shown in \cref{Fig1}(a). The cSMPD is based on two transmon qubits of angular frequency $\omega_{q_k}$ and self-Kerr $\chi_{q_kq_k}$, with $k\in\{1,2\}$. The first qubit is dispersively coupled with a strength $\chi_{q_1b}$ to an input resonator (named the buffer) of angular frequency $\omega_b$, and to a second resonator (named the memory) of angular frequency $\omega_m$ with a dispersive coupling $\chi_{q_1m}$. The memory acts as the output mode of the first SMPD and the input mode for the second. The second qubit is dispersively coupled to the memory with a strength $\chi_{q_2m}$, and to a third resonator (named the waste) of angular frequency $\omega_w$ with a dispersive coupling strength $\chi_{q_2w}$. Each qubit is capacitively coupled to a broad-band pump line through which microwave signals can be applied. The frequency and amplitude of these pump tones are denoted as $\omega_{p_k}$ and $\xi_k$, $k\in\lbrace 1,2 \rbrace$, the latter describing the pump-induced phase drop across each transmon junction. The qubits, buffer, memory, and waste modes are described respectively by the annihilation operators $\hat{q}_k$, $\hat{b}$, $\hat{m}$, and $\hat{w}$.

Under application of these pump tones, the qubits undergo 4WM processes which are at the heart of the cSMPD. As explained in~\cite{lescanne_irreversible_2020,balembois_cyclically2024,amay_thesis}, these processes can be described by an effective Hamiltonian

\begin{equation}\label{H4wm cascaded N=2}
    \dfrac{\hat{H}_{\text{4WM}}}{\hbar} = {g_{4,1}}\hat{b}\hat{q}_1^\dagger\hat{m}^\dagger+{g_{4,2}}\hat{m}\hat{q}_2^\dagger\hat{w}^\dagger+\text{H.c}
\end{equation}

\noindent
with $g_{4,1}=-\xi_1\sqrt{\chi_{q_1b}\chi_{q_1m}}$ and $g_{4,2}=-\xi_2\sqrt{\chi_{q_2m}\chi_{q_2w}}$. This description is valid when the following frequency matching conditions are satisfied

\begin{equation}\label{eq: freqmatchcond CPD N=2}
\begin{cases}
    \tilde{\omega}_b + \omega_{p_1} = \tilde{\omega}_{q_1} + \tilde{\omega}_m\\[10pt]
    \tilde{\omega}_m + \omega_{p_2} = \tilde{\omega}_{q_2} + \tilde{\omega}_w,
\end{cases}
\end{equation}

\noindent
where $\tilde{.}$ angular frequencies are the mode frequencies renormalized by the off-resonant microwave drives, $\tilde{\omega}_{q_k}=\omega_{q_k}- 2\chi_{q_kq_k}\abs{\xi_k}^2$ for $k\in\{1,2\}$, $\tilde{\omega}_b = \omega_b - \abs{\xi_1}^2\chi_{q_1b}$, $\tilde{\omega}_m = \omega_m - (1+\abs{\xi_1}^2)\chi_{q_1m} - \abs{\xi_2}^2\chi_{q_2m}$ and $\tilde{\omega}_w = \omega_w - (1+\abs{\xi_2}^2)\chi_{q_2w}$.  

As explained in the following section, the simultaneous action of the two 4WM processes under pumps 1 and 2 can be treated as an effective 4WM process between the buffer and the waste mode. In this cascaded 4WM process, both qubits Q1 and Q2 are excited, and a photon is converted from $\omega_b$ to $\omega_w$. This is the core process of the cSMPD device.

\begin{figure*}[t!]
    \centering
    \includegraphics[width=\textwidth]{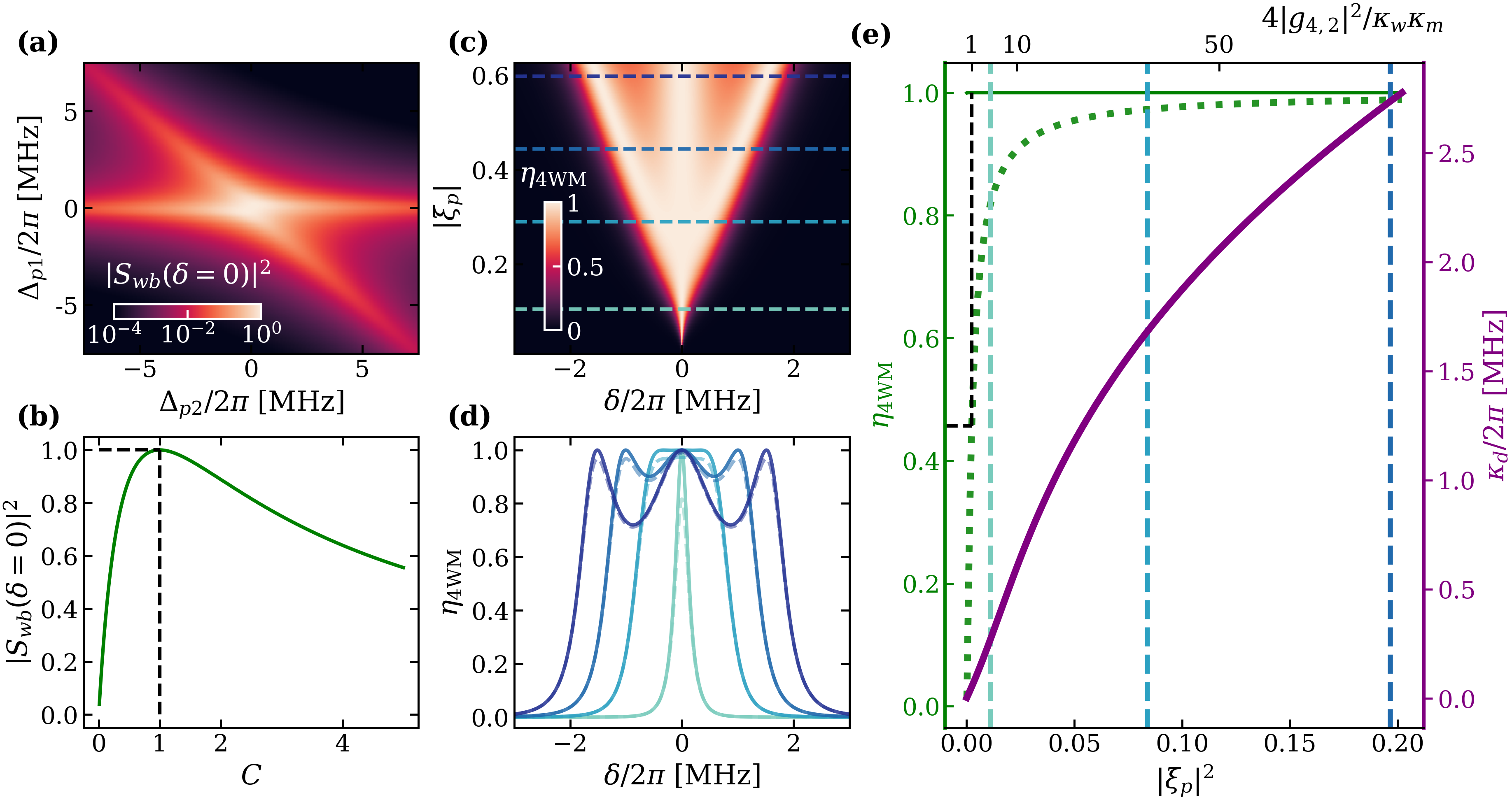}
        \caption{\textbf{Three-coupled-mode model.} In this figure, predictions of the three-coupled-mode model are shown, using the following fixed parameters $\chi_{q_1b}=\chi_{q_1m}=\chi_{q_2m}=\chi_{q_2w}=2\pi\times \SI{2}{\mega\Hz}$, $\kappa_b=\kappa_w=\SI{9.4e5}{\per\s}$ and $\Gamma_{2,k}^* = 0$. (a) Conversion efficiency at resonance on the buffer mode $\abs{S_{wb} (\delta=0)}^2$ as a function of the pump frequency detuning  $\Delta_{p1,2}/2\pi$. It approaches $1$ around $\Delta_{p_1}=\Delta_{p_2}= 0$. The 4WM amplitudes are set to $g_{4,1}=g_{4,2}=2\pi\times \SI{500}{\kilo\Hz}$ and $\kappa_m = 0$. The diagonal line corresponds to a second-order 4WM process. (b)  Conversion efficiency at resonance on the buffer mode is plotted as a green solid line as a function of the cooperativity $C$. It reaches $1$ when $C=1$. (c) Conversion efficiency $\abs{S_{wb} (\delta)}^2\equiv\eta_\mathrm{4WM}$ as a function of $\abs{\xi_p}$ and constant $\beta_\xi$,  and input drive frequency detuning $\delta/2\pi$, at $C=1$ (with $\kappa_m = 0$). (d) Blue solid lines show $\eta_\mathrm{4WM}$ for $\abs{\xi_p}^2 = \lbrace 0.01,\ 0.08,\ 0.19\rbrace$. Note the Lorentzian lineshape at low $\abs{\xi_p}^2$ and the multi-pole function at higher $\abs{\xi_p}^2$. Blue dashed lines show the same quantities, in presence of finite memory loss rate, $\kappa_m=\SI{3.1e5}{\per\s}$. (e) Conversion efficiency as a function of $\abs{\xi_p}$ with $\kappa_m=\SI{3.1e5}{\per\s}$ (dotted green line) or $\kappa_m=0$  (solid green line), with $C=1$. $\eta_\mathrm{4WM}$ tends towards $1$ when $4\abs{g_{4,2}}^2/\kappa_w \gg \kappa_m$ (see dashed black line representing the threshold). The purple line shows the detector bandwidth $\kappa_d$ as a function of $\abs{\xi_p}^2$.} 
        \label{figMAIN: fig2}
\end{figure*}

\subsection{Three-coupled-mode model}
\label{section: 3-mode model main}

In order to model this cascaded 4WM, it is useful to proceed to further approximations. In single-qubit SMPDs, the qubit degree of freedom can be eliminated in the weak-excitation limit, resulting in a two-coupled-cavity model that can be solved analytically \cite{balembois_cyclically2024, pallegoix_smpd2024}. Similarly, in the cSMPD, one can eliminate the two qubit degrees of freedom from Hamiltonian \cref{H4wm cascaded N=2}. This results in a Hamiltonian involving only the buffer, the memory, and the waste modes, with pairwise couplings $g_{4,1}$ and $g_{4,2}$. In this three-coupled-mode model, the successful detection of a photon is equivalent to the transmission of a photon through the $3$-mode network from buffer to waste described by the scattering coefficient, so that the conversion efficiency $\eta_\mathrm{4WM} =|S_{wb}|^2$. The latter is readily computed using input-output theory [see \cref{appendix: Coupled mode models} for more details], yielding 

\begin{multline}
\label{eq: general Swb expression}
    S_{wb}(\delta,\Delta_{p_1},\Delta_{p_2}, \xi_1, \xi_2)\\
    =\sqrt{\kappa_{w,\mathrm{ext}}\kappa_{b,\mathrm{ext}}}\cdot\dfrac{g_{4,1}g_{4,2}}{\lambda_b \lambda_m \lambda_w + \abs{g_{4,1}}^2\lambda_w+\abs{g_{4,2}}^2\lambda_b},
\end{multline}
where $\delta \equiv \tilde{\omega}_b-\omega_d$ is the detuning of the input microwave frequency from the cSMPD center detection frequency, $\Delta_{p_1} \equiv \omega_{p_1} - (\tilde{\omega}_{q1} + \tilde{\omega}_{m} - \tilde{\omega}_{b})$, and $\Delta_{p_2} \equiv \omega_{p_2} - (\tilde{\omega}_{q2} + \tilde{\omega}_{w} - \tilde{\omega}_{m})$ are the pump frequency detunings from the 4WM resonance conditions \cref{eq: freqmatchcond CPD N=2}, and $\lambda_b = i\delta + \kappa_b/2$, $\lambda_m=i(\Delta_m+\delta)+\tilde{\kappa}_m/2$, $\lambda_w=i(\Delta_w+\delta)+\tilde{\kappa}_w/2$, where $\tilde{\kappa}_m = \kappa_m + 2\Gamma_{2,1}^*$ and $\tilde{\kappa}_w = \kappa_w + 2(\Gamma_{2,1}^* + \Gamma_{2,2}^*)$ with $\Gamma_{2,k}^* = \Gamma_{1,k}/2 + \Gamma_{\phi, k}$. $\Gamma_{2,k}^* = 1/T_{2,k}^*$ is the decoherence rate of qubit $k$, $\Gamma_{1,k}$ its energy decay rate and $\Gamma_{\phi,k}$ its dephasing rate. Note that similar expressions are obtained in microwave to optical coherent transducers via a mechanical mode \cite{andrews_bidirectional_2014, Brubaker2022}. Also, for reasons that will become apparent below, we define $\beta_\xi \equiv \sqrt{\xi_1/\xi_2}$, and $\xi_p = \sqrt{\xi_1 \xi_2}$, so that $\xi_1 = \beta_\xi \xi_p$ and $\xi_2 = \xi_p/\beta_\xi $.
The theoretical cascaded 4WM efficiency $\eta_\mathrm{4WM}$ resulting from \cref{eq: general Swb expression} is plotted in \cref{figMAIN: fig2}(a) as a function of $\Delta_{p1}$ and $\Delta_{p2}$. The cascaded 4WM is strongest for $\Delta_{p1}=\Delta_{p2}=0$, as expected from the 4WM conditions. Finite efficiency is also predicted outside of this resonance condition, due to second-order processes (see horizontal and diagonal lines in \cref{figMAIN: fig2}(a)). In the following, we will concentrate on the resonant regime, and assume $\Delta_{p1}=\Delta_{p2}=0$. 

In this regime, an insightful simple expression can be obtained by retaining only the leading-order term in $\delta$, neglecting the internal memory loss rate $\kappa_m$ and the qubit decoherence in \cref{eq: general Swb expression}, yielding:

\begin{align}
    &\eta_\mathrm{4WM}=\abs{S_{wb}(\delta)}^2\approx\frac{4C}{\abs{1+C}^2}\frac{1}{1+\left(\dfrac{2\delta}{\kappa_d}\right)^2} \label{eq:14}\\
    &\kappa_d  = \frac{4\abs{g_{4,1}}^2}{\kappa_b} + \frac{4\abs{g_{4,2}}^2}{\kappa_w} \label{eq:15}\\
    &C = \dfrac{\kappa_w}{\kappa_{b}}\abs{\dfrac{g_{4,1}}{g_{4,2}}}^2. \label{eq: coop n=2}
\end{align}
\noindent

When $\delta = 0$, $\eta_\mathrm{4WM}$ is thus governed by the dimensionless parameter $C$, which appears as the cooperativity of the cascaded 4WM conversion process. We note that \cref{eq:14} can be interpreted as the transmission through a Fabry-Pérot resonator (the memory mode), coupled on one hand to the input line at a rate $4{\abs{g_{4,1}}^2}/\kappa_{b}$, and on the other hand to the output line at a rate $4{\abs{g_{4,2}}^2}/\kappa_{w}$. $C$ therefore appears as the ratio of these rates. Unit transmission is achieved when the two rates are equal, which corresponds to $C=1$ and to the unit efficiency condition of the cSMPD.

At constant $C$, \cref{eq:14} predicts that the cSMPD response has a Lorentzian profile, of full width at half maximum $\kappa_d$. Therefore, the detection bandwidth should be tunable by changing the pump amplitude $\xi_p$, which can be achieved at constant $C$ (and in particular in the most relevant regime $C=1$). This interesting feature is visible in \cref{figMAIN: fig2}(c,e). We set $C=1$ and consider infinitely coherent qubits ($\Gamma_{2,k}^*=0$). We plot $\eta_\mathrm{4WM}$ computed with the 3-coupled-mode model \cref{eq: general Swb expression} as a function of $\delta$ for various drive amplitudes. For small pump amplitudes, the response is indeed a Lorentzian with a width linearly increasing with $\abs{\xi_p}^2$. For larger pump amplitudes, strong coupling is reached, and the response markedly deviates from a Lorentzian profile. More information on this regime can be found in \cref{appendix: Coupled mode models}.

\subsection{Operational efficiency}
\label{sec: ope eff}

We now discuss the operational efficiency $\eta$ of an actual device. Various contributions can be identified, beyond the 4WM efficiency. Indeed, in microwave photon counting operation, the cSMPD is used in cyclic mode, similar to single-qubit SMPDs (see Fig.~\ref{Fig1}c). A detection cycle starts by a detection window during which both pumps are turned on (duration $T_d$), followed by simultaneous readout of the qubits (duration $T_\mathrm{RO}$), followed by conditional reset of both qubits in their ground state (duration $T_\mathrm{reset}$). The cyclic operation leads to additional efficiency reduction factors due to internal buffer losses, finite duty cycle, transients, finite qubit and memory lifetime, and finite readout fidelity of the qubit states. The operational efficiency $\eta$, defined as the probability to detect the two qubits in their excited state if one photon reaches the device input, writes:

\begin{equation}\label{eq: op. eff. cSMPD N=2}
    \eta = \eta_b\ \eta_{\text{cycle}}\ \eta_{\text{4WM}}\  \eta_{\mathrm{q}}\ \mathcal{F}_{\mathrm{RO}}.
\end{equation}

This formula is similar to the one encountered for single-qubit SMPDs~\cite{balembois_cyclically2024,pallegoix_smpd2024}. However, as explained below, the value of each term in \cref{eq: op. eff. cSMPD N=2} is expectedly lower because the losses now involve two qubits and the memory mode. Therefore, the cSMPD is expected to display a slightly reduced efficiency compared to single-qubit SMPDs. We now discuss \cref{eq: op. eff. cSMPD N=2} term by term.\\

\noindent\textbf{Buffer internal loss} -- Upon entering the buffer resonator, the photon can can be lost internally, before participating in the cascaded process. This can be modeled by $\eta_b = \kappa_\mathrm{b, ext}/\kappa_\mathrm{b}$.\\

\noindent\textbf{Duty cycle} -- The detector operates cyclically, as illustrated in \cref{Fig1}(c). Since the detector is inactive during readout and reset, $\eta_\mathrm{cycle} = T_d/(T_d + T_\mathrm{RO+reset})$, with $T_\mathrm{RO+reset} = (n+1) T_\mathrm{RO} + n T_\mathrm{reset}$ and $n$ the number of reset cycles.\\

\noindent\textbf{Transients} -- The impact of cyclic operation goes beyond the simple duty cycle factor $\eta_\mathrm{cycle}$. Compared to the steady-state formulas obtained in \cref{section: 3-mode model main}, transients also need to be considered. A detailed analysis is provided in \cref{appendix: Coupled mode models}, based on an analytical and numerical time-domain integration of the three-coupled-mode model. We outline here the main conclusions.

When the signal is pulsed synchronously with the pump tones, buffer and waste fields reach steady-state after a transient of duration $\sim \kappa_d^{-1}$, resulting in a reduced overall efficiency $\eta_\mathrm{4WM}$ compared to \cref{eq: general Swb expression}. This reduction is significant when $\kappa_d^{-1}$ is comparable to the detection time window $T_d$, which happens at the lowest pump powers. Analytical formulas for the reduced $\eta_\mathrm{4WM}$ can be found in \cref{appendix: Coupled mode models}. Transient effects are taken into account in the analysis of the measurements presented in \cref{section: cSMPD detectors metrics exp.} (\cref{fig: eff vs C cSMPD2} and \cref{figMAIN: fig3}a-c).

It is worth noting that the situation is more complex when the signal to be detected is always on. Indeed, the buffer may then reach steady-state already before the pump tones are switched on, depending on the frequency shift $|\xi_1|^2 \chi_{q_1 b} $ undergone by the buffer mode upon switching of the pump 1. If $|\xi_1|^2 \chi_{q_1 b} \ll \kappa_d$, as is the case in the measurements discussed in this work, then the transients have a negligible impact and the steady-state behavior is approximately recovered. In the opposite limit ($|\xi_1|^2 \chi_{q_1 b} \gg \kappa_d$), the transients are expected to reduce the efficiency similar to the case where the signal is pulsed.\\

\noindent\textbf{Qubit efficiency} -- The two-qubit excited state $\ket{e_1e_2}$ created by the cascaded 4WM process may decay before the readout. Since the photon arrives at a random time within the detection window, an analysis similar to Ref.~\cite{balembois_cyclically2024}
gives
\begin{equation}
\eta_q=\frac{T_\mathrm{ee}}{T_d}\left(1-e^{-T_d/T_\mathrm{ee}}\right),
\end{equation}
where $T_\mathrm{ee}$ is the decay time of $\ket{e_1e_2}$ under pump,
close to $(1/T_1^{(1)}+1/T_1^{(2)})^{-1}$. This is the dominant effect of qubit decoherence. Decoherence during the conversion itself is smaller and is already included in $\eta_\mathrm{4WM}$ (\cref{appendix: Coupled mode models}).\\

\noindent\textbf{Readout fidelity} -- The readout fidelity $\mathcal{F}_{\mathrm{RO}} = \Pi_{k=1,2} \mathcal{F}_{\mathrm{RO}}^{(k)} $, where $\mathcal{F}_{\mathrm{RO}}^{(k)}$ is qubit $k$ readout fidelity when the other qubit is excited.\\

\subsection{Dark counts : model}
\label{subsec: SMPD noise}

The cSMPD dark count rate $\alpha = \alpha_\mathrm{err} + \alpha_\mathrm{th}$ has two contributions: one from false positives (rate $\alpha_\mathrm{err}$), and the other one from thermal photons incoming at the detector input (rate $\alpha_\mathrm{th}$).

\subsubsection{False positives}

False positives happen when the two qubits are found in their excited state at the end of the detection cycle, whereas no photon actually impinged on the detector input during that time. The key insight of the cSMPD is that spurious qubit excitations of Q1 and Q2 are expected to be largely uncorrelated; therefore, two-qubit spurious excitations should be very rare. Owing to the correlated encoding, cSMPDs should therefore achieve much lower false positive error rates than single-qubit SMPDs. 

This can be quantified as follows. Consider single-qubit spurious excited state rates $\alpha_\mathrm{err}^{(k)}$, $k\in\{1,2\}$. Assuming these rates are uncorrelated, it is straightforward that the resulting two-qubit (i.e., cSMPD) error rate should be 

\begin{equation}
\label{eq:UncorrelatedErrors}
\alpha_\mathrm{err} = \alpha_\mathrm{err}^{(1)} \alpha_\mathrm{err}^{(2)} T_\mathrm{cycle}. 
\end{equation}

In the relevant limit where the number of spurious excitation events per cycle is small $\alpha_\mathrm{err}^\mathrm{(1,2)} T_\mathrm{cycle} \ll 1$, the resulting cSMPD error rate thus verifies $\alpha_\mathrm{err} \ll \alpha_\mathrm{err}^{(1,2)}$. In the experimental section, \cref{eq:UncorrelatedErrors} will moreover allow us to quantify the correlated nature of the various processes at play.

In the remainder of this section, we review established single-qubit spurious excitation processes. Three mechanisms can be distinsguished : qubit excitation through quasi-particles (rate $\alpha_\mathrm{qp}^{(k)}$), by application of the pumps (rate $\alpha_\mathrm{pumps}^{(k)}$), and by application of the readout pulses (rate $\alpha_\mathrm{ROs}^{(k)}$). In total, $\alpha_\mathrm{err}^{(k)} = \alpha_\mathrm{qp}^{(k)} + \alpha_\mathrm{pumps}^{(k)}+ \alpha_\mathrm{ROs}^{(k)}$.\\

\textbf{Quasiparticles contribution} - A number of experiments have established that superconducting qubits often exhibit an average excited state population $p_\mathrm{eq}$ larger than expected from the cryostat base temperature. This phenomenon is understood as being due to the generation of out-of-equilibrium quasiparticles in the qubit. The dominant contribution is in general absorption of infrared radiation at the qubit junction~\cite{serniak_hot_2018, connolly_coexistence_2023, kerschbaum_assessing_2026, lin_suppression_2026}. This process leads to uncorrelated qubit spurious excitations in multi-qubit devices. High-energy radiation (due for instance to cosmic rays) has also been established as a source of quasi-particle generation via phonon generation and recombination. The excitation rate is in general much lower than the infrared contribution; however, it can lead to spatial and temporal correlations in multi-qubit devices~\cite{vepsalainen_impact_2020, mcewen_resolving_2022,mcewen_resisting_2024,kurilovich_correlated_2025, larson_quasiparticle_2025, valenti_spatial_2025, yang_high-temporal-resolution_2026, celi_measuring_2026}. Mechanical vibrations have also been recently observed to lead to spurious qubit excitations in some setups~\cite{kono_mechanically_2024}, and to correlations in multi-qubit devices~\cite{binney_distinguishing_2026}.

In SMPD operation, a detection cycle ends with a reset operation that effectively prepares the two qubits in their ground states. Then, during the next detection cycle, the qubits relax towards equilibrium during the detection window at their energy relaxation rate, $1/T_1$. For each individual qubit $k\in\{1,2\}$, this leads to a false positive detection rate 

\begin{equation}\label{alpha_err}
    \alpha_\mathrm{qp}^{(k)} = \dfrac{\pth^{(k)}-\preset^{(k)}}{T_1^{(k)}}\eta_\mathrm{cycle}+\dfrac{\preset^{(k)}}{T_\mathrm{cycle}}
\end{equation}

\noindent 
where $p_{\mathrm{eq,reset}}^{(k)}$ is the qubit $k$ residual excited state population after reset~\cite{balembois_cyclically2024,pallegoix_smpd2024}. \\

\textbf{Pump contribution} - Application of the pump tones can increase the equilibrium population of the individual qubit above $p_\mathrm{eq}^{(k)}$, due to heating of the microwave components in the line, or to a variety of parametric processes~\cite{connolly_full_2025, dai_characterization_2026, xia_exceeding_2025}, leading to the extra term $\alpha_\mathrm{pumps}^{(k)}$. Modeling is difficult and likely depends strongly on the detailed pump frequency and amplitude. Qualitatively, one may expect that most pump-induced qubit excitation processes are also uncorrelated, and will therefore be strongly suppressed in the cSMPD operation. Note that pump-induced heating may also lead to an increase of the $\hat{b}$ mode average thermal population.\\

\textbf{Readout} - Qubit readout can also induce qubit excitations~\cite{khezri_measurement-induced_2022, shillito_dynamics_2022, cohen_reminiscence_2022, lescanne_escape_2019, sank_measurement-induced_2016, dumas_unified_2024}. This source of dark counts can be mitigated by performing a dispersive readout at the resonator frequency corresponding to the excited state of the qubit. Since the qubit predominantly resides in its ground state, this approach prevents the readout mode from being populated, thereby avoiding unwanted multi-photon transitions and preserving the qubit ground state integrity. Similar to pump-induced qubit excitation, readout-induced qubit excitation (contributing to error rate $\alpha_\mathrm{ROs}^{(k)}$) is also expected to be largely uncorrelated and therefore suppressed in cSMPD operation. 

\subsubsection{Thermal photons}

The cSMPD input line is characterized by mean photon number occupation per mode $n_\mathrm{th}$ at $\omega_b$, noted $n_\mathrm{th,b}$. This number can be converted into an effective temperature of the microwave field, $T_\mathrm{eff}$, which is often found to be higher than the cryostat base temperature (see below). These thermal photons lead to detector dark count rate 

\begin{equation}\label{alpha_th}
    \alpha_\mathrm{th} = \int_{-\infty}^{+\infty}\dd\nu\ n_\mathrm{th}(\nu)\ \eta(\nu),
\end{equation}

\noindent
that leads to an insightful simple expression for a Lorentzian lineshape $\alpha_\mathrm{th} = \eta\kappa_d  n_\mathrm{th,b}/4$ as discussed in~\cite{balembois_cyclically2024,pallegoix_smpd2024} (see \cref{Appendix_noise}). Note that significant deviation from the Lorentzian equation is expected in the strong coupling regime.

Contrary to false positives, the contribution from thermal photons is not expected to be suppressed in the cSMPD design, since they trigger an actual cascaded 4WM process. With the false positives contribution largely suppressed, thermal photons are expected to become the dominant dark count contribution in cSMPD devices. We will come back to this important point in the experimental results section.

\section{Device description}
\label{Device description}

\begin{figure}
    \centering
    \includegraphics[width=\linewidth]{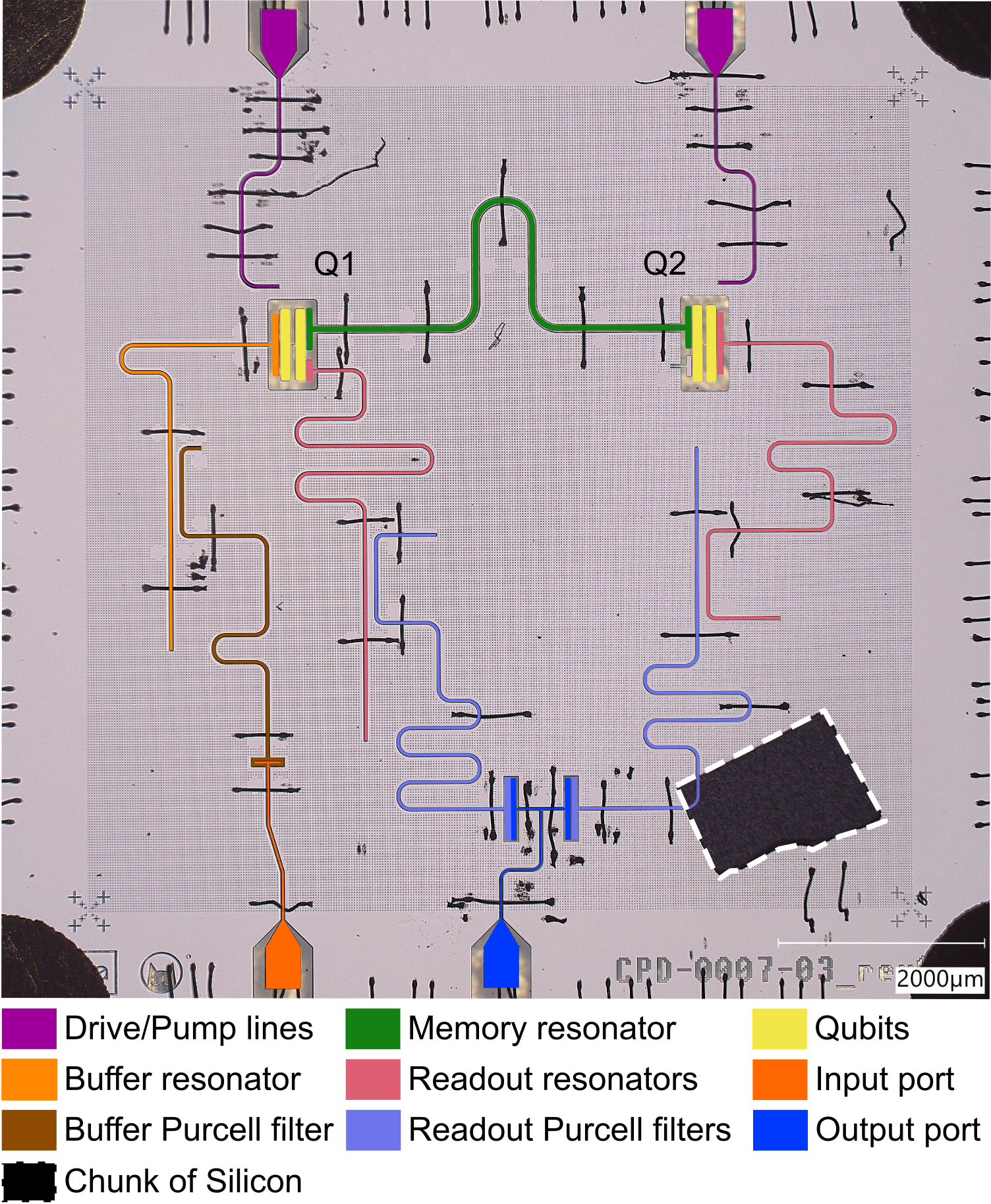}
    \caption{\textbf{Colored micrograph of the cSMPD2}. See legend and main text for details.}
    \label{fig: colored_micrograph}
\end{figure}

We now describe the device-level implementation of the cSMPD principle (see photograph of device cSMPD2 in \cref{fig: colored_micrograph}). The buffer and waste resonators are connected to the $50\Omega$ coaxial lines via Purcell filters, to protect the qubits from spontaneous emission. Qubit 1 readout is achieved via a dedicated resonator, while Qubit 2 readout is achieved via the waste resonator. Readout signals are sent and retrieved via one port, connected to the two readout resonators via Purcell filters. Simultaneous qubit readout is performed using frequency multiplexing. The ground-plane is pierced with small holes to trap residual magnetic flux. On-chip wire-bonds are used to maintain the ground-plane at the same electric potential throughout the device, thus suppressing spurious microwave modes.

The circuit is wire-bonded to a microwave printed circuit board inside a light-tight sample holder mounted on the bottom plate of a dilution refrigerator with base temperature $\sim$\SI{10}{\milli\K}, and connected to the driving and measuring circuit, a detailed wiring diagram of which is provided in \cref{wiring and setup}. In this manuscript, we report results from two different devices, for which two different Purcell filter designs were implemented. cSMPD1 is designed with notch filters on the cSMPD1 sample [see \cref{appendix: cSMPD1} for details], whereas cSMPD2 utilizes extra coplanar waveguide resonators. To adjust the circuit parameter $\kappa_w$ post-fabrication of cSMPD2, a silicon slab has been deposited on the readout 2 Purcell filter resonator.

The initial characterization involves determining frequencies, couplings, and coherence times using standard cQED techniques, with experimental details outlined in \cref{Appendix_ro,Appendix_InputPowCalibration,Appendix_Memoryloss calibration}. All circuit parameters are summarized in \cref{table: CPD-0002-02-A-and-W025-CPD-0007-03-01-C1-circuit-params} for the two devices. It is noteworthy that the two devices were designed with significantly different buffer frequencies: respectively $\SI{8.798}{\giga\Hz}$ for cSMPD1 and $ \SI{11.722}{\giga\Hz}$ for cSMPD2. This has important consequences for thermal photons, as discussed in Sections \cref{section: cSMPD detectors metrics exp.} and \cref{section: app thermometry}. The two devices also differ in their frequency arrangement, which determines which pump-activated processes are close to resonance (\cref{Appendix_high_order_RWA}).

The memory mode is not directly accessible via the measurement lines, and it therefore cannot be characterized by simple microwave reflectometry. Instead, we use the qubits to characterize it. More precisely, we apply a microwave tone at $\omega$ through the Q1 drive line, which couples weakly to the memory resonator through a small capacitance, while performing a Ramsey experiment on Q2. When $\omega = \omega_m$, increased dephasing on Q2 is observed. This measurement-induced-dephasing technique yields both $\omega_m$ and $\chi_{q_2 m}$. An analogous sequence is used to measure $\chi_{q_1 m}$. We finally determine $\kappa_m$ by preparing the Fock state $\ket{1}$ in the memory, and mapping its decay onto the Q2 state [see \cref{Appendix_Memoryloss calibration} for details].

\section{cSMPD tuning}
\label{section: cSMPD tuning}

Tuning the cSMPD requires determining the optimal parameters for the cascaded 4WM process. This is slightly more complex than for a single-qubit SMPD, as there are now $4$ parameters to set ($\omega_{p_{1,2}}$ and $\xi_{1,2}$) instead of two.

\subsection{Initial estimate of pump parameters}
\label{subsection: activate individual 4WM}

An approximate value of pump 1 frequency $\omega_{p_1}$ can be estimated using the first 4WM condition [see \cref{eq: freqmatchcond CPD N=2}]. An approximate value of pump $1$ amplitude $\xi_{1}$ is estimated by tuning the 4WM process converting buffer mode photons into qubit 1 readout resonator through qubit 1, injecting photons at the buffer input ~\cite{lescanne_irreversible_2020,balembois_cyclically2024,pallegoix_smpd2024}. Approximate values of $\omega_{p_2}$ and $\xi_{2}$ are further determined by tuning a 4WM process converting photons from the memory mode to the waste resonator via qubit 2, injecting photons into the memory mode via pump 2 input line (harnessing a small capacitive coupling between the two).

    \subsection{Setting the pump frequencies}
    \label{subsection: pump freq cascaded 4WM}

\begin{figure}[!t]
    \centering
    \includegraphics[width=8.4cm]{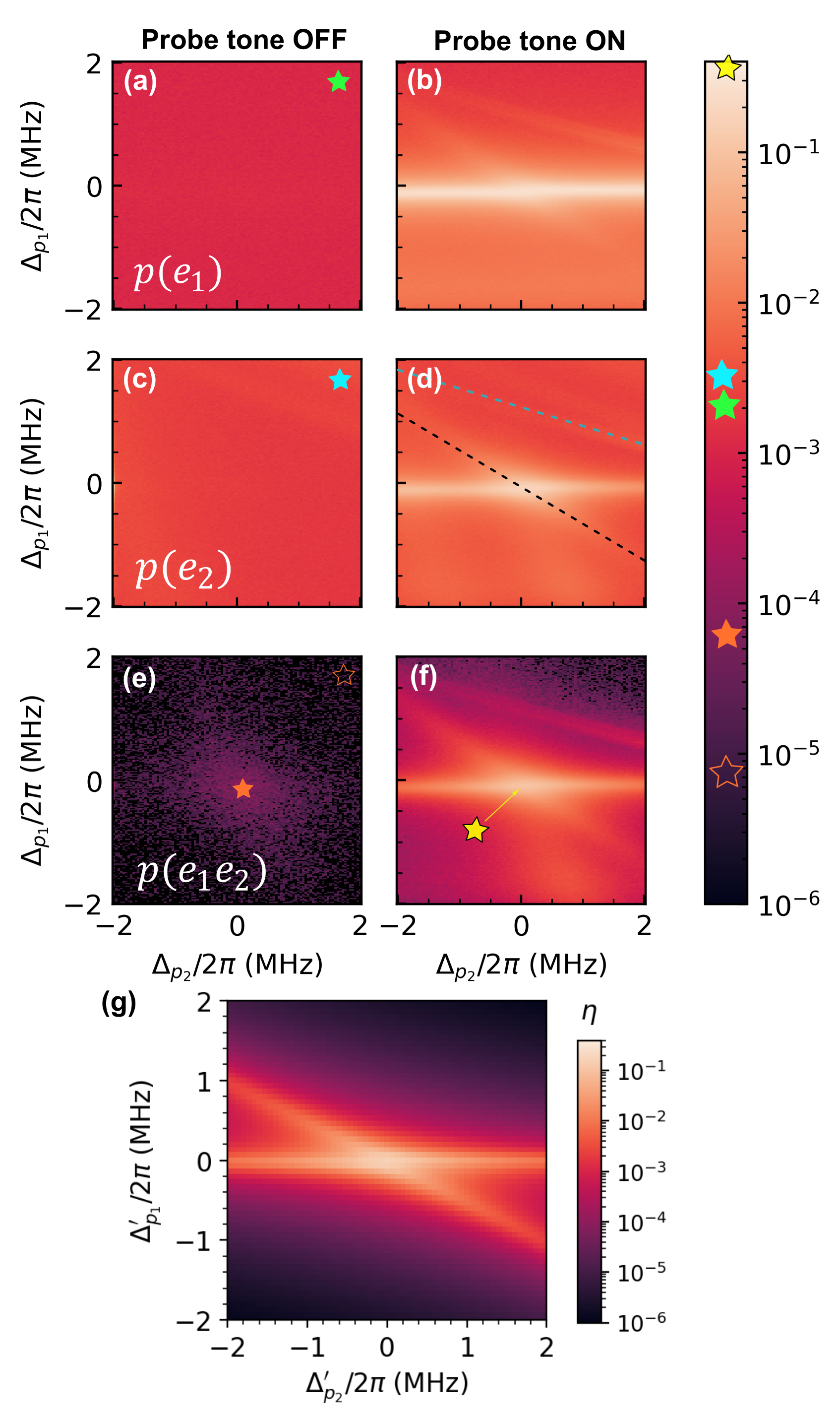}
    \caption{
        \textbf{Cascaded 4WM process: experiment (cSMPD1 data).}
        Measured probability (see logarithmic scale on the right) to find qubit Q1 (a,b), Q2 (c,d) and both of them (e,f) in their excited states, as a function of $\Delta_{p1}$ and $\Delta_{p2}$ at constant pump amplitudes, in presence (b,d,f) or absence (a,c,e) of a probe tone at $\omega_b$ applied at the buffer input. (b) Bright area shows the pump 1 frequency at which the first 4WM process occurs. (d) Central bright spot shows pump 2 frequency that maximizes the second 4WM process and extends horizontally due to memory losses (see text). Dashed lines points other bright excitation 'lines' corresponding to a virtual transition on the memory mode (black - see text) and an unidentified spurious process (cyan). (e) The probe-tone-off correlated excitation probability $p(e_1e_2)$ is orders of magnitude lower than the individual qubit excitation probabilities (green and cyan stars in panels a and c). It reaches $8\times 10^{-6}$ outside of the cascaded 4WM condition (open red star) and $7\times 10^{-5}$ at the 4WM maximum, due to thermal photons. (f) Correlated excitation in presence of probe tone, with working point marked by a yellow star. (g) 3-coupled-mode model based on \cref{eq: op. eff. cSMPD N=2} rescaled by independently measured efficiencies $\eta_\mathrm{cycle}$, $\eta_\mathrm{q}$, $\mathcal{F}_\mathrm{RO}$. See main text and \cref{sec: Kerr correction to 3 cav semiclassical model} for details.}
    \label{figMAIN: cSMPD1 pump maps}
\end{figure}

The next step is to observe the cascaded 4WM process. In that goal, we apply the two pump tones simultaneously during $\sim \SI{12}{\micro\s}$, at the pre-determined amplitudes $\xi_{1,2}$, sweeping their frequencies $\omega_{p_{1,2}}$ around the pre-determined values, while either sending [\cref{figMAIN: cSMPD1 pump maps}(a,c,e)] or not [\cref{figMAIN: cSMPD1 pump maps}(b,d,f)] a weak coherent probe tone to the buffer input at $\omega_b$. We then measure the final state of Q1 and Q2. Repeating this sequence a large number of times, we compute the individual qubit excited state probabilities $p(e_1)$ and $p(e_2)$ as well as the correlated excited state probability $p(e_1 e_2)$.

Results are plotted in \cref{figMAIN: cSMPD1 pump maps}(a-f) using a logarithmic color scale. The bright horizontal line visible in $p(e_1)$ determines the resonant condition for pump 1 ($\Delta_{p_1}=0$) to trigger the first part of the cascaded process, leading to Q1 excitation. The cascaded 4WM process is found within the $\Delta_{p_1}=0$ line, when pump 2 resonant frequency condition is moreover satisfied ($\Delta_{p_2}=0$), leading to a peak in $p(e_2)$ [see \cref{figMAIN: cSMPD1 pump maps}(d)]. The correlated nature of this process is confirmed by observing the same peak in $p(e_1 e_2)$ [see \cref{figMAIN: cSMPD1 pump maps}(f)], and by noting that at this point, $p(e_1,e_2) \gg p(e_1) p(e_2)$. 

Correlated qubit excitations are not only observed when $\Delta_{p_1} = \Delta_{p_2}=0$, due to finite bandwidth of the cascaded 4WM process (horizontal line) as well as to the existence of a second-order process in which the memory mode is only virtually excited [oblique line in \cref{figMAIN: cSMPD1 pump maps}(f)]. Due to ac-Stark shifts on both qubits, this second-order process does not take place exactly when $\Delta_{p_1} = - \Delta_{p_2}$, as is visible in \cref{figMAIN: cSMPD1 pump maps}(f). Finally, extra lines are visible in $p(e_1e_2)$ [see \cref{figMAIN: cSMPD1 pump maps}(f)], which signals a higher-order undetermined parametric process. The three-coupled-mode model, modified to take into account ac-Stark shifts [see \cref{sec: Kerr correction to 3 cav semiclassical model}], reproduces the main features of our measurements using measured circuit parameters, apart from the spurious process mentioned earlier [see \cref{figMAIN: fig4}(g)]. We also note a higher background at the lower-left corner of \cref{figMAIN: cSMPD1 pump maps}(e) compared to the expected one from the semi-classical model, which remains unexplained.

The measurements in \cref{figMAIN: cSMPD1 pump maps} reveal interesting properties of the cascaded 4WM process, which will be further confirmed in photon counting mode [see \cref{section: cSMPD detectors metrics exp.} below]. First, out of the 4WM condition, and in the absence of probe tone, we measure $p(e_1 e_2) \simeq 8 \cdot 10^{-6}$ (open red star in \cref{figMAIN: cSMPD1 pump maps}(e)), approaching $p(e_1) p(e_2)$ since $p(e_1) \simeq p(e_2) \simeq 3\cdot 10^{-3}$ (full green and blue stars in \cref{figMAIN: cSMPD1 pump maps}(a,c)). This indicates that the majority of qubit errors are uncorrelated, as already demonstrated for repetition codes~\cite{GoogleQuantumAI2021RepetitionCode}. They are therefore efficiently suppressed by majority voting. This efficient suppression is evidenced by the large difference in $p(e_1e_2)$ when the probe tone is switched on [$p(e_1e_2) \simeq 0.35$, yellow star in \cref{figMAIN: cSMPD1 pump maps}(f)] or off [$p(e_1e_2) \simeq 7\cdot 10^{-5}$, orange star in \cref{figMAIN: cSMPD1 pump maps}(e)]. Moreover, the value measured with probe tone off is 10 times larger than the value measured when the pump frequencies are detuned from the cascaded 4WM resonance [full and empty stars in \cref{figMAIN: cSMPD1 pump maps}(e)]. This indicates that the residual excitation when $\Delta_{p_1}=\Delta_{p_2}=0$ [orange star in \cref{figMAIN: cSMPD1 pump maps}(e)] is mostly caused by microwave photons impinging on the device input, and leading to a residual correlated qubit excitation by cascaded 4WM with this thermal noise.

\subsection{Setting the pump amplitudes}
\label{subsection: pump amps cascaded 4WM}
The next step consists in setting the pump amplitudes $\xi_{1,2}$. According to the three-coupled-mode model [see \cref{section: 3-mode model main}], $\beta_\xi$ fully determines the cooperativity $C$ of the cascaded 4WM process, independently of $\xi_p$. On the other hand, once $\beta_\xi$ is set such that $C=1$, $\xi_p$ determines the detector bandwidth $\kappa_d$. 

The experimental challenge in tuning the pump amplitudes arises from the fact that their optimal frequencies also depend on the amplitudes via the qubit ac-Stark shifts [see \cref{section: cascaded hamiltonian}]. A first method therefore consists in determining the pump optimal frequencies for each set of pump amplitudes $\xi_{1,2}$, by repeating the two-dimensional pump frequency sweep measurements described in \cref{subsection: pump freq cascaded 4WM}. Recording the maximum $p(e_1e_2)$ for each set of $\xi_{1,2}$ values then allows finding the unit-cooperativity condition, at the expense of long measurement times. This was achieved in the cSMPD1 device (data not shown).

\begin{figure}[t!]
    \centering
    \includegraphics[width=\linewidth]{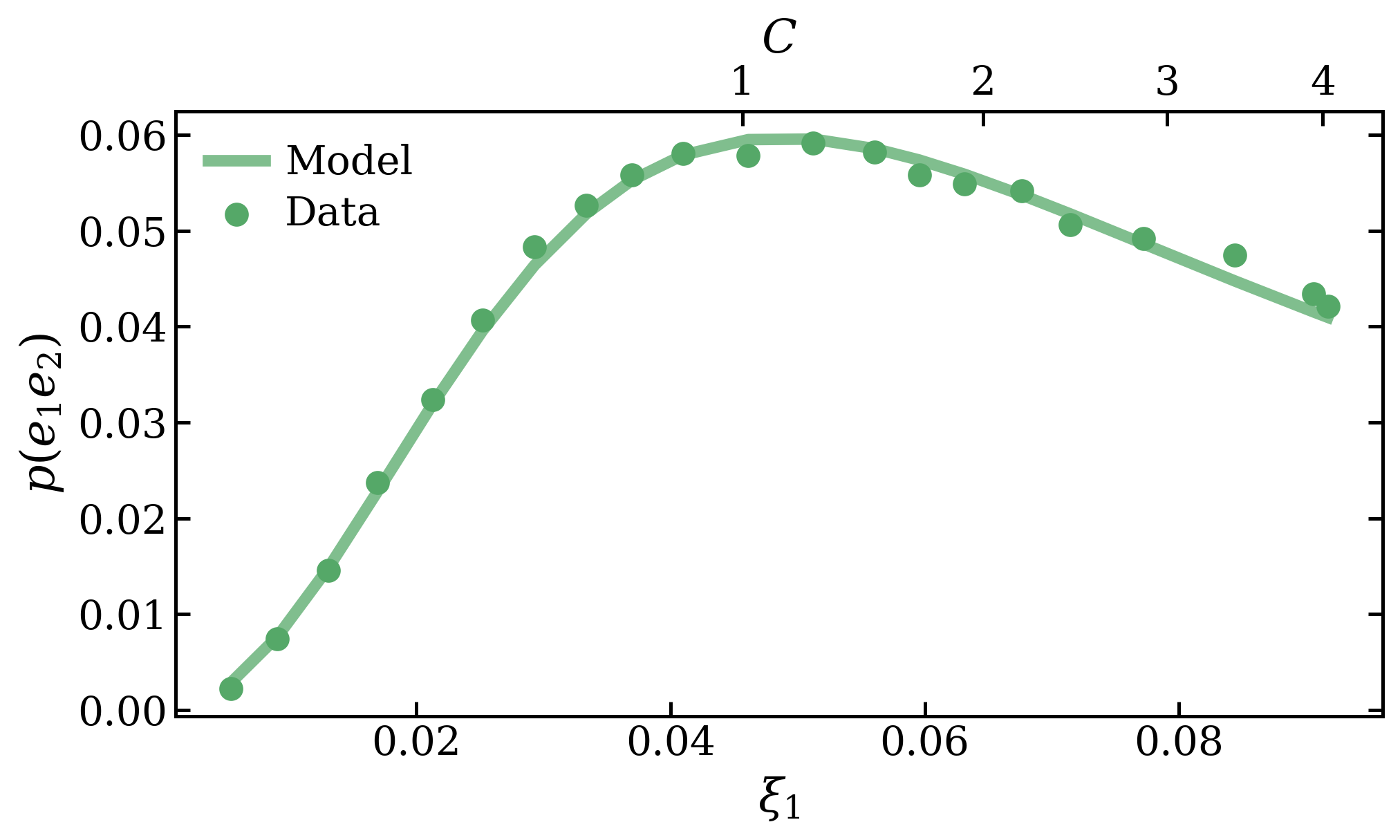}
    \caption{Measured (full green circles) correlated excitation probability $p(e_1e_2)$ as a function of $\xi_1$, with $\xi_2$ kept constant. Full green line is the three-coupled mode model prediction, from which the cooperativity $C$ can be extracted. The highest probability measured is reached around $C=1$, corresponding to $\beta_\xi=0.6$ and $\abs{\xi_p}=0.08$.}
    \label{fig: eff vs C cSMPD2}
\end{figure}

A second method consists in first measuring the qubit frequencies as a function of $\xi_{1,2}$, and using these calibrated values to directly obtain the optimal pump frequency shifts. This avoids performing two-dimensional pump frequency sweeps for each set of values $\xi_{1,2}$, and is therefore much faster. This method was applied in device cSMPD2. \cref{fig: eff vs C cSMPD2} shows the resulting $p(e_1e_2)$ as a function of $\xi_1$, while keeping the $\xi_2$ constant. The agreement with \cref{eq: op. eff. cSMPD N=2} is quantitative, supporting the validity of the three-mode-model. In the following, we set $\beta_\xi$ so that the cooperativity is $C=1$. This can be achieved independently of $\xi_p$, which controls the detector bandwidth, as demonstrated in the following.

\subsection{Cyclic operation}

Having determined the optimal pump parameters, we demonstrate cyclic operation of the cSMPD. The cycle is schematically described in \cref{Fig1}(c). It is the two-qubit version of the single-qubit SMPD cycle described in \cite{balembois_cyclically2024,pallegoix_smpd2024}. Parameters used for each device are reported in \cref{table: circuit-params-part2-merged}. A count is recorded if and only if both qubits are detected in their excited state. Resulting time traces are shown in \cref{figMAIN: fig4}(a). By recording these traces multiple times, the average number of counts per second, $\langle \dot{N} \rangle$ can be obtained, which allows characterizing the cSMPD properties as explained in the next section.

\section{Cascaded SMPD properties}
\label{section: cSMPD detectors metrics exp.}

We now characterize the cSMPD efficiency $\eta$, detection bandwidth $\kappa_d$, and dark count rate $\alpha$. In all the measurements reported below, the pump frequencies are tuned at their optimal values, and the pump amplitude ratio is tuned so that $C=1$ at $\xi_p = \SI{0.08}{}$, as explained in the previous section.

\subsection{Efficiency}
\label{section: photon counting benchmark}

\begin{figure}[h!]
    \centering
    \includegraphics[width=8.4cm]{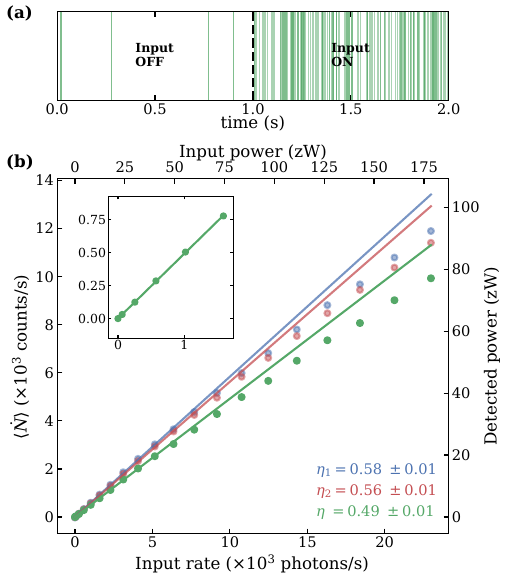}
    \caption{\textbf{Individual and Cascaded detector efficiencies calibration (cSMPD2 data).} The data has been taken for $\xi_p$ of $\SI{0.08}{}$ and $\beta_{\xi}$ of $\SI{0.6}{}$. (a) Count time trace without (left) or with (right) input photons to detect. (b) Measured click rate of Q1 (blue dots), Q2 (red dots), and both qubits (green dots), vs. applied photon flux on the buffer ($1$zW = $10^{-21}$ W = -180 dBm). Solid lines are the linear response fits at low power, the slopes indicating operational efficiencies. Inset: zoom at low photon numbers showing  $\eta = \SI{0.49 \pm 0.01}{}$, $\eta_1 = \SI{0.58 \pm 0.01}{}$, and $\eta_2 = \SI{0.56 \pm 0.01}{}$.}
    \label{figMAIN: fig4}
\end{figure}

To measure the cSMPD efficiency $\eta$, we record the number of counts per second $\langle \dot{N} \rangle$ as a function of the number of photons per second incoming at the cSMPD input. The latter is calibrated using Q1 as a probe, as explained in refs.~\cite{lescanne_irreversible_2020,balembois_cyclically2024,pallegoix_smpd2024} and in \cref{Appendix_InputPowCalibration}. As already explained, a count is defined as a detection of the two qubits in their excited states. An example curve is shown in~\cref{figMAIN: fig4} for $\omega = \omega_b$, and pump frequencies and amplitudes set to their optimal values to maximize efficiency (see below), using device cSMPD2. In this figure, we also plot the independent detection rate of Q1 and Q2. For all three curves, the low-power region is well fitted by a linear response, while saturation is observed at higher power ($>10^4$ photon/s). 

The cSMPD2 efficiency is found to be $\eta=0.49 \pm 0.01$. This number is comparable to efficiencies reached with single-qubit SMPDs~\cite{balembois_cyclically2024,pallegoix_smpd2024}, confirming that the cascaded SMPD process can be achieved with high efficiency. The efficiency computed using the independent qubit excited states detection is $0.58 \pm 0.01$ for Q1 and $0.56 \pm 0.01$ for Q2. We note that the cSMPD efficiency $\eta$ is larger than the product of the independent detection rates, as expected for a correlated process. Similar results are achieved for cSMPD1, although the maximum efficiency reached is significantly lower ($0.25$) due to shorter qubit relaxation times and degradation of these relaxation times under the action of the pump tones (see \cref{Appendix_high_order_RWA}).

\begin{figure}[t!]
    \centering
    \includegraphics[width=8.4cm]{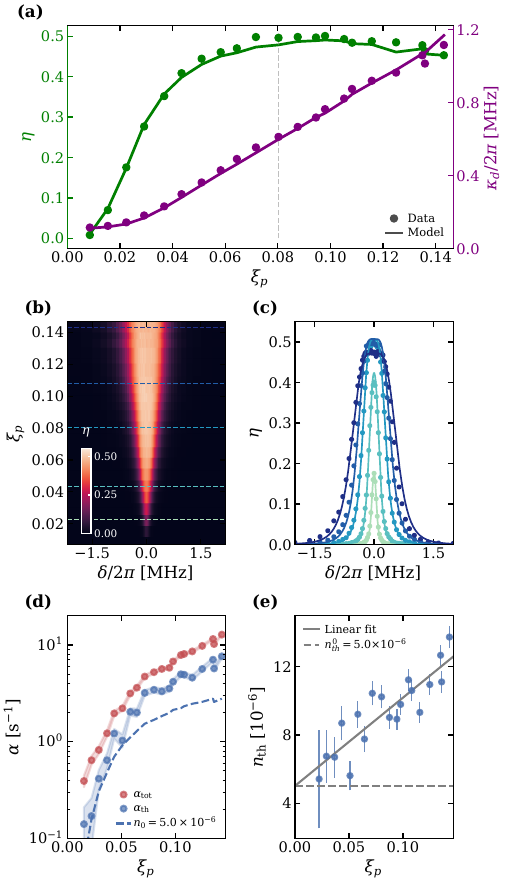}
    \caption{\textbf{Detector performance versus pump amplitudes (cSMPD2 data).} (a) Cascaded efficiency (green dots) and detector bandwidth $\kappa_d/2\pi$ (purple dots) versus $\abs{\xi_p}$. The lines are the three-coupled-mode model prediction with no fitting parameters [see \cref{appendix: T1s under pump}, \cref{appendix: 4wmcal under pump}, and \cref{table: circuit-params-part2-merged} for parameter details]. The dashed gray lines represent the optimum working point at $\abs{\xi_p}=0.08$. (b) Cascaded efficiency as a function of $\abs{\xi_p}$ and input drive frequency detuning $\delta/2\pi$. The colored dashed lines indicate the slices shown in Panel~(c) for $\abs{\xi_p} = \lbrace 0.02,\ 0.04,\ 0.08,\ 0.12,\ 0.15 \rbrace$. (c) Slices from Panel~(b) (dots) together with the three-coupled-mode model (lines, no fitting parameters). (d) Total dark-count rate (red dots) and its thermal contribution (blue dots) versus $\abs{\xi_p}$. The blue dashed line is a model of the thermal contribution (\cref{alpha_th}) with the interpolated thermal photon population with pumps switched off, namely $n_\mathrm{th} = 5\times10^{-6}$; for higher amplitudes the cryostat temperature increased, consequently increasing the dark-count rate [see main text and \cref{appendix: fridge heating under strong pumps} for temperature details]. (e) Estimated average number of thermal photons per mode versus $\abs{\xi_p}$.}
    \label{figMAIN: fig3}
\end{figure}

We now study $\eta$ as a function of pump amplitude $\abs{\xi_p}$ [see \cref{figMAIN: fig3}(a)]. We observe that $\eta$ increases with $\abs{\xi_p}$ until it saturates at $0.5$. This behavior is well reproduced by a model taking into account the buffer internal losses, the finite qubit and memory damping rates as well as their measured dependence on $\abs{\xi_p}$ [see \cref{appendix: cSMPD2 extended dataset} for extended dataset]. As explained in \cref{section: 3-mode model main}, the 4WM rate $4\abs{g_{4,2}}^2/\kappa_w$ is proportional to $\abs{\xi_p}^2$; this explains why low efficiency is observed when $4\abs{g_{4,2}}^2/\kappa_w$ is smaller than the memory damping rate $\kappa_m$. The efficiency $\eta$ reaches a plateau when $4\abs{g_{4,2}}^2/\kappa_w > \kappa_m$. It even somewhat decreases, due to increased memory and qubit damping rates under stronger pump amplitudes. 

Owing to the quantitative agreement with the model, we can identify that the limiting factor of $\eta$ are mainly finite duty cycle ($\eta_\mathrm{cycle} = 0.85$), and finite qubit relaxation times ($\eta_\mathrm{q}=0.83$), as well as the buffer internal losses ($\eta_b = \SI{0.76}{}$). Further improvements in fabrication~\cite{bal_systematic_2024, bland_2d_2025} make efficiencies of $0.8$ or higher likely achievable with cSMPDs in the near future.

\subsection{Bandwidth tunability}
    \label{section: detection bandwidth}

We now demonstrate that the cSMPD bandwidth can be tuned by varying the pump amplitude $\abs{\xi_p}$ [see \cref{section: 3-mode model main}]. For that, we measure $\eta$ as a function of the detuning $\delta$ between the probe frequency $\omega$ and the buffer frequency $\omega_b$. The probe signal is pulsed synchronously with the pump tones. 

A series of curves $\eta(\delta)$ are shown in \cref{figMAIN: fig3}(b) for various values of pump amplitudes $\xi_p$, using device cSMPD2. The bandwidth increase with $\abs{\xi_p}$ is evident. At low values of $\abs{\xi_p}$, the shape of the response is well fitted by a Lorentzian. At higher pump amplitude, the curve broadens and deviates from the Lorentzian shape. All the curves are quantitatively fitted with the expressions derived in the 3-coupled-mode model, taking into account the efficiency reduction caused by transients (see \cref{sec: ope eff} and \cref{appendix: Coupled mode models}). The Full-Width-Half-Maximum detection bandwidth $\kappa_d$ is plotted as a function of $\abs{\xi_p}$ in \cref{figMAIN: fig3}(a), together with the model prediction (solid line). The agreement is quantitative.

In these measurements, the cSMPD detection bandwidth $\kappa_d/2\pi$ was tuned from $\sim 100$\,kHz to $\sim 1$\,MHz, similar to what is achieved in single-qubit SMPD using a frequency-tunable Purcell filter~\cite{pallegoix_smpd2024}.  Note that increasing the pump amplitude beyond what is shown in~\cref{figMAIN: fig3} would not lead to much larger bandwidth than maximum value reached here, as the device response would deviate more and more from a single peak (rather approaching a sum of $3$ Lorentzian peaks \cite{amay_thesis}).

\subsection{Dark counts}
\label{subsection: Dark counts}

We now characterize the cSMPD dark counts. For that, counting traces are measured without sending a probe tone. From these traces, we extract the cSMPD (correlated $e_1e_2$ events) as well as single-qubit dark-count rates. In what follows, we discuss in-depth the results obtained with cSMPD2 biased at its maximum efficiency (i.e. $C=1$ and $\abs{\xi_p} = 0.08$ leading to $\eta = 0.5$, and $\kappa_d/2\pi = 0.6$\,MHz). Similar results were obtained with cSMPD1. Note that due to the low number of counts in the cSMPD data, long measurement times are necessary to obtain statistically-significant results. The data shown in \cref{tab:cSMPD2_counts} and \cref{figMAIN: DC breakdown} were obtained with typically $135$ s-long traces. 

\subsubsection{Dark count budget}

The total cSMPD2 measured dark count rate is $\alpha = 5.2 \pm 0.2 \, \mathrm{s}^{-1}$. In the same conditions, the single-qubit rates are much higher ($200-300\, \mathrm{s}^{-1}$). This establishes that in this device single-qubit dark counts are dominated by uncorrelated qubit spurious excitation events, as anticipated in \cref{subsec: SMPD noise}. Those are efficiently suppressed by correlated qubit detection. The resulting dark count rates are the lowest reported for microwave photon counting, which demonstrates the high potential of the cSMPD design for low-noise microwave detection.

Further insight into the dark count contributions is obtained with complementary measurements. By purposely applying the pump frequencies at a large detuning from the cascaded 4WM optimal frequencies while keeping their amplitudes unchanged, we suppress the thermal photon detection contribution. The resulting dark counts are then equal to the false positive contribution, $\alpha_\mathrm{err}$. We obtain $\alpha_\mathrm{err} = 1.8 \pm 0.1\, \mathrm{s}^{-1}$, from which we infer that $\alpha_\mathrm{th} = \alpha - \alpha_\mathrm{err} = 3.4 \pm 0.2\, \mathrm{s}^{-1}$. We observe that $\alpha_\mathrm{th}$ is larger than $ \alpha_\mathrm{err}$, implying that cSMPD2 dark counts are indeed dominated by the thermal photon contribution as anticipated [see \cref{subsec: SMPD noise}]. On the other hand, it is interesting to note that the single-qubit false positive rate is not measurably different from the total dark count contribution, as expected given the small value of $\alpha_\mathrm{th}$ compared to single-qubit false positives. This demonstrates that cSMPDs are capable to measure much lower microwave powers than state-of-the-art single-qubit SMPDs. 

\subsubsection{False positives contribution}

We now study in greater detail the false positive contribution, $\alpha_\mathrm{err}$. From the measured single-qubit false positive rate, we can infer the expected rate if those events are uncorrelated, using \cref{eq:UncorrelatedErrors}. We find an uncorrelated rate of $1.0 \, \mathrm{s}^{-1}$. The measured cSMPD false positive rate is larger ($1.8 \, \mathrm{s}^{-1}$), indicating a possible contribution of correlated false positives at an approximate rate of $0.8 \mathrm{s}^{-1}$, possibly caused by high-energy impacts or mechanical noise ~\cite{vepsalainen_impact_2020, mcewen_resolving_2022,mcewen_resisting_2024,kurilovich_correlated_2025, larson_quasiparticle_2025, valenti_spatial_2025, yang_high-temporal-resolution_2026, celi_measuring_2026}. Overall, in the present device, cSMPD false positives appear to be still dominated by uncorrelated single-qubit errors, although future devices might reach the limit imposed by correlated qubit errors.

Further insight is obtained by measuring the cSMPD dark counts with pumps switched off. We measure $\alpha_\mathrm{qp}+\alpha_\mathrm{ROs} = 0.6 \pm 0.1\,\mathrm{s}^{-1} $, implying that $\alpha_\mathrm{pumps} = 1.2 \pm 0.1 \, \mathrm{s}^{-1}$. Pump-induced qubit excitations are therefore the main source of cSMPD errors in this device. Pump-induced heating is also visible in the mixing chamber temperature (see \cref{appendix: fridge heating under strong pumps}).

The single-qubit rates are also strongly reduced by switching the pumps off. Applying \cref{eq:UncorrelatedErrors}, we would expect 2-qubit excitation events at the level of $\SI{0.2}{\per\s}$ from the measured single-qubit rates, significantly lower than the measured value. These figures indicate that the residual pump-off errors are in fact dominated by correlated qubit excitation events, at the level of $\SI{0.4}{\per\s}$, close to the number obtained in presence of the pumps. This observation is compatible with the residual correlated excitations being due to high-energy impacts or mechanical vibrations, and not to the application of the pump tones. The readout-induced correlated spurious excitation rate $\alpha_\mathrm{ROs}$ is likely very close to zero, although we cannot prove it rigorously.

In complement, a dark-count budget analysis has been conducted for various values of $\abs{\xi_p}$, the result being presented in \cref{appendix: dck vs pump amplitudes}. It overall confirms the validity of the above analysis for $\abs{\xi_p}=0.08$. An additional observation is that at much higher pump powers, the rate of correlated false positive events increases significantly. We attribute those to the Joule heating induced by the pumps, manifested by a sizeable increase of mixing-chamber temperature, giving rise to a variety of thermally-assisted processes causing false positives (see \cref{appendix: fridge heating under strong pumps} and \cref{appendix: dck vs pump amplitudes}).

In summary, the false-positive rate $\alpha_\mathrm{err}=\SI{1.8}{\per\s}$ originates for a large part from the residual contribution of uncorrelated single-qubit spurious excitations induced by the pumps, and from spurious correlated qubit excitations likely caused by phonon-induced quasiparticles.

\begin{table}[h!]
    \centering
    \setlength{\tabcolsep}{8pt}
    \caption{\textbf{Individual and correlated dark count budgets (cSMPD2 data).} Values are given at the optimal working point $\abs{\xi_p} = 0.08$.}
    \label{tab:cSMPD2_counts}
    \begin{tabular}{cccc}
        \toprule
         & $\alpha$ [$\mathrm{s}^{-1}$] & $\alpha_\mathrm{err}$ [$\mathrm{s}^{-1}$] & $\alpha_\mathrm{qp} + \alpha_\mathrm{RO}$ [$\mathrm{s}^{-1}$] \\
        \midrule
        Correlated & 5.2 $\pm$ 0.2   & 1.8 $\pm$ 0.1   & 0.6 $\pm$ 0.1  \\
        $Q_1$      & 316 $\pm$ 2     & 305 $\pm$ 2     & 149 $\pm$ 1    \\
        $Q_2$      & 220 $\pm$ 1     & 243 $\pm$ 2     & 85 $\pm$ 1     \\
        \bottomrule
    \end{tabular}
\end{table}

\begin{figure}[t!]
    \centering
    \includegraphics[width=0.485\textwidth]{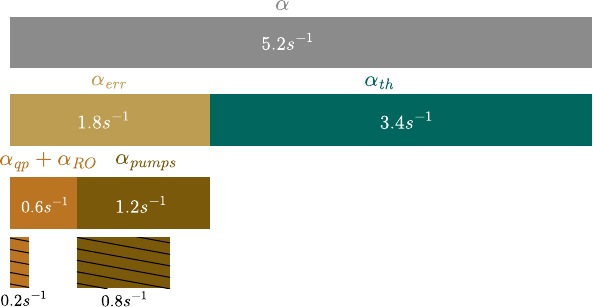}
    \caption{\textbf{Dark count rate breakdown (cSMPD2 data).} Values are given at the optimal working point $\abs{\xi_p}=0.08$. The measured total dark count rate $\alpha$ (gray) is divided into two contributions, $\alpha_\mathrm{err}$ (gold) and $\alpha_\mathrm{th}$ (green). The intrinsic dark count rate is divided into two measured contributions: $\alpha_\mathrm{qp}+\alpha_\mathrm{RO}$ (light brown) and $\alpha_\mathrm{pumps}$ (dark brown). The expected contributions of uncorrelated events (see \cref{eq:UncorrelatedErrors}) are represented with the same colors and stripes. }
    \label{figMAIN: DC breakdown}
\end{figure}

\subsubsection{Thermal photons contribution}

We now discuss the measured thermal photon contribution $\alpha_\mathrm{th}$. Using \cref{alpha_th}, we estimate the number reported above corresponds to a mean photon number per mode $n_\mathrm{th,b} = \SI{1e-5}{}$. Further insight is gained by repeating the measurements for various pump amplitude $\xi_p$. As shown in \cref{figMAIN: fig3}(d), $n_\mathrm{th,b}$ increases with $\xi_p$ approximately linearly, probably due to Joule heating in the lines. The extrapolation at $\xi_p=0$ indicates $n_\mathrm{th,b} = \SI{5e-6}{}$. Such low occupation number can only be measured owing to the cSMPD sensitivity, and is the lowest ever reported to our knowledge in circuit quantum electrodynamics. This number is approximately two orders of lower than previous measurements performed with SMPDs. This reduction is very likely linked to the higher buffer frequency $\omega_b$, from $\sim 7.5$\,GHz to $11.7$\,GHz in this device.

It can be converted to an effective temperature of the microwave field in the input line of $T_\mathrm{eff}=$\, \SI{46}{\milli\K}, significantly higher than the cryostat base temperature. Part of it (approximately a third) can be traced to insufficient isolation in the qubit readout output line, whereas the rest may be due to imperfect thermalization of various microwave components. Further discussion of this topic can be found in \cref{section: app thermometry}.

\subsection{Power sensitivity}

We now briefly summarize the cSMPD overall performance. At the expense of a slightly reduced efficiency compared to single-qubit SMPDs ($0.5$ compared to $\sim 0.75$), the cSMPD2 device achieves a large reduction in dark count rate ($5\,\mathrm{s}^{-1}$ compared to $\sim 40\,\mathrm{s}^{-1}$), while maintaining a sizeable detection bandwidth of $600$\,kHz. These numbers correspond to a power sensitivity of \SI{3.6e-23}{}\wpsrh at 11.72 GHz. An even higher power sensitivity is achieved by choosing a lower pump power, leading to slightly reduced efficiency but also lower dark count rates. A value of \SI{2.4e-23}{}\wpsrh is reached at $\xi_p = 0.04$, which is the highest reported for microwave photon counting (both working points are described in \cref{table: circuit-params-part2-merged}).

The remaining detector noise is dominated by the thermal photon contribution, which can likely be further reduced by better thermalization or simply slightly increasing the frequency. For instance, in the same conditions as demonstrated here, we anticipate that the thermal contribution would be suppressed by a factor 30 if the buffer frequency was at \SI{15}{\giga\Hz}. Reducing single-qubit errors will necessitate minimizing Joule heating in the pump lines, and improving the infrared shielding \cite{lin_suppression_2026}. Overall, we believe that cSMPD devices are amenable to power sensitivity better than \SI{1e-23}{}\wpsrh in the near future. Further improvements will require reducing correlated qubit spurious excitation events \cite{mcewen_resisting_2024, harrington_synchronous_2025}.

     \subsection{Temperature dependence}
     \label{sec: temperature effect cSMPD1}

\begin{figure}[h]
    \centering    
    \includegraphics[width=8.4cm]{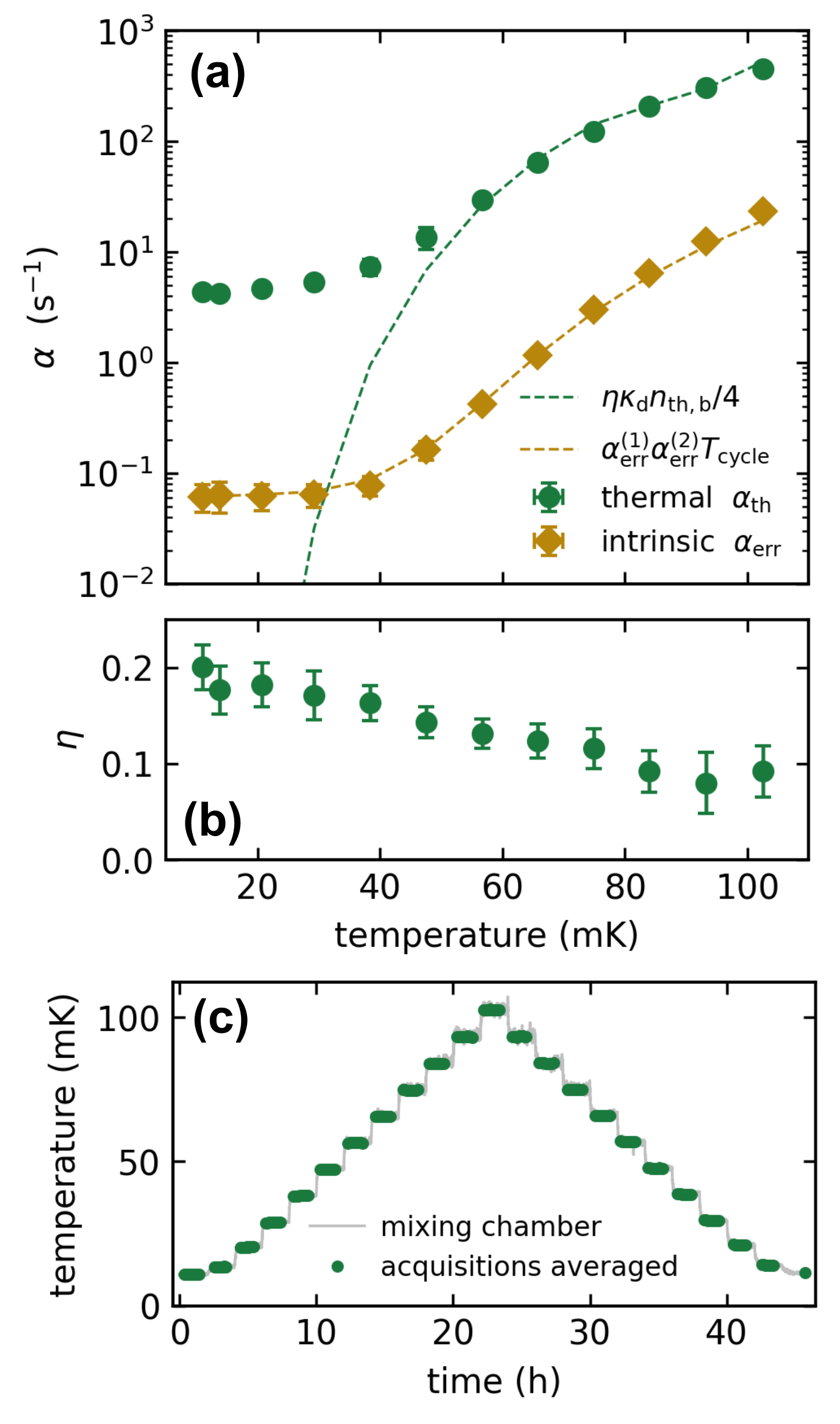}
    \caption{\textbf{Temperature dependence of the correlated dark count rate and efficiency (cSMPD1).} (a) Correlated dark count rates versus mixing chamber temperature: thermal contribution $\alpha_\mathrm{th}$ (green circles) and intrinsic contribution $\alpha_\mathrm{err}$ (mustard diamonds). The thermal contribution is obtained by subtraction, $\alpha_\mathrm{th} = \alpha - \alpha_\mathrm{err}$. Both rates saturate below $\sim\SI{40}{\milli\kelvin}$. Dashed green line: thermal dark count model~\cref{alpha_th}, evaluated with  $\kappa_\mathrm{d} = 2\pi\times\SI{220}{\kilo\hertz}$. Dashed mustard line: accidental coincidence model~\cref{eq:UncorrelatedErrors}, computed from the single qubit rates measured with detuned pumps from the cascaded 4WM resonance. (b) Cascaded two qubit efficiency. (c) Mixing chamber temperature over the 46h run (grey line); green dots mark the acquisitions retained for averaging. }
    \label{fig: temperature sweep}
\end{figure}

We finally study the cSMPD performance as a function of temperature, between \SI{11}{\milli\K} and \SI{93}{\milli\K}. The cryostat was stepped through successive setpoints and back down to base, dwelling two hours at each so that the microwave components could approach thermal equilibrium. Over the resulting \SI{46}{\hour} run, 391 measurements were recorded at \SI{7}{\minute} intervals, yielding 22 plateaux of $\sim$\SI{1.5}{\hour}. Each setpoint is therefore visited twice, once on warming and once on cooling; pooling the two branches gives the 11 points shown in \cref{fig: temperature sweep}, spanning \SI{11}{\milli\K} to \SI{93}{\milli\K}. Data were acquired with cSMPD1. The pump frequencies were chosen at the optimal values, and the amplitudes were chosen such that $C=1$ and $\xi_p$ such that the efficiency was maximum. At this working point, the detector bandwidth was set to $\kappa_d/2\pi = \SI{220}{\kilo\Hz}$.

As described in \cref{section: cSMPD detectors metrics exp.}, we measure the detector efficiency $\eta$, the total dark count rate $\alpha$, the false positive rate $\alpha_\mathrm{err}$, and compute $\alpha_\mathrm{th}=\alpha - \alpha_\mathrm{err}$ as a function of temperature. At $10$\,mK, $\alpha_\mathrm{th} \gg \alpha_\mathrm{err}$ as already discussed for cSMPD2, and $\eta=0.2$, lower than in cSMPD2 due to shorter qubit relaxation times. At higher temperature, $\eta$ decreases due to shorter qubit relaxation times, but remains close to $\sim 0.1$ up to $100$\,mK. 

The thermal contribution $\alpha_\mathrm{th}$ increases with temperature. The measurements are in quantitative agreement with the predictions of \cref{alpha_th}, as seen in \cref{fig: temperature sweep}(a). The background thermal noise at the lowest temperature corresponds to $n_\mathrm{th,b} = \SI{6.3e-5}{} $ (\SI{43.6}{\milli\K}). 

The false positive rate $\alpha_\mathrm{err}$ is also temperature dependent, as expected from the thermally assisted spurious excitation of the individual qubits. Computing the accidental coincidence rate from the single qubit data alone, \cref{eq:UncorrelatedErrors} reproduces the measured $\alpha_\mathrm{err}$ quantitatively over nearly three decades. The intrinsic dark counts of cSMPD1 are therefore entirely accounted for by uncorrelated single qubit errors, with no evidence of the additional correlated mechanism observed in cSMPD2. Panel (a) highlights a two-orders-of-magnitude difference between the dominant thermal dark count rate, and the intrinsic dark count rate at the cascaded level for cSMPD1 ($\omega_b/2\pi=\SI{8.798}{\giga\Hz}$). The extrapolation of the Bose-Einstein contribution to $\alpha_\mathrm{th}$ (green dashed line) intersects the intrinsic dark count background (mustard dotted line) near \SI{30}{\milli\kelvin}. This temperature provides an estimate below which intrinsic dark counts would start to contribute significantly to the dark count budget with microwave photons at thermal equilibrium.

This measurement demonstrates the ability of the cSMPD to perform power detection over a temperature range spanning from base temperature to \SI{100}{\milli\K}, with the lower bound limited by microwave line noise and the upper bound by the reduction in qubit relaxation times.

\begin{figure*}[t]
    \centering
    \includegraphics[width=\textwidth]{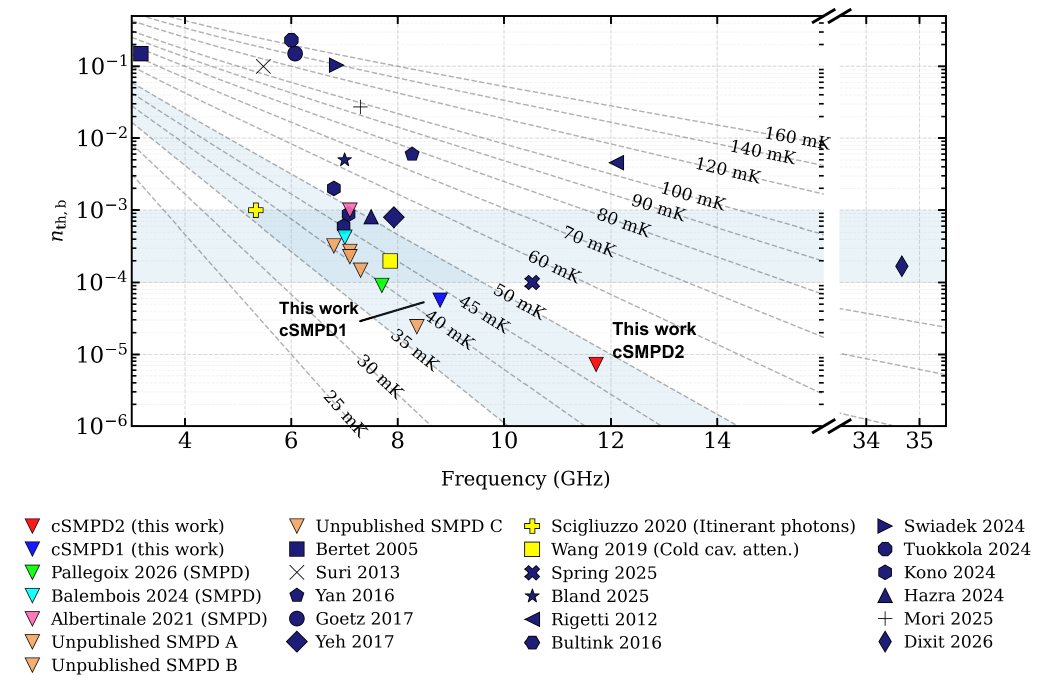}
    \caption{\textbf{Average number of photon in a cavity / per mode in a waveguide versus frequency.} The photon numbers reported by Bertet \textit{et al.}\cite{bertet_dephasing_2005}, Yan \textit{et al.} \cite {yan_flux_2016}, Yeh \textit{et al.} \cite{yeh_microwave_2017}, Wang \textit{et al.} \cite{wang_cavity_2019}, Bland \textit{et al.} \cite{bland_2d_2025}, Spring \textit{et al.} \cite{spring_fast_2025}, Bultink \textit{et al.} \cite{bultink_active_2016}, Swiadek \textit{et al.} \cite{swiadek_enhancing_2024}, Tuokkola \textit{et al.} \cite{tuokkola_methods_2024}, Kono \textit{et al.} \cite{kono_mechanically_2024}, Hazra \textit{et al.} \cite{hazra_benchmarking_2025}, Dixit \textit{et al.} \cite{dixit_millimeter_2026}, Mori \textit{et al.} \cite{mori_suppression_2025} (dark blue) are either directly reported or computed based on the induced dephasing rate measured with an $T_{2,\mathrm{echo}}$ or CPMG pulse sequence.  We also report two different methods to measure the average number of thermal photon per mode (itinerant photons): (i) based on scattering of the thermal field on a qubit as in Scigliuzzo \textit{et al.} \cite{scigliuzzo_primary_2020} (yellow cross), (ii) based 4WM-based SMPD in Albertinale \textit{et al.} \cite{albertinale_detecting_2021}, Balembois \textit{et al.} \cite{balembois_cyclically2024}, Pallegoix \textit{et al.} \cite{pallegoix_smpd2024} and this work (colored triangles pointing down). We included data from unpublished SMPD samples. Recent data tends to accumulate in the $10^{-3}-10^{-4}$ region and along the diagonal representing $\sim\SI{40}{\milli\K}$ (pale blue areas).}
    \label{figMAIN: fig8}
\end{figure*}

\section{Application to primary microwave thermometry}
\label{section: app thermometry}

One application of microwave photon counting is primary microwave thermometry, by measurement of the occupation number per mode $n_\mathrm{th}$. This topic is relevant to astrophysics and quantum computing. 

In astrophysics, high-resolution x-ray spectroscopy is commonly achieved by detecting the temperature increase of an absorber. While Transition-Edge Sensors are generally used, microwave thermometry may provide an alternative method.

In superconducting quantum computing, thermal noise has been shown to induce qubit dephasing via uncontrolled frequency shifts~\cite{bertet_dephasing_2005,gambetta_qubit-photon_2006}. It is therefore desirable to monitor the microwave temperature of the many coaxial cables reaching the sample. This phenomenon has so far been poorly studied, by lack of a sufficiently sensitive detector. Usual methods to characterize microwave mode occupation number rely on superconducting qubits. In one class of experiments, the qubit is dispersively coupled to a resonator, itself coupled to the line of interest; fluctuations of the photon number in the cavity cause extra qubit dephasing which can be quantitatively linked to $n_\mathrm{th}$~\cite{wang_cavity_2019} (a phenomenon used in this work to calibrate the photon rate at the SMPD input). In a second class of experiments, the qubit is directly inserted in the line of interest so that it resonantly exchanges energy with the microwave bath; microwave transmission determines its average excited state population which mirrors $n_\mathrm{th}$ \cite{scigliuzzo_primary_2020}. Both methods lack time resolution and sensitivity, being unable to reliably measure occupation numbers below $\sim 10^{-4}$ at best.

cSMPD devices, on the other hand, are well-suited for such studies. As demonstrated in this work, cSMPDs can readily access occupation numbers as low as $n_\mathrm{th}\sim 10^{-6}$. As a primary thermometer, the power sensitivity is $S_\mathrm{err}=\hbar \omega \sqrt{ \alpha_\mathrm{err} }/\eta $. It reaches $\SI{1.6e-23}{\W} /\sqrt{\mathrm{Hz}}$ for cSMPD1, $\SI{2.1e-23}{\W} /\sqrt{\mathrm{Hz}}$ for cSMPD2 at the best efficiency working point and $\SI{1.7e-23}{\W} /\sqrt{\mathrm{Hz}}$ for cSMPD2 at the best sensitivity working point. 

To obtain insight into the physical origin of the low-temperature residual microwave photon occupation number, we have compiled $n_\mathrm{th}$ measurements reported in the literature or inferred from published qubit coherence data, as well as measurements from our laboratory using SMPD and cSMPD devices (some unpublished). The plot shows $n_\mathrm{th}$ as a function of the detection frequency (see \cref{figMAIN: fig8}). Several features can be noted. First, microwave temperatures are systematically higher than the base cryostat temperature of $10$\,mK (the lowest reported temperature being $37$\,mK). Second, the lowest values of $n_\mathrm{th}$ strongly decrease with increasing frequency, as expected for a black-body radiation source. These observations suggest that despite present-day best efforts, the thermalization of coaxial cables inner conductors and of microwave components at the mixing-chamber plate is not perfect, resulting in microwave emission from a black-body with an effective temperature of $\sim 40 \pm 5$\,mK. One simple method for minimizing $n_\mathrm{th}$ is therefore to increase the operation frequency, as demonstrated by the cSMPD2 device which achieves the lowest reported value of $n_\mathrm{th}$ by operating at slightly higher frequency than previous devices.

\section*{Conclusion}

In conclusion, we have introduced a new microwave photon counter architecture, the cascaded Single-Microwave Photon Detector (cSMPD). The cSMPD encodes photon detection on the state of two transmon qubits, facilitating rejection of uncorrelated single-qubit errors. The detection bandwidth is moreover tunable with the pump power. We demonstrate false-positive error rates at the $\SI{1}{\per\s}$ level, while retaining detection efficiencies as high as $0.5$ in the best devices. Combined with operation at a frequency sufficiently high to minimize the thermal contribution, this results in a record power sensitivity for microwave signals. This detector will likely find applications in single-spin magnetic resonance, dark-matter detection, and primary thermometry.

\section*{Data availability}
The data supporting the findings of this study, as well as the code used for the analysis, are available upon reasonable request.

\section*{Acknowledgements}

We acknowledge technical support from P.~S\'enat, D. Duet, P.-F.~Orfila and S.~Delprat. We are grateful for fruitful discussions within the Quantronics group, A\&B teams, K. Mølmer, Y. Nojiri, F. Rautschke and S. Lopes. We acknowledge support from the Agence Nationale de la Recherche (ANR) through the MOLEQUBE (ANR-23-CE47-0011) project. We acknowledge support of the R\'egion Ile-de-France through the DIM QUANTIP, from the AIDAS virtual joint laboratory, and from the France 2030 plan under the ANR-22-PETQ-0003 grant. This project has received funding from the European Union Horizon 2020 research and innovation program under the project OpenSuperQ100+, and from the European Research Council under the grant no. 101042315 (INGENIOUS). We acknowledge IARPA and Lincoln Labs for providing the Josephson Traveling-Wave Parametric Amplifier.

\section*{Author Contributions}
A.S.M., L.S. and E.F. designed the sample. J.S., G.C. and L.P. fabricated the samples. A.S.M. and L.S. performed the experiment with the support of J.O'S. . A.S.M. and L.S. performed the analysis with the support of P.B. and E.F. L.C. performed the high-order RWA analysis. A.S.M. wrote the manuscript with inputs from all co-authors. P.B. and E.F. supervised the work.

%%%%%%%%%%%%%%%%%%%%%%%%%%%%%%%%%%%%%%%%%%%%%
%% HERE START THE SUPPLEMENTARY MATERIAL
%%%%%%%%%%%%%%%%%%%%%%%%%%%%%%%%%%%%%%%%%%%%%
\newpage
\appendix

\section*{SUPPLEMENTARY MATERIAL}

\section{Nanofabrication}
\label{appendix fab}

The sample was fabricated using a 430-\SI{}{\micro\meter}-thick 2-inch c-plane sapphire substrate, with a 200-nm thick tantalum (Ta) layer sourced from STAR Cryoelectronics. The resonators, coplanar waveguides, ground plane, and capacitors were patterned into the Ta layer via direct laser photolithography and Reactive Ion Etching (RIE). The process began with an oxygen plasma treatment to clean the wafer, followed by spin-coating of S1805 photoresist. The wafer was then exposed using a maskless aligner ($\mu$MLA Heidelberg). After development in Microposit MF319 and rinsing with deionized water, the Ta was etched in an RIE system using a fluorine-based plasma.
After the removal of residual resist, the Al/AlO$_x$/Al Josephson junctions were fabricated using the double-angle shadow evaporation technique with the Dolan-bridge-style geometry. The resist mask was composed of a bilayer stack: a methyl methacrylate (MMA) EL13 copolymer resist and a poly(methyl methacrylate) (PMMA) A3 resist. To mitigate charging effects during electron beam writing, a conductive coating of Electra 92 was spin-coated atop the resist layers. Electron beam lithography was conducted at an acceleration voltage of 30 kV to define the Josephson junctions. The anti-charging layer was removed by immersion in water, followed by development of the patterns in a 3:1 IPA:DI water solution at 6°C. Before evaporation, an oxygen plasma treatment was performed to clean the surface underneath the bridge. The Josephson junctions were deposited using a Plassys electron-beam evaporator equipped with a separated load-lock. The wafer was pumped overnight to ensure proper vacuum conditions. The process started with argon ion milling to remove the tantalum oxide, followed by the deposition of a 35-nm layer of aluminum. The chamber was then flooded with $O_2$ to establish a static oxygen background at 20 mbar for 10 minutes to oxidize the aluminum, forming a tunnel barrier of AlO$_x$. A second aluminum layer, 70 nm thick, was then deposited, and a second oxidation step was carried out under $O_2$ at a pressure of 100 mbar for 20 minutes, capping the aluminum with a layer of pure aluminum oxide. The wafer was subsequently immersed in an N-methylpyrrolidone (NMP)-based solvent at 80°C for thorough lift-off, followed by cleaning in acetone, isopropanol, and DI water.
Finally, the wafer was coated with protective resist before being diced into individual chips using a DAD3350 Disco dicer. The chips were then cleaned by sonicating in NMP, acetone, isopropanol, and DI water.

\section{Experimental setup}
\label{wiring and setup}

\begin{figure*}[t]
    \centering
\includegraphics[width=\linewidth]{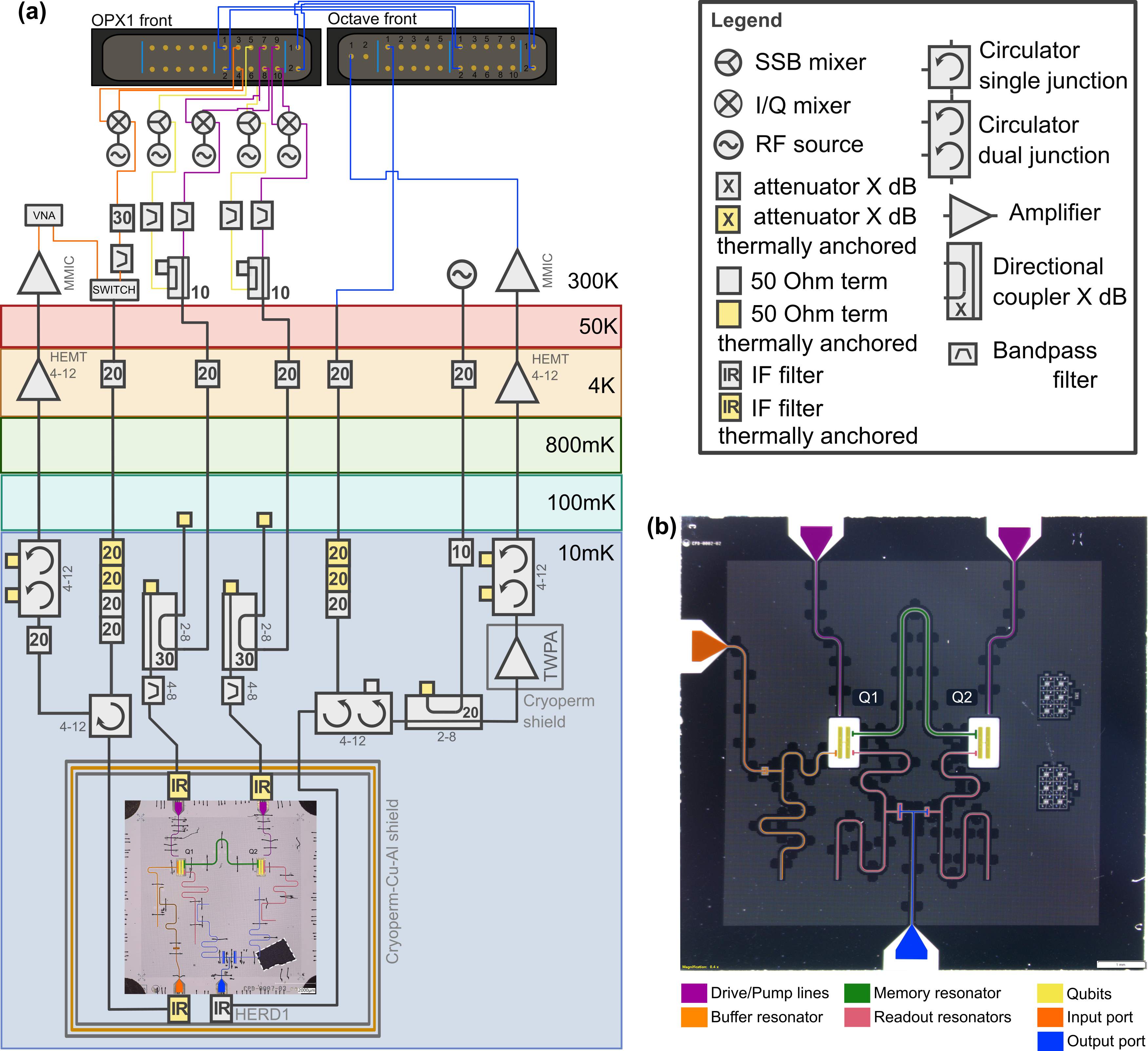}
\caption{\textbf{Experimental setup and cSMPD1 colored micrograph.} (b) See \cref{appendix: cSMPD1} for details.}
\label{fig: wiring of the setup and cSMPD1}
\end{figure*}

See \cref{fig: wiring of the setup and cSMPD1}(a).  The room-temperature setup for the qubit drive lines, qubit pump lines, and buffer drive line consists of five microwave sources, three AWG I/Q pairs, and two single AWG channels. The AWG used is the OPX1 from Quantum Machines. All microwave pulses required for the buffer mode, qubit drives, and pumps in the experiment are generated by mixing the AWG output on I/Q mixers (for the buffer and qubit pumps) or Single Sideband (SSB) mixers (for qubit drives), with their respective high frequency local oscillators (LOs).

The read-out operation integrates the OPX1 with an Octave module (Quantum Machines), which handles modulation, optional RF amplification, demodulation, filtering, and IF amplification. The Octave module was added to the setup after calibration, which is why all qubit and buffer control signals are still managed using discrete microwave components on a microwave table. In addition, a sixth local oscillator is dedicated to pumping the JTWPA.

A single line is used for both driving and off-resonantly pumping individual qubits. These control signals are combined at room temperature using a directional coupler. To ensure signal integrity, each control tone is filtered at room temperature with tunable band-pass filters. Proper filtering of the pump lines is particularly crucial to prevent strong power LO leakage or unwanted spurious sidebands through the I/Q mixers. We employ manually adjustable tunable band-pass filters from the WTBCX7-SS series by Wainwright Instruments GmbH, with bandwidths ranging from 50 to 150 MHz. Power splitting and attenuation of pump tones occur at different temperature stages via directional couplers, with most of the pump power dissipated at higher temperature levels (100 mK and 300 K). Lines directly connected to qubits are terminated with well-thermalized 50$\Omega$ loads attached to the directional couplers.

Read-out is performed using a reflection setup. The input line is connected to a double circulator anchored on the 10 mK flange. The multiplexed readout signal is scattered off the readout resonators and subsequently amplified through a JTWPA (gain $\sim$18 dB) housed within a dedicated magnetic shield. Further amplification is provided by a HEMT amplifier at 4K (gain $\sim$40 dB) and a low noise room temperature amplifier at 300K (gain $\sim$30 dB). Four isolators ensure the qubits remain well shielded from HEMT noise.

The buffer resonator is driven through a carefully filtered and attenuated input line. To facilitate characterization with a Vector Network Analyzer (VNA), an additional output line is included in the cryogenic setup. At room temperature, the VNA output port is combined with the RF input line via a switch.

The chip is packaged in a proprietary sample holder from Alice \& Bob made of an OFHC copper base and a lid made, by default, with Aluminum 2024 (see \cref{Appendix_sh} for variations in the lid materials and their impact on the performance of the detector). The sample holder is thermally anchored to an OFHC copper plate, enclosed within a three-layer shield (Aluminum/OFHC Copper/Cryoperm). To mitigate infrared radiation, IR filters are installed on all input lines inside the shield. These include a combination of Bluefors IR filter prototypes, Quantum Microwave IR filters, home-made IR filters, and HERD-1 filters (Sweden Quantum).

\section{cSMPD1 design, a Purcell filter-less design}
\label{appendix: cSMPD1}
 
See \cref{fig: wiring of the setup and cSMPD1}(b) legend for circuit elements and \cref{table: CPD-0002-02-A-and-W025-CPD-0007-03-01-C1-circuit-params} for parameter details. 

To ease the design process based on eigenmode simulations in HFSS ANSYS, cSMPD1 was designed without single-pole filters on the buffer and readout modes. Instead, the Purcell filters are implemented as positional notch filters~\cite{sunada_fast_2022}, shown to be efficient when the notch frequency coincides with the qubit mode frequency. In practice, this filtering strategy was not found to be very effective for cSMPD1, for several reasons: (i) if the qubit, the resonator, or both are detuned from the target frequency (within a precision of a few tens of MHz), the filter efficiency is degraded; (ii) the AC Stark shift induced by the 4WM pumps can displace the qubit mode outside the notch, reducing $T_1$ and thus the efficiency; (iii) spurious resonances activated by the 4WM pumps can reduce qubit $T_1$, as these weakly-activated terms in the dynamics are no longer suppressed by the filtering function of a dedicated single-pole filter. The efficiency of cSMPD1 was limited by reduced $T_1$ under 4WM pumping, which motivated the implementation of single-pole filters on the buffer and readout modes in cSMPD2.

Filtering does not protect against nearly resonant conversions driven by the pumps; only the choice of mode and pump frequencies can avoid them. \Cref{Appendix_high_order_RWA} finds several such conversions in cSMPD1, which may have contributed to its reduced lifetimes under pump.

\section{Circuit parameters}
\label{Appendix_paramters}

See \cref{table: CPD-0002-02-A-and-W025-CPD-0007-03-01-C1-circuit-params} and \cref{table: circuit-params-part2-merged}.

\begin{table*}[h!]
    \centering
    \begin{tabular}{|c|c|c|}
        \toprule
        \toprule
        Sample & \textbf{cSMPD1} & \textbf{cSMPD2} \\
        \toprule
        \toprule
        \multicolumn{3}{|c|}{\textbf{Qubit 1}} \\
        \midrule
        \midrule
        $\omega_{q_1}/2\pi$ & \SI{6.614}{\giga\Hz} & \SI{5.578}{\giga\Hz}\\
        $2\chi_{q_1q_1}/2\pi$ & $\sim$ \SI{-240}{\mega\Hz} & \SI{-225}{\mega\Hz}\\
        $\chi_{q_1b}/2\pi$ & \SI{-1.8}{\mega\Hz} & \SI{-5.1}{\mega\Hz}\\
        $\chi_{q_1m}/2\pi$ & $\sim$ \SI{-2}{\mega\Hz} & \SI{-3.4}{\mega\Hz} \\
        $\chi_{q_1r_1}/2\pi$ & \SI{-1.4}{\mega\Hz} & \SI{-1.1}{\mega\Hz} \\
        $T_1$ & \SI{30}{\micro\second} & \SI{55\pm15}{\micro\second} \\
        $T_2^*$ & \SI{44}{\micro\second} & \SI{115}{\micro\second} \\
        $\pth$ & $2.6\times10^{-3}$ & $\SI{5.1e-3}{}$ \\
        $\preset$ (after integrating $T_\mathrm{RO}$) & $4.5\times10^{-4}$ &  $5.1\times10^{-5}$ \\
        $\mathcal{F}_{\mathrm{RO}_1}$ & 0.84 & 0.95 \\
        $\omega_{p_1}/2\pi$ & \SI{5.8946}{\giga\Hz} & \SI{3.7586}{\giga\Hz} \\
        \midrule
        \midrule
        \multicolumn{3}{|c|}{\textbf{Qubit 2}} \\
        \midrule
        \midrule
        $\omega_{q_2}/2\pi$ & \SI{6.284}{\giga\Hz} & \SI{5.503}{\giga\Hz}\\
        $2\chi_{q_2q_2}/2\pi$ & $\sim$ \SI{-240}{\mega\Hz} & \SI{-189}{\mega\Hz}\\
        $\chi_{q_2m}/2\pi$ & $\sim$ \SI{-2}{\mega\Hz} & \SI{-2.3}{\mega\Hz}\\
        $\chi_{q_2w}/2\pi$ & \SI{-1.8}{\mega\Hz} & \SI{-10.1}{\mega\Hz}\\
        $T_1$ & 30-50\SI{}{\micro\second} & \SI{70\pm15}{\micro\second} \\
        $T_2^*$ & \SI{40}{\micro\second} & \SI{112}{\micro\second} \\
        $\pth$ & $3.1\times10^{-3}$ & $\SI{3.2e-3}{}$\\
        $\preset$ (after integrating $T_\mathrm{RO}$) & $3.3\times10^{-4}$ & $1.5\times10^{-4}$ \\
        $\mathcal{F}_{\mathrm{RO}_2}$ & 0.88 & 0.95 \\
        $\omega_{p_2}/2\pi$ & \SI{5.6342}{\giga\Hz} &\SI{3.1903}{\giga\Hz} \\
        \midrule
        \midrule
        \multicolumn{3}{|c|}{\textbf{Readout 1}} \\
        \midrule
        \midrule
        $\omega_{r_1}^{g}/2\pi$ & \SI{7.650}{\giga\Hz} &  \SI{7.902}{\giga\Hz} \\
        $\kappa_{\mathrm{tot}}$ & \SI{3.3e6}{\per\second} & \SI{6.9e6}{\per\s} \\
        \midrule
        \midrule
        \multicolumn{3}{|c|}{\textbf{Readout 2 / Waste}} \\
        \midrule
        \midrule
        $\omega_{r_2}^{g}/2\pi\equiv\omega_{w}^{g}/2\pi$ & \SI{7.462}{\giga\Hz} & \SI{7.594}{\giga\Hz} \\
        $\kappa_{\mathrm{tot}}\equiv\kappa_w$ & \SI{3.4e6}{\per\second} & \SI{2.7e7}{\per\s} \\
        \midrule
        \midrule
        \multicolumn{3}{|c|}{\textbf{Buffer}} \\
        \midrule
        \midrule
        $\omega_{b}^{g}/2\pi$ & \SI{8.798}{\giga\Hz} & \SI{11.722}{\giga\Hz}\\
        $\kappa_{\mathrm{tot}}\equiv\kappa_b$ & \SI{5.8e6}{\per\second} &  \SI{2.3e6}{\per\second} \\
        $\kappa_{b,\mathrm{int}}$ & \textit{not measured} & \SI{5.6e5}{\per\s} \\
        \midrule
        \midrule
        \multicolumn{3}{|c|}{\textbf{Memory}} \\
        \midrule
        \midrule
        $\omega_{m}^{gg}/2\pi$ & \SI{8.095}{\giga\Hz} & \SI{9.910}{\giga\Hz} \\
        $\kappa_m$ & \SI{6.8e4}{\per\second} & \SI{9.0e4}{\per\second} \\
        $T_2^*$ & \textit{not measured} & \SI{21.1}{\micro\s} \\
        \bottomrule
        \bottomrule
    \end{tabular}
    \caption{First column contains cSMPD1 circuit parameters. Second column contains cSMPD2 circuit parameters.}
    \label{table: CPD-0002-02-A-and-W025-CPD-0007-03-01-C1-circuit-params}
\end{table*}

\begin{table*}[t]
    \centering
    \begin{tabular}{|c|c|c|c|}
        \toprule
        \toprule
        & \textbf{cSMPD1 } & \textbf{cSMPD2} & \textbf{cSMPD2} (highest sensitivity) \\
        \midrule
        \midrule
        $S$ & \SI{5.9\pm0.6}{}$\times10^{-23}$ \wpsrh & \SI{3.6\pm0.1}{}$\times10^{-23}$ \wpsrh & \SI{2.4\pm0.1}{}$\times10^{-23}$ \wpsrh\\
        $S_\mathrm{err}$ & \SI{1.6}{}$\times10^{-23}$\wpsrh  & \SI{2.1}{}$\times10^{-23}$\wpsrh  & \SI{1.7}{}$\times10^{-23}$\wpsrh \\
        $\eta$ & \SI{0.25\pm 0.02}{} & \SI{0.5\pm0.01}{} & \SI{0.35\pm0.01}{} \\
        $\alpha$ & \SI{6.4\pm 0.7}{\per\second} & \SI{5.2\pm 0.2}{\per\second} & \SI{1.2\pm 0.1}{\per\second}\\
        $\kappa_d/2\pi$ & \SI{220}{\kilo\Hz} & \SI{600}{\kilo\Hz} & \SI{230}{\kilo\Hz}\\
        $\kappa_d/2\pi$ (tunability range) & 140-240 \SI{}{\kilo\Hz} & $\sim$ 100-1000 \SI{}{\kilo\Hz} & $\sim$ 100-1000 \SI{}{\kilo\Hz}\\
        \midrule
        $\beta_\xi$ & 1.02 & 0.6 & 0.6\\
        $\xi_p$ & 0.07 & 0.08 & 0.04\\
        $g_{4,1}/2\pi$ & \SI{-130\pm5}{\kilo\Hz} & \SI{-200\pm5}{\kilo\Hz} & \SI{-93\pm5}{\kilo\Hz}\\
        $g_{4,2}/2\pi$ & \SI{-125\pm7}{\kilo\Hz} & \SI{-635\pm7}{\kilo\Hz} & \SI{-292\pm7}{\kilo\Hz}\\
        $C$ & $0.62$ & $1.1$ &  $1.1$\\
        $\eta_\mathrm{4WM}$ & 0.95 & 0.96 & 0.84\\
        $\eta_{\mathrm{q}}$ & $0.57$ & 0.83 & 0.84\\
        $\mathcal{F}_{\mathrm{RO}}$ & 0.74 & 0.9 & 0.9\\
        $\alpha_{\mathrm{th}}$ & \SI{5.9\pm 0.7}{\per\second} & \SI{3.4\pm 0.2}{\per\second} & \SI{0.64\pm 0.12}{\per\second}\\
        $\alpha_{\mathrm{err}}$ & \SI{0.5\pm 0.3}{\per\second} & \SI{1.8\pm 0.1}{\per\second} & \SI{0.58\pm 0.07}{\per\second}\\
        $\alpha_\mathrm{qp}+\alpha_{\mathrm{ROs}}$ & $\sim\ 10^{-2}$ \SI{}{\per\second} & \SI{0.6\pm 0.1}{\per\second} & \SI{0.3\pm 0.05}{\per\second}\\
        $\alpha_{\mathrm{pumps}}$ & $\sim$ \SI{0.5\pm 0.3}{\per\second} & \SI{1.2\pm 0.1}{\per\second} & \SI{0.28\pm 0.08}{\per\second}\\
        $T_d$ & \SI{13}{\micro\second} & \SI{11.6}{\micro\second} & \SI{11.6}{\micro\second}\\
        $T_{\mathrm{RO}}$ & \SI{1500}{\nano\second} & \SI{1000}{\nano\second} & \SI{1000}{\nano\second}\\
        $T_{\mathrm{reset}}$ & \SI{128}{\nano\second} & \SI{128}{\nano\second} & \SI{128}{\nano\second}\\
        cycle rate & $\sim$ \SI{65000}{\per\second} & $\sim$ \SI{73600}{\per\second} & $\sim$ \SI{73800}{\per\second}\\
        $\eta_{\mathrm{cycle}}$ (average) & 0.78 & 0.85 & 0.86\\
        $T_{1,1}$ (under pump) & \SI{30}{\micro\second} & \SI{54}{\micro\second} & \SI{62}{\micro\second}\\
        $T_{1,2}$ (under pump) & \SI{15}{\micro\second} & \SI{73}{\micro\second} & \SI{77}{\micro\second}\\
        $T_{2,1}^*$ (under pump) & \textit{not measured} & \SI{89}{\micro\second} & \SI{98}{\micro\second}\\
        $T_{2,2}^*$ (under pump) & \textit{not measured} & \SI{72.4}{\micro\second} & \SI{116}{\micro\second}\\
        $\kappa_m$ (under pump) & \SI{3.7e5}{\per\second} & \SI{9.5e4}{\per\second} & \SI{9.3e4}{\per\second}\\
        \bottomrule
        \bottomrule
    \end{tabular}
    \caption{\textbf{cSMPD1 and cSMPD2 parameters and performance metrics at the optimal working point.} cSMPD1 data are different from the data in \cref{sec: temperature effect cSMPD1}. The cSMPD1 data shown in this table are part of a dataset where the pump amplitude were swept.}. 
    \label{table: circuit-params-part2-merged}
\end{table*}

\section{Multiplexed dispersive readout and threshold estimations}
\label{Appendix_ro}

Individual qubits are read out dispersively by scattering an off-resonant microwave signal and analyzing the amplified scattered response. The qubit state information is encoded in the amplitude and phase of the complex scattered signal, expressed as \( I(t) + jQ(t) \), where \( I(t) \) and \( Q(t) \) represent the time-dependent in-phase and quadrature components, respectively. The scattered readout signal is digitized using an ADC, and the I-Q plane is rotated by a fixed angle through multiplication of the integration weights by a rotation matrix. This procedure enables encoding the qubit state information onto a single quadrature, \( I \).

The primary limitation of a (c)SMPD is its dark count rate. We
therefore bias the readout towards a high ground-state assignment
fidelity, by choosing the threshold that minimizes the ratio of the
false-positive probability to the squared detection probability, in
analogy with the sensitivity of \cref{eq: E1}:
\begin{equation}\label{eq: E1}
V_\mathrm{th}:=\underset{V}{\mathrm{argmin}}\;
\frac{P(I\geq V\,|\,g)}{P(I\geq V\,|\,e)^2},
\end{equation}
where $P(I\geq V\,|\,g)$ [resp.\ $P(I\geq V\,|\,e)$] is the probability
that a qubit prepared in $\ket{g}$ (respectively $\ket{e}$) yields a signal $I\geq V$. We also define a reset threshold $V_\mathrm{th,reset}:=2V_0-V_\mathrm{th}$, symmetric to $V_\mathrm{th}$
with respect to the maximum-fidelity threshold $V_0$, so that
$V_\mathrm{th,reset}<V_0<V_\mathrm{th}$. At the end of a detection
cycle:
\begin{itemize}
  \item $I\geq V_\mathrm{th}$: the qubit is assigned to $\ket{e}$ and
        reset;
  \item $V_\mathrm{th,reset}<I<V_\mathrm{th}$: the outcome is ambiguous
        and the qubit is read out again;
  \item $I\leq V_\mathrm{th,reset}$: the qubit is assigned to $\ket{g}$
        and the next detection cycle starts.
\end{itemize}

\begin{figure}[h]
\centering
\includegraphics[width=8.45cm]{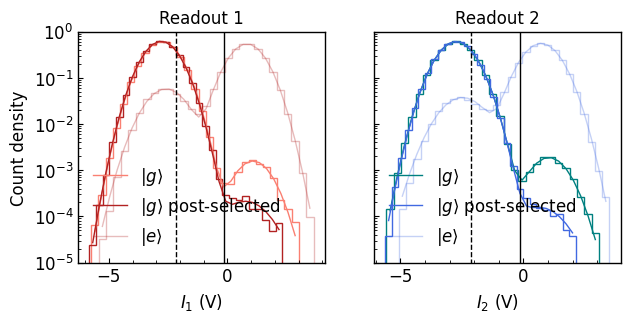}
\caption{\textbf{Multiplexed readout histograms with/without active preparation of the qubit ground state. (cSMPD1 data)}. Experimental histograms (staircase lines) of readout quadrature $I_i$ and corresponding bi-gaussian fit (solid lines) for readout $i=0$ (a) and readout $i=1$ (b), with qubit $i$ initially at equilibrium (legend $\ket{g}$), actively reset to $\ket{g}$ (legend $\ket{g}$ post-selected), and after a $\pi$-pulse excitation starting from equilibrium (legend $\ket{e}$).  Initialization for the next pulse sequence is ensured by waiting $\sim10$ $T_1$ qubit relaxation time. Equilibrium probabilities extracted from ground state fits: (a)  $2.6\times10^{-3}$ (salmon) and $4.5\times10^{-4}$ (dark red); (b) $3.1\times10^{-3}$ (teal) and $3.3\times10^{-4}$ (dark blue).
}
\label{fig: multiplexed ro}
\end{figure}

\section{Input power calibration}
\label{Appendix_InputPowCalibration}

The calibration of the photon flux is carried in-situ at the chip level. We measure the dephasing and the AC-Stark shift of the qubit induced by the illumination of the input resonator (buffer) with a weak coherent tone. The qubit, influenced by the coherent drive, experiences a frequency shift corresponding to the average photon number in the cavity while photon number fluctuations lead to a qubit dephasing (also dependent on the photon number). The dynamics followed by the system is, in the rotating frame of the coherent drive:

\begin{equation}
    \dfrac{\hat{H}}{\hbar}=\delta\hat{b}^\dagger\hat{b}+\dfrac{\omega_{q_1}}{2}\hat{\sigma}_{z,1}-\dfrac{\chi_{q_1b}}{2}\hat{b}^\dagger\hat{b}\hat{\sigma}_{z,1}+\epsilon_d\left(\hat{b}^\dagger+\hat{b}\right)
\end{equation}

\noindent
where $\epsilon_d$ is the coherent drive amplitude and $\delta=\omega_b-\omega_d$. Following the method in~\cite{gambetta_qubit-photon_2006}, the complex dephasing rate is the sum of two terms: $\delta\omega$ (real part, the frequency shift) and $\delta\gamma$ (imaginary part, the dephasing rate). Those two quantities are linked to the coherent complex amplitudes $\alpha_g$ and $\alpha_e$ of the photon field inside the input resonator induced by the coherent drive by:

\begin{equation}\label{eq: complex ACstakshift}
    \delta\omega + i\delta\gamma = -\chi_{q_1b}\alpha_g\bar{\alpha}_e=\dfrac{-4\chi_{q_1b}\abs{\epsilon_d}^2}{(\kappa_b+i\chi_{q_1b})^2+4\delta^2}
\end{equation}

\noindent
The coherent complex amplitudes are defined by:

\begin{equation}\label{eq: alpha_ge in buffer mode}
    \alpha_{g/e}=\dfrac{\epsilon_d}{\dfrac{\kappa_b}{2}+i\left(\delta\pm \dfrac{\chi_{q_1b}}{2}\right)}
\end{equation}

\begin{figure}[h!]
    \centering
        \includegraphics[width=\linewidth]{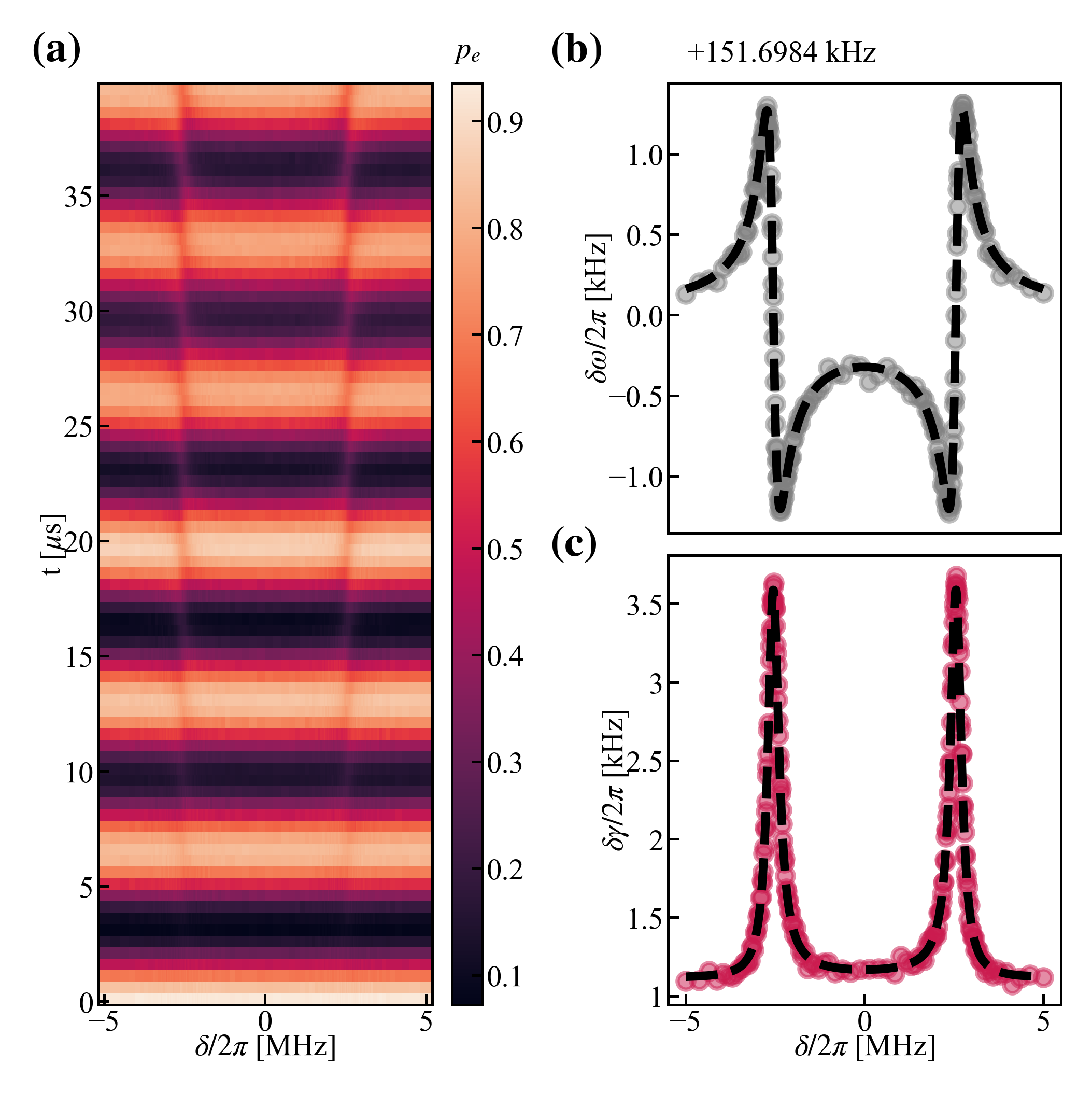}
        \caption{\textbf{Measurement induced dephasing on the buffer-Q1 circuit of cSMPD2.} (a) Qubit excitation probability $P_e$ after a Ramsey sequence (see text) as a function of coherent probe detuning $\delta$. (b, c) Qubit frequency shift $\delta \omega$ and dephasing rate $\delta y$ (black lines) deduced from a fit of each time traces of panel (a) by exponentially decaying cosine function. Dotted black lines correspond to a fit of the \cref{eq: complex ACstakshift} + \cref{eq: alpha_ge in buffer mode} model, yielding the dispersive shift $\chi_{q_1b}/2\pi$=\SI{-5.1}{\mega\Hz}, the buffer decay rate $\kappa_{b}/2\pi$=\SI{374}{\kilo\Hz} and $\epsilon_d/2\pi$=\SI{21.5}{\kilo\Hz}. The input power is calibrated to $P_\mathrm{in}=\SI{1.02e4}{\text{photon}\per\s}$.}
        \label{fig: inputpow_cal}
\end{figure}

The complex AC-Stark shift is experimentally measured by repeatedly performing Ramsey experiment on the qubit while varying the weak coherent drive frequency applied on the buffer. See \cref{fig: inputpow_cal} for illustration.

The power of the coherent drive at the input of the buffer resonator can be computed using an input-output relation:

\begin{equation}\label{eq: Pin input output}
    \dfrac{P_\mathrm{in}}{\hbar\omega_d} = \dfrac{\abs{\epsilon_d}^2}{\kappa_b-\kappa_{b,\mathrm{int}} }
\end{equation}

\noindent
where $\kappa_{b,\mathrm{int}}$ is the internal loss rate of the buffer resonator. We use $\kappa_{b,\mathrm{int}}=\SI{5.6e5}{\per\s}$ for cSMPD2 (see \cref{table: circuit-params-part2-merged}). For cSMPD1, $\kappa_{b,\mathrm{int}}$ was not measured and is set to zero, which yields an upper bound on the calibrated efficiency.

\section{Memory loss rate and coherence measurements}
\label{Appendix_Memoryloss calibration}

The methods utilized to characterize the memory mode lifetime and coherence are different for the cSMPD1 and cSMPD2 devices. We only measured $T_1$ for the cSMPD1, while we measured $T_1$ and $T_2^*$ for cSMPD2.

\subsection{cSMPD1 memory characterization}

\begin{figure}[h]
    \centering
    \includegraphics[width=8.4cm]{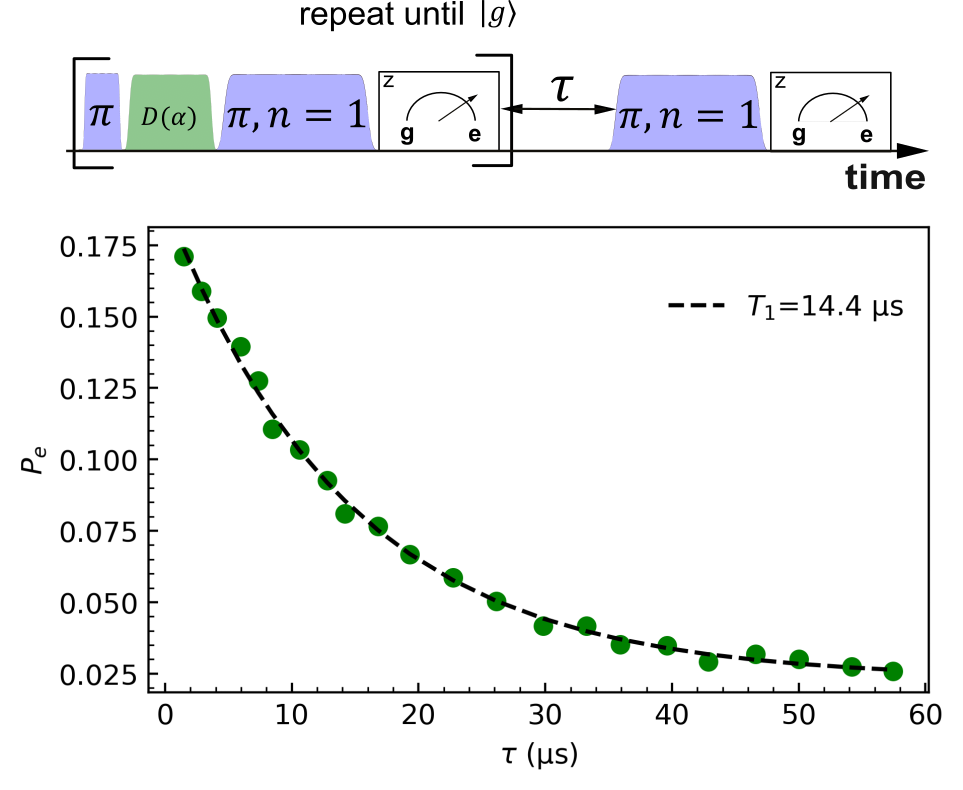}
    \caption{\textbf{Bare memory $T_1$ measurement (cSMPD1 data)}. (top) Pulse sequence initializing the memory in Fock state $\ket{1}$, using one of the qubit as an ancilla (see text). (bottom) Probability $P_e$ of re-exciting the qubit conditionally to the presence of a memory photon, as a function of delay $\tau$ after preparation. Dashed black exponential fit yields $T_1=\SI{14.4}{\micro\second}$.
    }
    \label{fig: memoryT1}
\end{figure}

The bare $T_1$ of the memory resonator is measured by preparing memory Fock state $\ket{1}$ using  one of the two qubits as an ancilla (see \cref{fig: memoryT1}), and measuring its decay. The preparation consists in repeating a nonselective excitation of the qubit ($\pi$-pulse), a small coherent excitation of the memory (also through the qubit drive line), a selective reset ($\pi$-pulse at $\omega_q-\chi_{q_2}$) of the qubit conditioned on the presence of one memory photon, and a qubit dispersive readout, until the qubit is found in its ground state $\ket{g}$, which means that the system state has been collapsed on $\ket{1,g}$. This preparation is followed by a variable delay time $\tau$ before re-exciting the qubit if the memory photon is still present ($\pi$-pulse at $\omega_q-\chi_{q_2}$), and finally by a qubit readout. The whole sequence is repeated for each $\tau$ to obtain the probability $P_e(\tau)$ of qubit excitation, which maps the energy decay of the memory and yields $T_1$ for the memory mode.

\subsection{cSMPD2 memory characterization}

Measuring $T_1$ and $T_2^{*}$ of the memory requires preparing it in the Fock
state $\ket{1_m}$ or in the superposition $(\ket{0_m}+\ket{1_m})/\sqrt{2}$, and
monitoring its decay after a variable delay $\tau$. Both the preparation and the
measurement use a qubit as an ancilla.

\begin{figure}[h]
    \centering
    \includegraphics{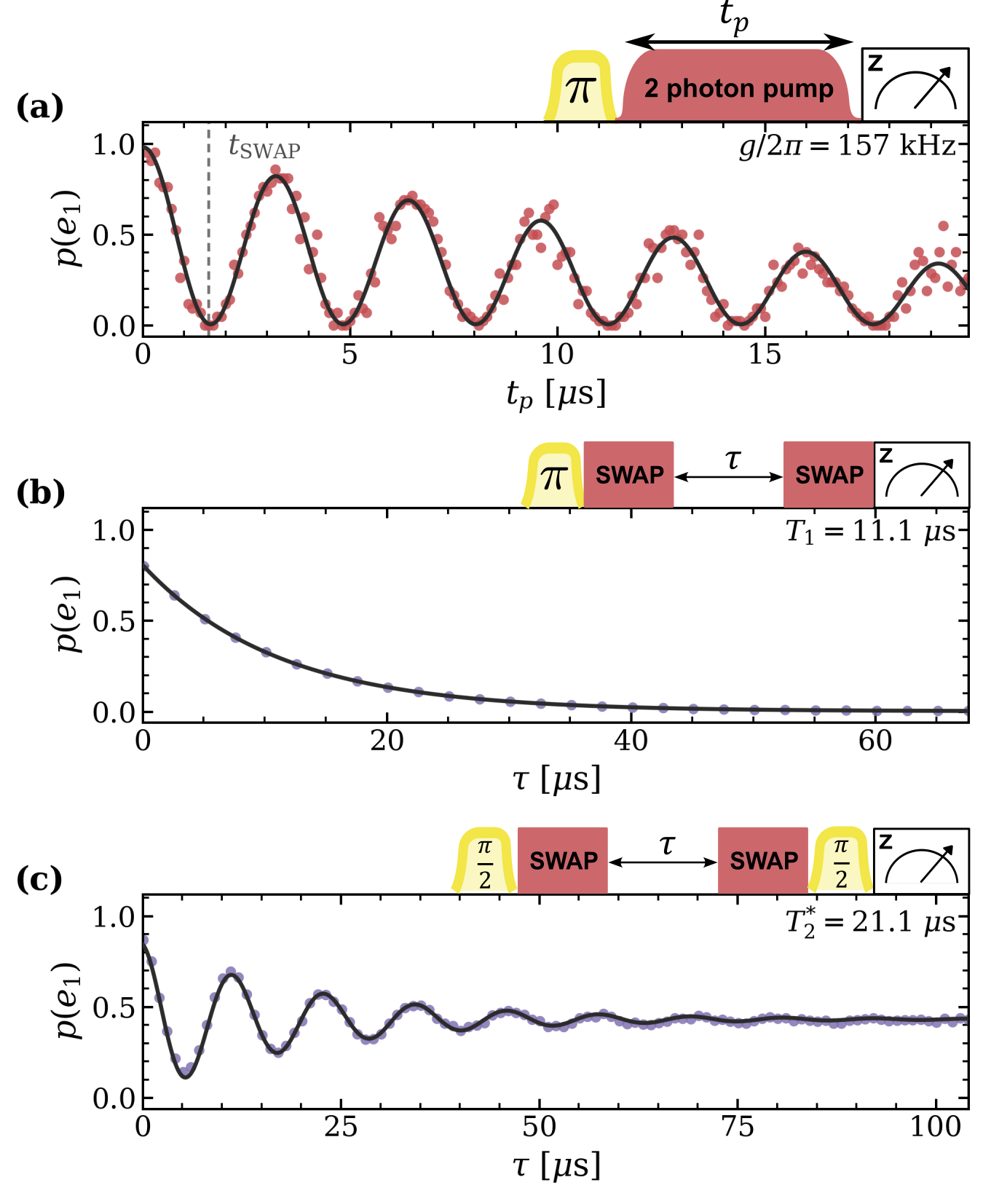}
    \caption{\textbf{Parametric SWAP gate and memory coherence (cSMPD2 data).}
    In all panels the upper drawing is the pulse sequence, $p(e_1)$ is the
    excited-state population of Q1 measured by dispersive readout, markers are
    data and solid lines are fits.
    (a)~Vacuum Rabi oscillation between Q1 and the memory. A $\pi$ pulse
    prepares Q1 in $\ket{e}$; a two-photon pump of duration $t_p$ at
    $\omega_p = (\omega_m - \omega_q)/2$ then activates the Jaynes--Cummings
    interaction. The pump frequency is
    $f_p = \SI{2.16652}{\giga\hertz}$. The fit to
    $p(e_1) = e^{-t_p/\tau_p}\cos^2(g t_p)$ gives a coupling
    $g/2\pi = \SI{157}{\kilo\hertz}$ and an envelope decay
    $\tau_p = 2/(1/T_{1,q}+1/T_{1,m})=\SI{17.9}{\micro\second}$. The dashed line marks the full-swap
    time $t_\mathrm{SWAP} = \pi/2g = \SI{1.60}{\micro\second}$, at which the
    excitation is entirely transferred to the memory.
    (b)~Energy relaxation of the memory. A $\pi$ pulse followed by a SWAP
    prepares the Fock state $\ket{1_m}$; after a delay $\tau$ a second SWAP maps
    the memory back onto Q1 for readout. An exponential fit gives
    $T_{1,m} = \SI{11.1}{\micro\second}$.
    (c)~Ramsey interferometry on the memory. A $\pi/2$ pulse and a SWAP prepare
    $(\ket{0_m} + \ket{1_m})/\sqrt{2}$; after a delay $\tau$ a second SWAP and a
    final $\pi/2$ pulse close the sequence. The fit to an exponentially damped
    oscillation gives $T_{2,m}^{*} = \SI{21.1}{\micro\second}$, consistent with
    the relaxation limit $2T_{1,m} = \SI{22.2}{\micro\second}$ and indicating
    negligible pure dephasing of the memory mode.}
    \label{fig: memory qubit exchange}
\end{figure}

We first prepare Q1 in $\ket{e}$ with a $\pi$ pulse, or in
$(\ket{g}+\ket{e})/\sqrt{2}$ with a $\pi/2$ pulse, and then transfer its state to
the memory through the parametric Jaynes--Cummings interaction
$\hat{H}_\mathrm{int} = \hbar g\,\hat{q}\,\hat{m}^\dagger + \mathrm{h.c.}$,
activated by a low-frequency two-photon pump. The interaction is resonant when
$2\omega_p + \omega_q = \omega_m$. \cref{fig: memory qubit exchange}(a) shows the
resulting vacuum Rabi oscillation, from which we extract
$g/2\pi = \SI{157}{\kilo\hertz}$ and a full-swap time
$t_\mathrm{SWAP} = \pi/2g = \SI{1.60}{\micro\second}$.

After the delay $\tau$, a second SWAP maps the memory state back onto Q1. For the
$T_1$ measurement, Q1 is then read out dispersively
[Fig.~\ref{fig: memory qubit exchange}(b)]. For the $T_2^{*}$ measurement, a
final $\pi/2$ pulse on Q1 completes the Ramsey sequence before readout
[Fig.~\ref{fig: memory qubit exchange}(c)]. Repeating the sequence for different
$\tau$ yields $p(e_1)$ as a function of delay, giving
$T_1 = \SI{11.1}{\micro\second}$ and
$T_2^{*} = \SI{21.1}{\micro\second}$. The latter is close to the
relaxation limit $2T_1$, so the coherence of the memory is limited by energy
decay rather than by pure dephasing.

\section{Noise Equivalent Power (NEP)}
\label{Appendix_NEP}

A frequently used figure of merit is the noise equivalent power (NEP), conventionally defined as the incident signal power required to produce a signal equal to the noise level for a given detection bandwidth~\cite{richards_bolometers_1994}. Another way to phrase it is to say that the NEP is the input power requested to obtain a signal-to-noise ratio (SNR) equal to 1 for some integration time $t$. It is expressed in W/$\sqrt{\mathrm{Hz}}$. 

For a SMPD device, the scenario is as follows: the signal is generated by the input power $P$ applied to the buffer resonator. The number of photons impinging on the resonator over a time window $t$ can be expressed as $Pt/\hbar\omega_b$. The detector has an operational efficiency $\eta$ and experiences false positives, characterized by its dark count rate $\alpha$.

Two quantities can be defined: $S_{\mathrm{OFF}}=\alpha t$ representing the signal measured when only noise is recorded and $S_{\mathrm{ON}} = \eta Pt/\hbar\omega_b + \alpha t$  representing the signal measured when a signal of interest is applied. The signal of interest can be expressed as $S_{\mathrm{ON}}-S_{\mathrm{OFF}}=\eta Pt/\hbar\omega_b$ and the noise, which can be modeled by Poissonian statistics, is directly given by $\sqrt{S_{\mathrm{ON}}}$. Since $S_{\mathrm{OFF}}$  can be independently measured, the noise of interest can be reduced to $\sqrt{S_{\mathrm{ON}}}$. Therefore, the detection SNR is given by:

\begin{equation}\label{eq: SNR detection}
    \mathrm{SNR} = \dfrac{S_{\mathrm{ON}}-S_{\mathrm{OFF}}}{\sqrt{S_{\mathrm{ON}}}}=\dfrac{\eta Pt/\hbar\omega_b}{\sqrt{\eta Pt/\hbar\omega_b + \alpha t}}
\end{equation}

The NEP can be calculated by solving for $P$ while taking
$\mathrm{SNR}=1$, which gives the minimum detectable power
\begin{equation}
P_\mathrm{min}(t)=\hbar\omega_b\,\frac{1+\sqrt{1+4\alpha t}}{2\eta}\,
\frac{1}{t},
\tag{H2}
\end{equation}
and the corresponding noise equivalent power
$\mathrm{NEP}=P_\mathrm{min}\sqrt{t}$. In the previous experiment, the
typical integration time was chosen to ensure $\alpha t\gg1$, resulting
in:
\begin{equation}
\mathrm{NEP}=\hbar\omega_b\,\frac{\sqrt{\alpha}}{\eta},
\tag{H3}
\end{equation}
which is independent of $t$ and coincides with the power sensitivity
$S$ of \cref{eq: sensitivity}.

\section{Detector efficiency measurement with small coherent states}
\label{subsubsec: on the average amplitude approx}

We recall in this section the argument stated in Appendix E of Ref.~\cite{lescanne_irreversible_2020}. We also recall that the efficiency of the detector is defined as the probability to detect a Fock state $\ket{1}$ impinging on the input cavity. However we do not use a dedicated single photon source producing Fock state $\ket{1}$ to calibrate $\eta$, but rather only small coherent state. The following argument bridges the gap between our experiment procedure and the formal definition of $\eta$.

On one hand, a source emitting randomly single photons with a small probability $\epsilon$ can be described with a density matrix $\hat{\rho}$:

\begin{equation}\label{eq: sinle photon source}
    \rho = (1-\epsilon)\ketbra{0}+\epsilon\ketbra{1}
\end{equation}

Illuminated by such a source, the photon detector produces a count with a probability proportional to $\epsilon$, $p_e=\eta\epsilon$. One can on the other hand describe the weak coherent state of amplitude $|\psi_0|\ll 1$ experimentally present in the input resonator ($\hat{r}_0$ mode) as:

\begin{align}
    \ket{\psi_0}&=e^{-\frac{|\psi_0|^2}{2}}(\ket{0}+\psi_0\ket{1}+...)\\
    &\approx\sqrt{1-|\psi_0|^2}\ket{0}+\psi_0\ket{1}
\end{align}

The statistical mixture of such coherent states with randomized phase reads, keeping only terms or order $o(\abs{\psi_0}^2)$:

\begin{equation}\label{eq: stat mixt fock states}
    \hat{\rho}_{\psi_0} \approx \left(1-\abs{\psi_0}^2\right)\ketbra{0}+\abs{\psi_0}^2\ketbra{1}
\end{equation}
Finally, we identify coefficients of \cref{eq: sinle photon source,eq: stat mixt fock states}. The statistical mixture produced experimentally is equivalent to an intermittent single photon source that has a probability to emit Fock state $\ket{1}$ of $\epsilon=\abs{\psi_0}^2$, provided that the average amplitude follows $\abs{\psi_0}^2\ll 1$. The probability to measure a count becomes $p_e=\eta\epsilon=\eta\abs{\psi_0}^2$.

\section{Coupled mode model} 
\label{appendix: Coupled mode models}

The steady-state physics of the SMPD/cSMPD can be described by a semiclassical analysis based on a one-dimensional chain of modes with nearest-neighbor couplings. In this section, we derive the photon number transmission coefficient $\abs{S_{31}(\omega)}^2$. This scattering parameter can be mapped to the conversion efficiency $\eta_{\mathrm{4WM}}$ and allows to derive an analytical expression of the detector bandwidth as well as to understand the effect of resonator losses.

The analytical description is constructed as follows: we first introduce a formalism based on a transfer matrix approach for an arbitrarily long one-dimensional chain of coupled modes, then we show this model is equivalent to a semi-classical analysis performed on a set of input-output quantum equations based on the 4WM Hamiltonian. The scattering coefficient is expressed in terms of an optimization criterion: the cooperativity.

\subsection{Transfer matrix approach applied to a 1D chain of linearly coupled modes}
\label{sec: Transmission coefficient of a 1D chain}

The condition for achieving unit transmission in a system composed of a cascade of single-mode cavities can be analytically determined for an arbitrary number of cavities. In this section, we derive a general analytical expression for the transmission coefficient $S_{N1}$ ($N$ being the index of the last cavity) of such a system and subsequently specialize the result to the case of the cSMPD architecture, which involves three coupled cavities.

\begin{figure}[h]
    \centering
    \includegraphics[width=0.9\linewidth]{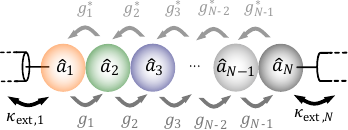}
    \caption{\textbf{1D chain of single mode cavities with nearest neighbor couplings.} Only the two ends of the chain are connected to the environment. Port $1$ and Port $N$ correspond to the leftmost and rightmost ports of the chain, respectively.}
    \label{fig:1Dchain of modes}
\end{figure}

We consider a 1D chain of $N$ single mode cavities with individual angular frequencies $\omega_k$ and energy loss rate $\kappa_k$. Nearest neighbor cavities are coupled by a Jaynes-Cummings Hamiltonian with a strength $g_{k,k+1}\equiv g_k$. We consider couplings such that $g_{k,k+1}=g_{k+1,k}^*$. A cavity field is described by the annihilation operators $\hat{a}_k$. The system is represented in \cref{fig:1Dchain of modes}. The Hamiltonian of the full 1D cavity chain is
\begin{multline}\label{eq: 1D cav chain ham}
    \frac{\hat{H}}{\hbar}=\Delta_N\hat{a}_N^\dagger\hat{a}_N+\\ \sum_{k=1}^{N-1}\left[\Delta_k\hat{a}_k^\dagger\hat{a}_{k}+\left(g_k\ \hat{a}_k\hat{a}_{k+1}^\dagger+g_k^*\ \hat{a}_k^\dagger\hat{a}_{k+1}\right)\right].
\end{multline}

In the following, we probe the system by sending a monochromatic tone on port 1, i.e on the cavity 1, at angular frequency $\omega$. In \cref{eq: 1D cav chain ham} we consider a rotating frame around the drive angular frequency, hence $\Delta_k=\omega_k-\omega$. The field operators of the cavities follows a coupled set of quantum Langevin equations that can be casted in a matrix form:
\begin{equation}\label{eq: diff eq 1D chain}
    \mathbf{\dot{\hat{a}}} = -\mathbf{M} \mathbf{\hat{a}} + \mathbf{\hat{d}}
\end{equation}
where $\mathbf{\hat{a}}=\left(\hat{a}_1,\dotso,\hat{a}_N\right)^\top$ is the vector of time-dependent field annihilation operators, $\mathbf{\hat{d}}=\left(\sqrt{\kappa_{\mathrm{ext},1}}\hat{a}_{\mathrm{in},1},\dotso,\sqrt{\kappa_{\mathrm{ext},N}}\hat{a}_{\mathrm{in},N}\right)^\top$ is the external drive vector, and $\mathbf{M}$ reads \cref{eq: M matrix qfield} and represents the linear coupling as well as damping rates. $\mathbf{M}$ is a tridiagonal matrix with complex entries. 
\begin{equation}\label{eq: M matrix qfield}
    M_{k,k}=i\Delta_k+\frac{\kappa_k}{2},\qquad M_{k,k+1}=ig_k,\qquad M_{k+1,k}=ig_k^*
\end{equation}
We consider a single monochromatic continuous coherent drive on the port 1 of the chain. It allows to drastically simplify the problem as all other resonators, except the first one, only exhibit quantum fluctuations. We note $\expval{\hat{a}_k(t)}=\alpha_k(t)$. The drive term becomes $\expval{\mathbf{\hat{d}}(t)}=\left(\sqrt{\kappa_\mathrm{ext,1}}\alpha_\mathrm{in,1}(t),0,\dotso,0\right)^\top$, and since we are only interested in the steady state of the system, we have $\expval{\mathbf{\dot{\hat{a}}}(t)}=0$. The matrix form \cref{eq: diff eq 1D chain} becomes a linear system of equations $\mathbf{M} \mathbf{a} = \mathbf{d}$. It is worth noting that the use of coherent states to probe the system is not only helpful for building intuition about its response, but also relevant to our photon-counting problem. We show in \cref{subsubsec: on the average amplitude approx} that sufficiently weak coherent states can approximate an incoherent single-photon source.

The output field of last cavity satisfies $\alpha_{\mathrm{out},N}=\sqrt{\kappa_{\mathrm{ext},N}}\alpha_N$, thus the transmission coefficient in amplitude, as a function of the input detuning from the first cavity $\delta$, is
\begin{equation}\label{eq: s21 with coef inverse}
    S_{N1}(\delta)=\frac{\alpha_{\mathrm{out},N}}{\alpha_{\mathrm{in},1}}=\sqrt{\kappa_{\mathrm{ext},1}\kappa_{\mathrm{ext},N}}\cdot\left[\mathbf{M^{-1}}\right]_{N,1}.
\end{equation}
We now derive a closed-form expression for the inverse matrix elements by expressing them recursively in terms of determinants of principal submatrices of the tridiagonal matrix $\mathbf{M}$. Let $D_k$ denote the determinant of the upper left $k\times k$ submatrix, with $D_0=1$. The sequence $\lbrace D_k\rbrace$ satisfies the following recursion:
\begin{equation}\label{eq: det rec}
D_{k+1}=M_{k+1,k+1}\cdot D_k-M_{k,k+1}M_{k+1,k}\cdot D_{k-1}
\end{equation}
The Thomas algorithm, a simplified Gaussian elimination scheme, provides a standard method for inverting a tridiagonal matrix. We directly obtain
\begin{align}
    \left[\mathbf{M^{-1}}\right]_{N,1}
    & =\frac{(-1)^{N+1}\prod_{k=1}^{N-1}M_{k,k+1}}{D_N} \\
    & =\frac{(-i)^{N-1}\prod_{k=1}^{N-1}g_{k}}{D_N},
\end{align}
and the transmission coefficient in amplitude follows
\begin{equation}\label{eq: s21 amp 1D cav chain generic}
    S_{N1}(\delta)=\sqrt{\kappa_{\mathrm{ext},1}\kappa_{\mathrm{ext},N}}\cdot\frac{(-i)^{N-1}\prod_{k=1}^{N-1}g_{k}}{D_N}.
\end{equation}
The structure of \cref{eq: s21 amp 1D cav chain generic} is reminiscent of well-known forms encountered in signal processing and filter design. Moreover, this expression enables us to derive a condition on the coupling parameters $g_k$ required to achieve unit power transmission through the coupled cavity system, a property that was exploited in Ref.\cite{yan_broadband_2023} to design a broadband Purcell filter. We explicitly address the condition for unit transmission in \cref{sec: three cav system linear}. \\

\textbf{Transient analysis} --- The steady-state analysis is sufficient as long as the transmission linewidth sets a timescale much shorter than those we are investigating. As we will see in the following section, this is not always the case in our experiments, and an extension of the steady-state analysis is therefore needed. Here, we analyze the mean-field solutions of \cref{eq: diff eq 1D chain}, obtained by setting $\boldsymbol{\alpha}=\langle\hat{\mathbf{a}}\rangle$. For a time-independent $\mathbf{M}$, the general solution reads

\begin{equation}\label{eq: diff eq 1D general solution}
    \boldsymbol{\alpha}(t) = e^{-\boldsymbol{M}t}\mathbf{\boldsymbol{\alpha}}(0) + \int_0^t d\tau e^{-\boldsymbol{M}(t-\tau)}\boldsymbol{d}(\tau).
\end{equation}

\noindent Given the initial intracavity fields $\boldsymbol{\alpha}(0)$, we can evaluate \cref{eq: diff eq 1D general solution} and infer, from the input-output relations at the boundaries, the output field into each cavity bath:
\begin{equation}\label{eq: langevin boundary cond}
    \boldsymbol{\alpha}_\mathrm{in} + \boldsymbol{\alpha}_\mathrm{out} = \sqrt{\boldsymbol{\kappa}_\mathrm{ext}} \boldsymbol{\alpha},
\end{equation}

\noindent where $\sqrt{\boldsymbol{\kappa}_\mathrm{ext}}$ is the diagonal matrix with entries $\left[\sqrt{\boldsymbol{\kappa}_\mathrm{ext}}\right]_{i,i} = \sqrt{\kappa_{\mathrm{ext},i}}$. \Cref{eq: diff eq 1D chain} can also be analyzed in the frequency domain. Taking its Fourier transform and including the boundary conditions, the relation between the output and input fields reads

\begin{align*}\label{eq: full input-output frequency domain}
    \boldsymbol{\alpha}_\mathrm{out} (\omega) &= \left[\sqrt{\boldsymbol{\kappa}_\mathrm{ext}}\left[\mathbf{M} - j\omega \mathbb{\mathbf{I}}\right]^{-1}\sqrt{\boldsymbol{\kappa}_\mathrm{ext}} - \mathbb{\mathbf{I}}\right]\mathbf{a}_\mathrm{in} (\omega) \\
    &\equiv S(\omega) \boldsymbol{\alpha}_\mathrm{in}(\omega).
\end{align*}

It is worth noting that, for a constant monochromatic coherent drive on port 1, $\boldsymbol{\alpha}_\mathrm{in}(\omega) = \left(2\pi a_\mathrm{in}\delta(\omega), 0, \dots, 0\right)^\top$. Only $S(0)$ then contributes to the output, and the transmission through the chain reduces to the steady-state result of \cref{eq: s21 with coef inverse}.

\subsection{3-mode system transmission coefficient based on the transfer matrix approach}
\label{sec: three cav system linear}

\cref{eq: s21 amp 1D cav chain generic} is applied to a three cavity system. We define $\lambda_k=i\Delta_k+\kappa_k/2,\ k\in\interval{1}{N}$. The determinant becomes
\begin{align}
    D_3 &= \lambda_3(\lambda_2\lambda_1+\abs{g_1}^2) + \lambda_1\abs{g_2}^2\\
    &= \lambda_1\lambda_2\lambda_3 + \lambda_3 \abs{g_1}^2 + \lambda_1\abs{g_2}^2
\end{align}
and thus the power transmission coefficient follows
\begin{equation}\label{eq: cc model n=3}
\abs{S_{31}(\omega)}^2 = \kappa_{\mathrm{ext},1}\kappa_{\mathrm{ext},3}\cdot \frac{\abs{g_1g_2}^2}{\abs{\lambda_1\lambda_2\lambda_3 + \lambda_3 \abs{g_1}^2 + \lambda_1\abs{g_2}^2}^2}.
\end{equation}
Instead of directly solving $\abs{S_{31}(\omega)}^2=1$ for $g_1$ and $g_2$, we can solve the trivial two-mode model and then extend the result to the relevant three-mode problem. The determinant of the two-cavity system follows $D_2=\lambda_1\lambda_2+\abs{g_1}^2$. The finite frequency detuning case as well as the internal losses of the cavities has no practical interest for our discussion, hence we set $\Delta_k=0$ and $\kappa_{\mathrm{ext,}k}=\kappa_k$. The unit absolute amplitude transmission requires the condition $\sqrt{\kappa_1\kappa_2}\abs{g_1} = \kappa_1\kappa_2/4 + \abs{g_1}^2$ to be fulfilled. The discriminant is zero, hence it admits a single degenerate root. We find $\abs{g_1}=\sqrt{\kappa_1\kappa_2}/2$, often casted as $\kappa_1 = 4\abs{g_1}^2/\kappa_2$. The unit transmission is achieved if and only if the input coupling rate matches the effective dissipation of the rest of the 1D chain seen by the first cavity.

Next, we extend the two-cavity optimal rate matching condition to the three-mode model. We proceed to a 1D chain model reduction from the right end side of the system. We suppose an energy input rate $\kappa_2$ on the second cavity. The optimal rate matching condition becomes $\kappa_2 = 4\abs{g_2}^2/\kappa_3$ between cavity 2 and 3. We can continue for cavity 1 and 2: the optimal rate matching condition follows $\kappa_1 = 4\abs{g_1}^2 / \kappa_2$, hence
\begin{equation}\label{eq: 3 cav matching cond}
    \kappa_1 = \kappa_3\abs{\frac{g_1}{g_2}}^2
\end{equation}

The section \cref{sec: 3-mode input-output} shows the previous results could have been obtained directly from input-output theory.

\subsection{3-mode system transmission coefficient from input-output theory}
\label{sec: 3-mode input-output}

\subsubsection{Input-output theory}

We consider the simplified 4WM energy conservation equations where the indices 1,2,3 are dropped for the mode labels \textit{b,m,w} to match the main text notations:
\begin{align}
    \omega_b + \left(\omega_{p_1}-\Delta_{p_1}\right) &= \omega_{q_1} + \left(\omega_{m}-\Delta_{m}\right)\label{eq: 2.1}\\
    \left(\omega_m - \Delta_{p_1}\right) + \left(\omega_{p_2}-\Delta_{p_2}\right) &= \omega_{q_2} + \left(\omega_{w}-\Delta_{w}\right)\label{eq: 2.2}
\end{align}
where we neglected every Kerr terms as they only contributes to frequency renormalization and do not contribute significantly to the 4WM dynamics that we want to model. 

We allow some arbitrary detunings on both pump ($\Delta_{p_1}, \Delta_{p_2}$) and introduce a memory detuning $\Delta_m$ and a waste detuning $\Delta_w$ to compensate for them.
4WM terms are time-independent, thus resonant, as long as $\Delta_m=\Delta_{p_1}$ and $\Delta_w = \Delta_{p_1} + \Delta_{p_2}$. In addition, we move to a rotation frame with respect to the input signal with $\delta = \omega_b - \omega_\mathrm{in}$. The time-independent Hamiltonian dynamics followed by the two-qubit three-mode system can be formally represented, when discarding every Kerr terms, by 

\begin{multline}\label{eq: H CPD}
\frac{\hat{H}}{\hbar}=\delta\hat{b}^\dagger\hat{b}+(\delta+\Delta_m)\,\hat{m}^\dagger\hat{m}
+(\delta+\Delta_w)\,\hat{w}^\dagger\hat{w}\\
+g_{4,1}\hat{b}\hat{q}_1^\dagger\hat{m}^\dagger+g_{4,1}^*\hat{b}^\dagger\hat{q}_1\hat{m}
+g_{4,2}\hat{m}\hat{q}_2^\dagger\hat{w}^\dagger+g_{4,2}^*\hat{m}^\dagger\hat{q}_2\hat{w},
\end{multline}
where $\hat{q}_k$ is the bosonic annihilation operator of transmon $k$. Next, we express the set of input-output equations for the buffer mode $\hat{b}$ mode as well as on the non-local operators $\hat{\sigma}_1\hat{m}$ and $\hat{\sigma}_1\hat{\sigma}_2\hat{w}$.

\begin{widetext}
\begin{equation}\label{eq: CPD HL}
\begin{cases}
\begin{aligned}
\partial_t\hat{b}=\;&-i\big[\delta\hat{b}+g_{4,1}^*\hat{q}_1\hat{m}\big]
-\frac{\kappa_b}{2}\hat{b}
+\hat{F}_b\\[8pt]
\partial_t\big(\hat{q}_1\hat{m}\big)=\;&-i\Big[(\delta+\Delta_m)\hat{q}_1\hat{m}
+g_{4,1}\hat{b}\big(1+\hat{q}_1^\dagger\hat{q}_1+\hat{m}^\dagger\hat{m}\big)
+g_{4,2}^*\hat{q}_1\hat{q}_2\hat{w}\Big]
-\frac{\tilde{\kappa}_m}{2}\hat{q}_1\hat{m}
+\hat{F}_{q_1m}
-i\zeta_1(t)\,\hat{q}_1\hat{m}\\[8pt]
\partial_t\big(\hat{q}_1\hat{q}_2\hat{w}\big)=\;&-i\Big[(\delta+\Delta_w)\hat{q}_1\hat{q}_2\hat{w}
+g_{4,2}\hat{q}_1\hat{m}\big(1+\hat{q}_2^\dagger\hat{q}_2+\hat{w}^\dagger\hat{w}\big)
+g_{4,1}\hat{b}\hat{m}^\dagger\hat{q}_2\hat{w}\Big]
-\frac{\tilde{\kappa}_w}{2}\hat{q}_1\hat{q}_2\hat{w}\\
&+\hat{F}_{q_1q_2w}-i\big(\zeta_1(t)+\zeta_2(t)\big)\hat{q}_1\hat{q}_2\hat{w}
\end{aligned}
\end{cases}
\end{equation}
\end{widetext}
where $g_{4,1}=-\xi_1\sqrt{\chi_{q_1b}\chi_{q_1m}}$,
$g_{4,2}=-\xi_2\sqrt{\chi_{q_2m}\chi_{q_2w}}$, $\hat{F}_b =\sqrt{\kappa_{b,\mathrm{ext}}}\,\hat{b}_\mathrm{in}
+\sqrt{\kappa_{b,\mathrm{int}}}\,\hat{b}_\mathrm{int}$, $\hat{F}_{q_1m}=\hat{q}_1\big(\sqrt{\kappa_{m,\mathrm{ext}}}\,\hat{m}_\mathrm{in}
+\sqrt{\kappa_{m,\mathrm{int}}}\,\hat{m}_\mathrm{int}\big)
+\sqrt{\Gamma_{1,1}}\,\hat{q}_{\mathrm{in},1}\hat{m}$ and $\hat{F}_{q_1q_2w}=\hat{q}_1\hat{q}_2\big(\sqrt{\kappa_{w,\mathrm{ext}}}\,\hat{w}_\mathrm{in}
+\sqrt{\kappa_{w,\mathrm{int}}}\,\hat{w}_\mathrm{int}\big)
+\big(\sqrt{\Gamma_{1,1}}\,\hat{q}_{\mathrm{in},1}\hat{q}_2
+\sqrt{\Gamma_{1,2}}\,\hat{q}_1\hat{q}_{\mathrm{in},2}\big)\hat{w}$. The transmons are subject
to energy relaxation at rates $\Gamma_{1,k}$, described by the vacuum
input fields $\hat{q}_{\mathrm{in},k}$, and to pure dephasing at rates
$\Gamma_{\varphi,k}$, described by classical white frequency noises
$\zeta_k(t)$ coupled to $\hat{q}_k^\dagger\hat{q}_k$, with
$\langle\zeta_k(t)\zeta_{k'}(t')\rangle=2\Gamma_{\varphi,k}\delta_{kk'}\delta(t-t')$
(It\^o convention). The non-local operators thus acquire the effective
damping rates $\tilde{\kappa}_m=\kappa_m+2\Gamma^*_{2,1}\approx\kappa_{m,\mathrm{int}}+2\Gamma^*_{2,1}$
and $\tilde{\kappa}_w=\kappa_w+2(\Gamma^*_{2,1}+\Gamma^*_{2,2})\approx\kappa_{w,\mathrm{ext}}$,
with $\Gamma^*_{2,k}=\Gamma_{1,k}/2+\Gamma_{\varphi,k}=1/T^*_{2,k}$.
The transmon anharmonicity and the dispersive couplings only act on
states where a qubit is doubly excited, or shift the frequencies of the non-local amplitudes by $\chi_{q_1m}$ and $\chi_{q_2w}$, which are
absorbed in $\Delta_m$ and $\Delta_w$.

\cref{eq: CPD HL} is supported on the infinite-dimensional Hilbert space
$\mathcal{H}_b\otimes\mathcal{H}_{q_1}\otimes\mathcal{H}_m\otimes\mathcal{H}_{q_2}\otimes\mathcal{H}_w$.
Starting from all modes in their ground state, a single photon entering
the buffer can only be converted, step by step, into a single excitation
of $q_1$ and of the memory, then of $q_2$ and of the waste. The
corresponding states span the subspace
\begin{multline}\label{eq: CPD subspace}
\mathcal{K}_\mathrm{cSMPD}=\big\{ \\
\ket{0_b,g,0_m,g,0_w},
\ket{1_b,g,0_m,g,0_w}, 
\ket{0_b,e,1_m,g,0_w},\\
\ket{0_b,e,0_m,g,0_w},
\ket{0_b,e,0_m,e,1_w},\ket{0_b,e,0_m,e,0_w}\\
\big\},
\end{multline}
in which each qubit holds at most one excitation, so that the
restriction of each transmon to its two lowest levels is exact. Qubit
relaxation, however, drives the system out of $\mathcal{K}_\mathrm{cSMPD}$.
The relaxation of $q_1$, of $q_2$, or of both, followed by the remaining conversion and waste decay processes, populates the five additional states
\begin{multline}\label{eq: CPD subspace relax}
\mathcal{K}_\mathrm{relax}=\big\{ \\
\ket{0_b,g,1_m,g,0_w},
\ket{0_b,g,0_m,e,1_w},
\ket{0_b,e,0_m,g,1_w},\\
\ket{0_b,g,0_m,g,1_w},
\ket{0_b,g,0_m,e,0_w}\\
\big\},
\end{multline}
and we project \cref{eq: CPD HL} onto the enlarged subspace
$\mathcal{K}'=\mathcal{K}_\mathrm{cSMPD}\cup\mathcal{K}_\mathrm{relax}$. None of the states of $\mathcal{K}_\mathrm{relax}$ contains a buffer photon, and none contains both an excited $q_1$ and a memory photon, or both excited qubits and a waste photon. Every operator entering \cref{eq: CPD HL}, namely $\hat{b}$, $\hat{q}_1\hat{m}$, $\hat{q}_1\hat{q}_2\hat{w}$, $\hat{b}(1+\hat{q}_1^\dagger\hat{q}_1+\hat{m}^\dagger\hat{m})$, $\hat{q}_1\hat{m}(1+\hat{q}_2^\dagger\hat{q}_2+\hat{w}^\dagger\hat{w})$ and $\hat{b}\hat{m}^\dagger\hat{q}_2\hat{w}$, therefore has vanishing matrix elements on $\mathcal{K}_\mathrm{relax}$: these states give no
contribution to the projected equations, which reduce to those obtained on $\mathcal{K}_\mathrm{cSMPD}$ alone. They are only populated incoherently, through the relaxation noise terms of \cref{eq: CPD HL}. Qubit decoherence thus only enters the coherent dynamics through the effective rates $\tilde{\kappa}_m$ and $\tilde{\kappa}_w$. The projection of \cref{eq: CPD HL} onto $\mathcal{K}'$ reads:

\begin{equation}\label{eq: CPD  restricted}
    \begin{cases}
        \begin{aligned}
            \partial_t\hat{b}=&-i\left[\delta\hat{b} + g_{4,1}^*\hat{q}_1\hat{m}\right]-\dfrac{\kappa_b}{2}\hat{b}+\hat{F}_b\\[10pt]
            \partial_t\left(\hat{q}_1\hat{m}\right)=&-i\left[\left(\delta + \Delta_m\right)\hat{q}_1\hat{m}+g_{4,1}\hat{b}+g_{4,2}^*\hat{q}_1\hat{q}_2\hat{w}\right] \\ &
            -\dfrac{\tilde{\kappa}_m}{2}\hat{q}_1\hat{m}+ \hat{F}_{q_1 m}\\[10pt]
             \partial_t\left(\hat{q}_1\hat{q}_2\hat{w}\right)=&-i\left[\left(\delta + \Delta_w\right)\hat{q}_1\hat{q}_2\hat{w} +g_{4,2}\hat{q}_1\hat{m}\right]\\ &-\dfrac{\tilde{\kappa}_w}{2}\hat{q}_1\hat{q}_2\hat{w}  + \hat{F}_{q_1q_2w}
        \end{aligned}
    \end{cases}
\end{equation}

\noindent In the following subsections we analyze the semi-classical steady-state and transient responses governed by the set of quantum Langevin equations in the restricted single propagating excitation subspace, given in \cref{eq: CPD restricted}. The steady-state analysis recovers the coupled cavity model results presented in \cref{sec: Transmission coefficient of a 1D chain}. This agreement is central to understanding the behavior of the cSMPD, particularly its efficiency in transferring an itinerant photon across its constituent cavities via frequency conversion, as well as its bandwidth. The transient analysis then provides the corrections needed to bring the steady-state description closer to the experimental setting.

\subsubsection{Semi-classical analysis}

\textbf{Steady-state analysis} --- We adopt an average amplitude approach for all operators, denoting $\langle\hat{b}\rangle=\beta$, $\langle\hat{b}_{\mathrm{in}}\rangle=\beta_{\mathrm{in}}$, $\langle\hat{q}_1\hat{m}\rangle=\mu$ and $\langle\hat{q}_1\hat{q}_2\hat{w}\rangle=\nu$. The drive terms are drastically simplified: in a real-case scenario, only the input resonator is explicitly driven by a calibrated weak coherent tone, and all other resonators only exhibit quantum fluctuations. Focusing on the study of the average amplitude of the modes simplifies the problem by averaging to zero all drive terms except the first one on the mode $b$. Importantly, we have $\langle\hat{q}_1\hat{m}_{\mathrm{in}}\rangle = \langle \hat{q}_1\hat{q}_2\hat{w}_{\mathrm{in}} \rangle = 0$ as only the buffer mode is driven. \cref{eq: CPD restricted} yields:

\begin{equation}\label{eq: CPD HL mean-field}
\begin{cases}
\partial_t\beta=-i\left[\delta\beta+g_{4,1}^*\mu\right]-\dfrac{\kappa_b}{2}\beta+\sqrt{\kappa_{b,\mathrm{ext}}}\,\beta_\mathrm{in}\\[10pt]
\partial_t\mu=-i\left[(\delta+\Delta_m)\mu+g_{4,1}\beta+g_{4,2}^*\nu\right]-\dfrac{\tilde{\kappa}_m}{2}\mu\\[10pt]
\partial_t\nu=-i\left[(\delta+\Delta_w)\nu+g_{4,2}\mu\right]-\dfrac{\tilde{\kappa}_w}{2}\nu
\end{cases}
\end{equation}

In order to obtain the steady-state result, we set $\partial_t\beta = \partial_t\mu = \partial_t\nu = 0$. We can now evaluate the transmission coefficient of the coupled cavity model, $S_{wb} = \nu_\mathrm{out}/\beta_\mathrm{in}$.

We leverage the \cref{eq: cc model n=3} to directly express the transmission coefficient of the two-qubit cSMPD by defining $\lambda_b = i\delta + \kappa_b/2$, $\lambda_m= i(\Delta_m+\delta)+\tilde{\kappa}_m/2$ and $\lambda_w=i(\Delta_w+\delta)+\tilde{\kappa}_w/2$:

\begin{multline}\label{eq: general S21 expression}
    S_{wb}(\delta,\Delta_{p_1},\Delta_{p_2}, \xi_1, \xi_2) = \\\sqrt{\kappa_{w,\mathrm{ext}}\kappa_{b,\mathrm{ext}}}\cdot\dfrac{g_{4,1}g_{4,2}}{\lambda_b \lambda_m \lambda_w + \abs{g_{4,1}}^2\lambda_w+\abs{g_{4,2}}^2\lambda_b}
    .
\end{multline}

The analytical method presented in this section can be trivially extended to a one-dimensional chain of $N$ modes to derive \cref{eq: s21 amp 1D cav chain generic}.\\

\noindent\textbf{Transient analysis} --- The cSMPD operates cyclically, with the pumps switched on only during a detection window of duration $T_d$. The intra-cavity fields must therefore build up anew in each cycle, which affects the measured efficiency. To quantify this effect, we evaluate \cref{eq: diff eq 1D general solution} with $\boldsymbol{\alpha} = (\beta, \mu, \nu)^\top$ and the drive $\mathbf{d}(\tau) = \left(\sqrt{\kappa_{b,\mathrm{ext}}}\,\beta_\mathrm{in}\,\Theta(\tau), 0, 0\right)^\top$, where $\Theta$ is the Heaviside step function. The dynamical matrix reads

\begin{equation}\label{eq: diff eq cSMPD matrix form}
    \mathbf{M} = \begin{pmatrix}
            i\delta + \dfrac{\kappa_b}{2} & ig^*_{4,1} & 0 \\
            ig_{4,1} & i\left( \delta + \Delta_m \right) + \dfrac{\tilde{\kappa}_m}{2} & ig^*_{4,2}\\
            0 & ig_{4,2} & i\left( \delta + \Delta_w \right) + \dfrac{\tilde{\kappa}_w}{2}
        \end{pmatrix}.
\end{equation}

\begin{figure}[h]
    \centering
    \includegraphics[width=8.4cm]{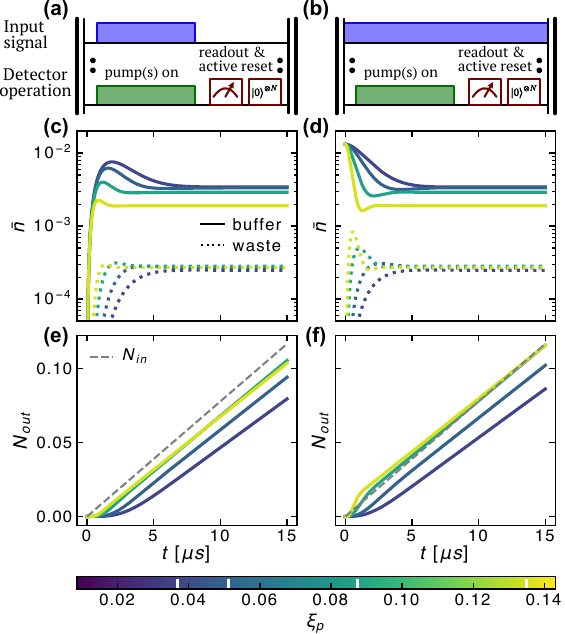}
    \caption{\textbf{Transient dynamics for the cSMPD2 parameters.}
(a,b) Pulse sequences of the two calibration protocols: (a) synchronous, with the input tone and pumps switched on and off together; (b) continuous, with the input tone always on while the pumps are cycled. Each pump pulse is followed by readout and active reset.
(c,d) Intra-cavity populations $\bar n$ of the buffer, $|\beta(t)|^2$ (solid lines), and of the waste, $|\nu(t)|^2$ (dotted lines).
(e,f) Cumulative number of photons emitted by the waste, $N_{\mathrm{out}}(t)=\int_0^t d\tau\,|\nu_{\mathrm{out}}(\tau)|^2$, compared with the incident photon number $N_{\mathrm{in}}(t)=|\beta_\mathrm{in}|^2 t$ (gray dashed line). Left and right columns correspond to protocols (a) and (b), respectively. Colors encode the pump amplitude $\xi_p$ defined in the main text, with the ratio of the two pump amplitudes held fixed. White ticks on the color bar mark the plotted values.}
    \label{fig: transient theory}
\end{figure}

\noindent Since the pumps are off between detection windows, no photons are converted into the memory and waste modes, so that $\mu(0) = \nu(0) = 0$. The initial buffer field, however, depends on the measurement protocol. We consider the two protocols used to calibrate the device, sketched in \cref{fig: transient theory}(a,b).

In the synchronous protocol, the input tone is pulsed together with the pumps, and the counts are averaged over many cycles. The buffer is then empty at the start of each window, and $\boldsymbol{\alpha}(0) = \mathbf{0}$. In the continuous protocol, the input tone remains on while the detector is cycled, which more closely resembles experiments in which the source emits photons in long wave packets~\alex{\cite{braggio_quantum-enhanced_2024, osullivan_nuclearspinregister2024, travesedo_all-microwave_2024, albertinale_detecting_2021, wang_single-electron_2023}}. During readout and reset, with the pumps off, the input tone populates the isolated buffer. Since $\kappa_b \gg 1/T_\mathrm{RO}, 1/T_\mathrm{reset}$, the buffer reaches its steady state before the next window begins, and

\begin{equation}\label{eq: continuous mode starting state}
    \boldsymbol{\alpha}(0) = \left( \frac{\sqrt{\kappa_{b,\mathrm{ext}}}\,\beta_\mathrm{in}}{\kappa_b/2 + i\delta},\, 0,\, 0 \right)^\top .
\end{equation}

\noindent \Cref{fig: transient theory}(c,d) shows the resulting buffer and waste populations for the cSMPD2 parameters and several pump amplitudes $\xi_p$. In the synchronous protocol, the buffer fills up and its population is progressively transferred along the chain to the waste until a steady state is reached. In the continuous protocol, the preloaded buffer instead partially empties at the start of the window: coupling to the rest of the chain opens an additional decay channel, so its steady-state population is lower than that of the isolated buffer. Both protocols converge to the same steady state, which is reached faster at higher pump amplitudes owing to the faster 4WM conversion. The steady-state waste population sets the output photon flux through \cref{eq: langevin boundary cond}. If the cooperativity were exactly unity at every pump amplitude, this population would be the same for all of them; its small spread reflects the slight deviation of the cooperativity from unity at the experimental pump amplitudes (see \cref{appendix cooperativity general}).

\Cref{fig: transient theory}(e,f) shows the cumulative number of photons emitted by the waste, $N_\mathrm{out}(t)$. Once the steady state is reached, $N_\mathrm{out}$ grows linearly with a slope that is the same for both protocols, and whose ratio to the incident photon flux gives the steady-state efficiency. For a cooperativity close to unity, this slope matches that of $N_\mathrm{in}(t)$. The two protocols differ only in their transients. In the synchronous protocol, $N_\mathrm{out}$ always remains below $N_\mathrm{in}$, the offset corresponding to the photons needed to fill the chain. This offset, and hence the efficiency loss over a finite detection window, decreases with increasing pump amplitude. In the continuous protocol, the photons accumulated in the buffer during readout and reset are released at the start of the window, so that $N_\mathrm{out}$ can exceed $N_\mathrm{in}$ at high pump amplitudes. This apparent excess vanishes when $N_\mathrm{in}$ is counted over the full cycle rather than only over the detection window.

\subsection{Qubit decay kernel}
To fully model the detector response, the energy relaxation of the qubits must also be included. In the cyclic operation, the qubits can be excited at any time during the detection window and remain idle until they are read out at its end. During this interval they can decay, thereby losing the information about the incoming photon, which reduces the efficiency. We model this effect using the input and output photon fluxes, $\left|\beta_\mathrm{in}\right|^2$ and $\left|\nu_\mathrm{out}\right|^2$, respectively. Each photon emitted into the waste is accompanied by the excitation of both qubits, so that in the weak-input regime $\left|\nu_\mathrm{out}(t)\right|^2$ can be identified with the probability per unit time $p(e_1e_2; t)$ of exciting both qubits at time $t$. The probability that the qubits remain excited until readout is then $e^{-(T_d - t)/T_{ee}}$, where $T_{ee}$ is the lifetime of the joint excited state. Combining these, the detector efficiency reads

\begin{equation}\label{eq: qubit decay kernel}
    \eta = \frac{\int_0^{T_d}dt\, \left|\nu_\mathrm{out}(t)\right|^2 e^{-(T_d - t)/T_{ee}}}{\int_0^{T_\mathrm{cycle}}dt\, \left|\beta_\mathrm{in}(t)\right|^2}.
\end{equation}

\noindent For a constant input, and neglecting the transient, this reduces to

\begin{equation}\label{eq: qubit decay kernel steady state}
    \eta = \frac{T_d}{T_\mathrm{cycle}}\, \frac{\left|\nu_\mathrm{out}\right|^2}{\left|\beta_\mathrm{in}\right|^2}\, \frac{T_{ee}}{T_d}\left(1- e^{-T_d/T_{ee}}\right) \equiv \eta_\mathrm{cycle}\, \eta_\mathrm{4WM}\, \eta_\mathrm{q},
\end{equation}

\noindent which is the expression used in the main text, \cref{eq: op. eff. cSMPD N=2}.

\section{Cooperativity}
\label{appendix cooperativity general}

\subsection{Definition and application to the 1D chain of coupled modes}

The concept of \textit{cooperativity}, analogous to its role in quantum electrodynamics (QED), serves as a key parameter in determining the operating regime of the one-dimensional chain of cavities—and, by extension, the operating regime of our photon detector. We define the cooperativity~\cite{kimble_strong_1998, clerk_hybrid_2020} as the ratio of the effective engineered dissipation rate of the entire transmission line $\gamma$ to the coupling rate with the input mode $\kappa_1$. Note that this definition is generic for arbitrarily long SMPD/cSMPD chains, but in the main text we defined the cooperativity, for simplicity, as the ratio of dissipation rates experienced by the memory mode rather than the buffer mode. It is expressed as
\begin{equation}\label{eq:coop_general}
    C=\dfrac{\gamma}{\kappa_1}.
\end{equation}
Based on the \cref{eq: s21 amp 1D cav chain generic}, we can derive an analytical expression for the cooperativity as a function of all coupling rates of the 1D cavity chain. We suppose the ideal scenario where all cavities are resonant, and the probe tone is perfectly tuned $\Delta_k=0$. Moreover, internal losses are neglected $\kappa_{k,\mathrm{ext}}=\kappa_k$ and $\forall k \in\interval{2}{N-1}\ \kappa_k=0$. 
\begin{equation}\label{eq: generic cooperativty expression}
C=
\begin{cases}
    N=2:\quad \dfrac{4\abs{g_1}^2}{\kappa_{1}\kappa_N}\\[10pt]
    \forall N> 2:\quad f(\kappa_1,\kappa_N)\abs{\dfrac{\displaystyle\prod_{k=1}^{N-1}g_{2k+1}}{\displaystyle\prod_{k=1}^{N-1}g_{2k}}}^2\\
    \ \mathrm{with } f(\kappa_1, \kappa_N)=\dfrac{\kappa_N}{\kappa_1} \mathrm{if\ \textit{N}\ is\ odd, else } \dfrac{4}{\kappa_1\kappa_N}
\end{cases}
\end{equation}

\noindent
We can apply \cref{eq: generic cooperativty expression} to our three-cavity system. \cref{eq: 3 cav matching cond} directly leads to $\gamma=\kappa_3\abs{g_1/g_2}^2$, and thus
\begin{equation}\label{eq: coop 3 cav}
    C = \frac{\kappa_3}{\kappa_1}\abs{\frac{g_1}{g_2}}^2
\end{equation}

The cooperativity serves as a metric for quantifying how well the dissipation rate at the input of the 1D cavity chain is matched to the effective dissipation of the remainder of the system—which, in the general case, can be abstracted as a load. This concept is directly analogous to impedance matching in radio-frequency engineering: if the dissipation rates are not matched, energy is not transmitted efficiently through the system, and a fraction of it is reflected. There are three possible regimes:

\begin{itemize}
    \item $C\ll1$: photon number transmission is small due to weak coupling or extremely bad matching.
    \item $C=1$: ideal matching/critical coupling is achieved.
    \item $C\gg1$: over-coupling suppresses the transmission and the reflection dominates.
\end{itemize}

From an operational perspective, the cooperativity can be interpreted as the dimensionless parameter in terms of which the transmission coefficient can be expressed. \cref{eq: s21 amp 1D cav chain generic}, modulus squared and factorized at the numerator and the denominator by $1/\displaystyle\prod_{k=1}^{N-1}\abs{g_{2k}}^2$, can be generally casted as
\begin{multline}\label{eq: general S21 Nqubit with f}
\abs{S_{N1}(\omega)}^2 \\=\dfrac{4C}{\abs{1+C+\dfrac{\kappa_{1,\mathrm{in}}}{\kappa_{1,\mathrm{ext}}}+f(\Delta_1,\dotso,\Delta_N,\kappa_1,\dotso,\kappa_N)}^2}
\end{multline}
where $f$ is a complicated polynomial. All terms of $f$ are proportional to frequency detunings or intermediate resonators internal losses. In the idealized case where all detunings are zero, the input photon is on resonance and the resonators exhibit no internal loss, \cref{eq: general S21 Nqubit with f} yields to:
\begin{equation}\label{eq: generic S21 limit when C=1}
\abs{S_{N1}(\omega=0)}^2 = \dfrac{4C}{\abs{1+C}^2}\xrightarrow{C\rightarrow 1}1
\end{equation}
It is important to emphasize that when detunings and internal loss rates are non-zero, the interference symmetry underlying ideal transmission is broken. Finite frequency detunings introduce phase mismatches that accumulate along multiple reflection paths, thereby preventing complete destructive interference of the reflected components. Internal losses, in turn, lead to irreversible energy dissipation within the system, so that cooperativity is no longer the sole parameter to optimize. In this regime, minimizing energy loss rates becomes equally crucial to maintaining unit transmission. This interplay plays a central role in the operation of the cSMPD.

\subsection{Effect of the cooperativity $C$ on the internal dynamics of the detector}

To accurately model the propagating nature of the itinerant electromagnetic field subject to detection, we adopt the theoretical framework developed in~\cite{motzoi_precise_2018, kiilerich_input-output_2019, kiilerich_quantum_2020, khanahmadi_multimode_2023}.

We begin by describing the emission of an itinerant single photon and its subsequent propagation along a transmission line. From the perspective of a quantum system, being driven by a traveling pulse is equivalent to being driven by the output field of a leaky cavity. To faithfully model this process, it is necessary to implement a non-reciprocal dynamical scheme that ensures the emitted photon propagates unidirectionally, i.e without being reflected back toward the emitter.

\subsubsection{Auxiliary cavity with a constant leak rate}
\label{sec: Generation of an itinerant single photon}

\begin{figure}[h]
    \centering
    \includegraphics[width=8.4cm]{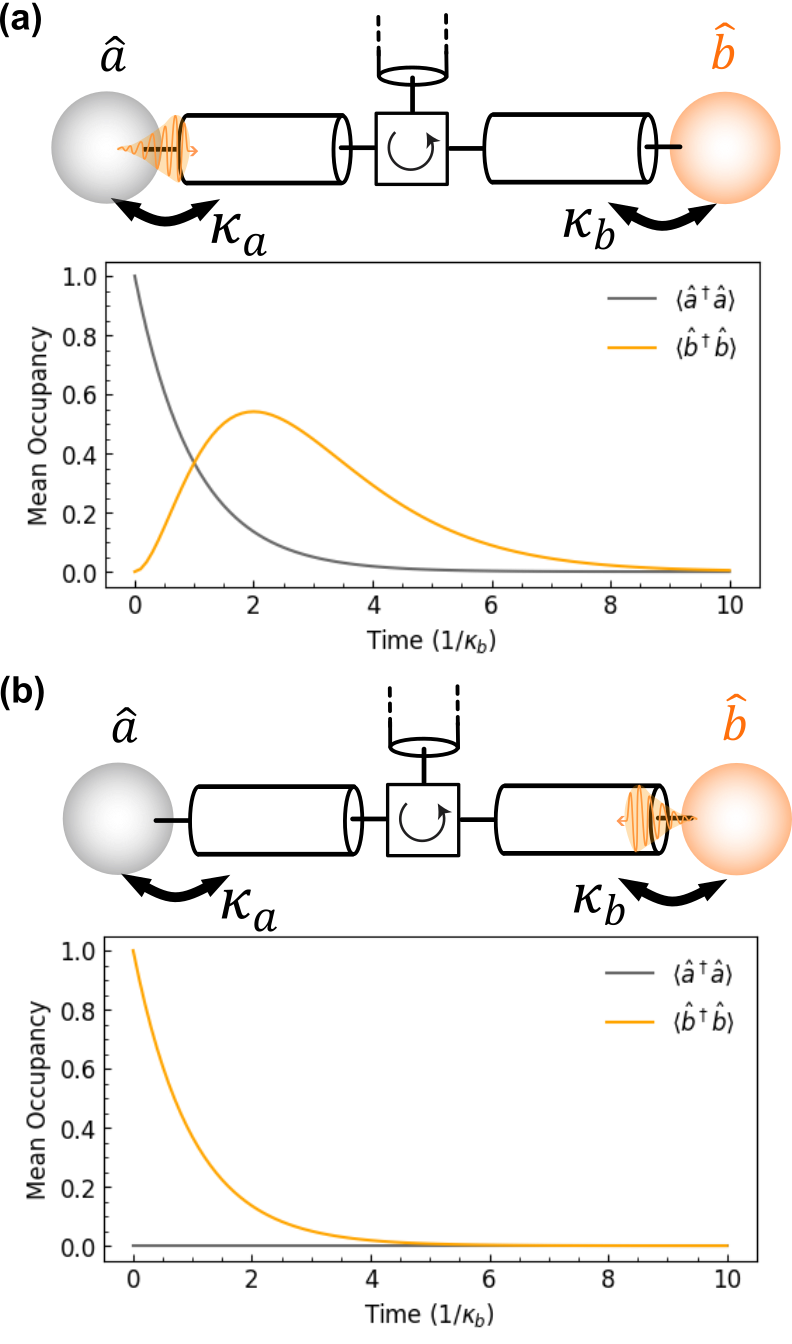}
    \caption{\textbf{Simulation of the non-reciprocal dynamics \cref{eq: master eq ab reduced}} with two modes $\hat{a}$ and $\hat{b}$, with $\kappa_a=\kappa_b$ and $\omega_a=\omega_b$. Mode occupancy are tracked as a function of time. Modes are truncated to two levels. (a) Initialization in $\ket{1_a,0_b}$. (b) Initialization in $\ket{0_a,1_b}$.}
    \label{fig: quantum circulator}
\end{figure}

The minimal model to describe such dynamics is to consider a quantum emitter-receiver system composed of a single mode cavity $\hat{a}$ coupled to its environment at a rate $\kappa_a$ (emitter) and a single mode cavity $\hat{b}$ coupled to its environment at a rate $\kappa_b$ (receiver). To simplify the model, we consider zero frequency detuning between the two modes, but the model can be trivially extended to finite detunings. The interaction between the two cavities can be modeled by a Jaynes-Cummings Hamiltonian \cite{GardinerZoller_quantumNoise}
\begin{equation}\label{eq: Hjc ab model}
    \frac{\hat{H}_{ab}}{\hbar} = i\frac{\sqrt{\kappa_a\kappa_b}}{2}\left(\hat{a}^\dagger\hat{b}-\hat{a}\hat{b}^\dagger\right)
\end{equation}
and a correlated dissipative part
\begin{equation}\label{eq: correlatev L ab}
    \hat{L}_{ab} = \hat{L}_a+\hat{L}_b
\end{equation}
where $\hat{L}_a = \sqrt{\kappa_a}\hat{a}$ and $\hat{L}_b=\sqrt{\kappa_b}\hat{b}$. We insist on the fact that the correlated dissipation is at the heart of the non-reciprocity. \cref{eq: correlatev L ab} can be interpreted as an input-output relation that represents the output field of the mode $\hat{b}$, the output field being the interfering contributions of the input field ($\hat{L}_a$) and the emission of the mode $\hat{b}$ itself ($\hat{L}_b$). The density matrix describing the state of the system follows a master equation
\begin{equation}\label{eq: master eq ab}
    \dot{\rho}=-i[\hat{H}_{ab}, \rho] + \mathcal{D}[\hat{L}_{ab}]\rho.
\end{equation}
After some straightforward algebra, \cref{eq: master eq ab} reduces to

\begin{multline}\label{eq: master eq ab reduced}
\dot{\rho}=\mathcal{D}[\hat{L}_a]\rho+\mathcal{D}[\hat{L}_b]\rho\\
+\sqrt{\kappa_a\kappa_b}\left(\hat{a}\rho\hat{b}^\dagger+\hat{b}\rho\hat{a}^\dagger
-\hat{b}^\dagger\hat{a}\rho-\rho\hat{a}^\dagger\hat{b}\right).
\end{multline}

\cref{eq: master eq ab reduced} comprises the individual damping of each mode together with a cross term that couples them. In this cross term, the contributions of
the Hamiltonian~\cref{eq: Hjc ab model} and of the correlated dissipator~\cref{eq: correlatev L ab} add up for
the terms $\hat{b}^\dagger\hat{a}\rho$ and $\rho\hat{a}^\dagger\hat{b}$,
which transfer excitations from $\hat{a}$ to $\hat{b}$, and cancel
exactly for the terms $\hat{a}^\dagger\hat{b}\rho$ and
$\rho\hat{b}^\dagger\hat{a}$, which would transfer them back from
$\hat{b}$ to $\hat{a}$. This destructive interference is the origin of
the non-reciprocity. It is made explicit by the equations of motion of
the mode amplitudes,
$\partial_t\langle\hat{a}\rangle=-(\kappa_a/2)\langle\hat{a}\rangle$ and
$\partial_t\langle\hat{b}\rangle=-(\kappa_b/2)\langle\hat{b}\rangle
-\sqrt{\kappa_a\kappa_b}\,\langle\hat{a}\rangle$: mode $\hat{a}$ drives
mode $\hat{b}$, whereas $\hat{b}$ has no effect on the dynamics of
$\hat{a}$.

The model is illustrated in \cref{fig: quantum circulator}. We simulate the dynamics of the coupled modes under two different initial conditions, expressed in the Fock basis: (i) $\ket{1_a,0_b}$, and (ii) $\ket{0_a,1_b}$. In the first case, we observe that the photon initially present in mode $\hat{a}$ propagates towards mode $\hat{b}$, populating it, and is subsequently lost to the environment. Furthermore, the population of mode $\hat{a}$ decreases monotonically, indicating that any back-propagation of the photon from mode $\hat{b}$ to mode $\hat{a}$ is suppressed due to a destructive interference mechanism. To reinforce this observation, we consider the system initialized in the state $\ket{0_a,1_b}$ and monitor its evolution. The excitation in mode $\hat{b}$ decays at a rate $\kappa_b$, while mode $\hat{a}$ remains unpopulated, consistently in its vacuum state throughout the evolution.

\subsubsection{Auxiliary cavity with a time-dependent leak rate}
\label{sec: Pulse shaping}

The quantum state launching protocol described in \cref{sec: Generation of an itinerant single photon} can be generalized to the emission of quantum pulses with arbitrary temporal envelopes. A precise modeling of pulse shaping necessitates a more refined treatment of the auxiliary cavity mode $\hat{a}$, which is initialized in a quantum state, and the temporal output mode $\hat{a}_u$, into which the state is emitted. This section outlines how the time-dependent coupling $g_u(t)$ between the auxiliary cavity and the quantum system $\hat{b}$ is engineered to model the interaction with a wave packet characterized by the envelope $u(t)$.

We consider the single mode cavity described by the annihilation operator $\hat{a}$ and its interaction with an input field $\hat{c}_\mathrm{in}(t)$. The interaction term reads 
\begin{equation}\label{eq: v coupling}
\frac{\hat{V}}{\hbar} = i\left[g(t)\hat{a}^\dagger \hat{c}_\mathrm{in}(t)-g^*(t)\hat{a}\hat{c}^\dagger_\mathrm{in}(t)\right].
\end{equation}  

The input-output equation followed by field operator of the auxiliary cavity is
\begin{equation}\label{eq: inout aux cav}
    \dot{\hat{a}}(t) = -\frac{\abs{g(t)}^2}{2}\hat{a}(t) - g(t)\hat{c}_\mathrm{in}(t).
\end{equation}
The general solution of equation \cref{eq: inout aux cav} is
\begin{equation}\label{eq: 1.47}
\begin{split}
\hat{a}(t) ={}& e^{-\frac{1}{2} \int_0^t \dd{t'}\, |g(t')|^2}\, \hat{a}(0) \\
&- \int_0^t \dd{t'}\, g(t')\, e^{-\frac{1}{2} \int_{t'}^t \dd{t''}\, |g(t'')|^2}\, \hat{c}_{\mathrm{in}}(t')
\end{split}
\end{equation}
The input-output relation connecting the incoming and outgoing fields at time $t$ that spatially overlap with the auxiliary cavity is $\hat{c}_\mathrm{out}(t)^\dagger=\hat{c}_\mathrm{in}^\dagger(t)+g(t)\hat{a}^\dagger(t)$. We represent the wave packet $u(t)$ that irradiate the quantum system under study as a weighted integral over the temporal mode of the bath, i.e
\begin{equation}
    \hat{c}_u^\dagger(t)=\int\dd{t} u(t)\hat{c}^\dagger_\mathrm{out}(t).
\end{equation}
Based on \cref{eq: 1.47}, the envelope of the temporal output mode populated by the  auxiliary cavity follows
\begin{equation}
    u(t)=g_u^*(t)e^{-\frac{1}{2} \int_0^t \dd{t'}\, |g(t')|^2} \longrightarrow g_u(t)=\frac{u^*(t)}{\sqrt{1-\int_0^t\dd{t'}\abs{u(t')}^2}}
\end{equation}
The envelope function is normalized to 1, i.e $\int\dd{t}\abs{u(t)}^2=1$. The time-dependent coupling $g_u(t)$ is required to emit the pulse shape $u(t)$. It is equivalent to modulate in time the transparency coefficient of the single sided auxiliary cavity mirror to sculpt its output field. We note that we can retrieve the \cref{sec: Generation of an itinerant single photon} results by taking $g(t)=\sqrt{\kappa_a}$.

We can now describe the join state of the quantum system under study (here the cavity mode $\hat{b}$), and the field annihilation operator $\hat{a}_u(t)$ (acting on the temporal mode with envelope $u(t)$) by a density matrix following a master equation with an interaction Hamiltonian
\begin{equation}
    \frac{\hat{H}_\mathrm{ab}}{\hbar}=i\frac{\sqrt{\kappa_b}}{2}\left[g_u(t)\hat{a}_u^\dagger\hat{b}-g^*_u(t)\hat{a}_u\hat{b}^\dagger\right]
\end{equation}
and a correlated dissipative part
\begin{equation}
    \hat{L}_{ab} = g_u^*(t)\hat{a}_u + \sqrt{\kappa_b}\hat{b}
\end{equation}

\begin{figure}[h]
    \centering
    \includegraphics[width=8.4cm]{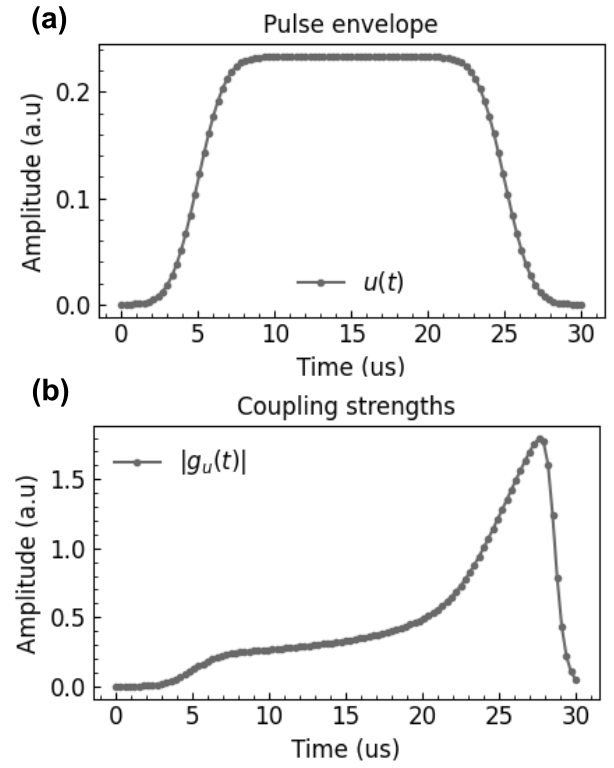}
    \caption{\textbf{Input pulse used in the cSMPD numerical simulation.} The gaussian error function rising and lowering time are \SI{6}{\micro\s}, the flat time is \SI{20}{\micro\s} long. (a) Propagating pulse driving the detector. (b) Coupling coefficient of the auxiliary cavity to generate the pulse on panel (a).}
    \label{fig: pulse erf rise}
\end{figure}

Having established a consistent framework for modeling a quantum drive, we now return to the 4WM cascaded single photon detector dynamics.

\subsection{Quantum simulation of the 3-coupled mode model}

We simulate the open quantum system detection dynamics across different cooperativity regimes with QuTip \cite{Qutip,Qutip2}, and show that a cooperativity of $C=1$ leads to optimal detection performances. 

The modes are considered two-level systems. The detection window is a square with a duration $T_d=\SI{30}{\micro\s}$. We use the quantum state launching tool described in the previous sections based on a leaky auxiliary cavity initialized in Fock state $\ket{1}$, while any other modes are in their vacuum states. We neglect internal loss rates as well as equilibrium thermal population and dephasing rates. We chose an experimentally relevant pulse shape for the cSMPD efficiency numerical characterization, a square pulse with gaussian error function rising and lowering part (the flat part is $\SI{20}{\micro\s}$ long, and the rising/lowering $\SI{6}{\micro\s}$ long), illustrated on \cref{fig: pulse erf rise}. We define the detector efficiency as  $\eta\equiv \min\left[P_{e,1}(t=T_d),\ P_{e,2}(t=T_d)\right]$, with $P_{e,k}(t)=\langle\hat{q}_k^\dagger(t)\hat{q}_k(t)\rangle,\ k\in\lbrace 1,2\rbrace$. 

We set symmetrical coupling rates to the environment on the buffer and waste resonator $\kappa_b=\kappa_w=\SI{3.1e6}{\per\s}$, and choose $\abs{g_{4,1}}/2\pi=\SI{300}{\kilo\Hz}$. Without lack of generality, we also choose the auxiliary cavity coupling $g_u$ to be real valued and define $g_u(t) = \sqrt{\kappa_a(t)}$. Finally, we set the strength $g_{4,2}$ of the second 4WM process based on the chosen cooperativity values.

\begin{figure*}[ht!]
    \centering
    \includegraphics[width=\textwidth]{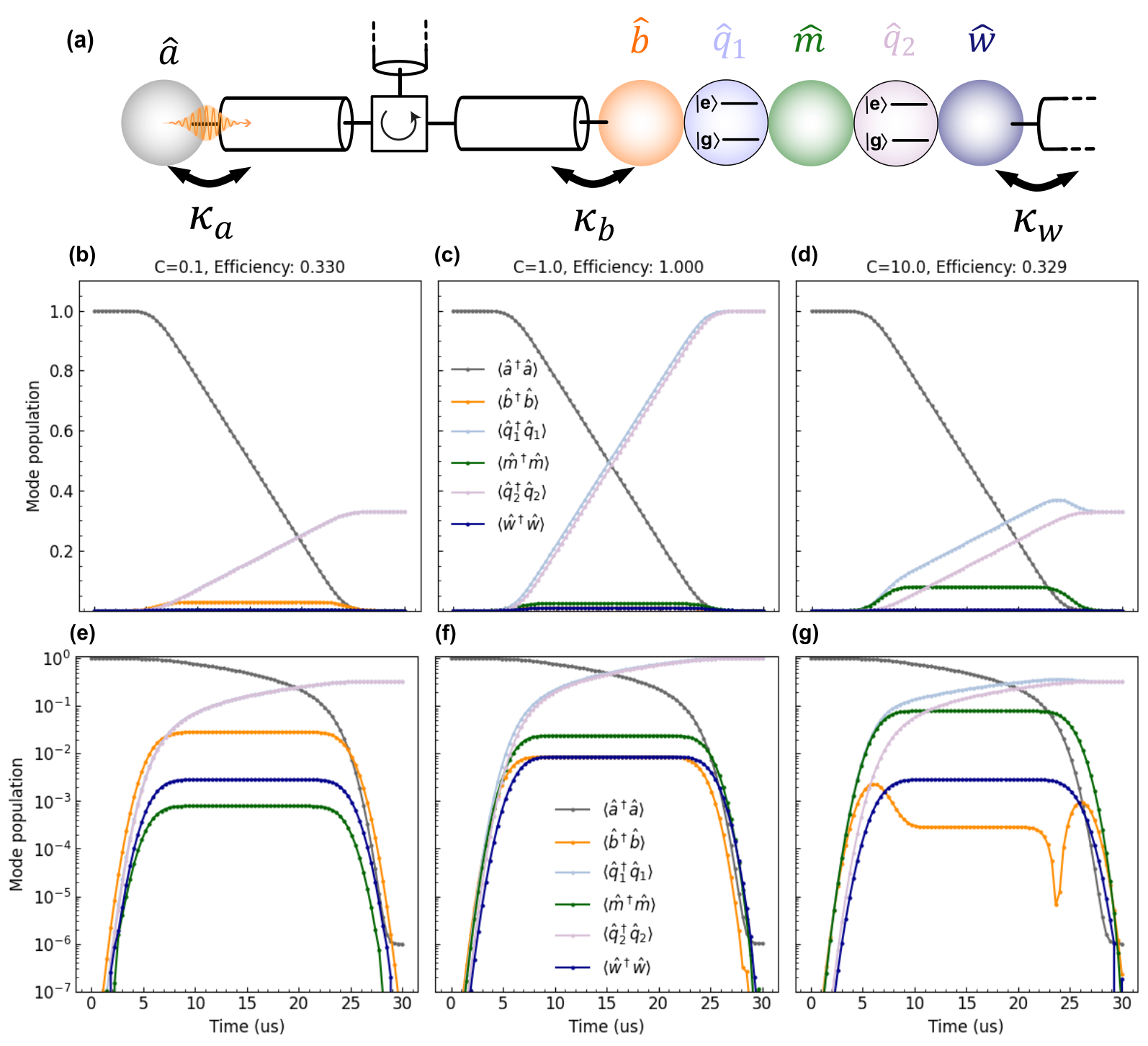}
    \caption{\textbf{Numerical quantum simulation of a two-qubit cSMPD in different cooperativity regimes.} The simulated system is constituted of the quantum state launching tool described in \cref{sec: Pulse shaping} and the effective 4WM dynamics of the 3-coupled mode system. We use $\kappa_b=\kappa_w=\SI{3.1e6}{\per\s}$ ($2\pi\times\SI{1}{\mega\Hz}$), $\abs{g_{4,1}}/2\pi=\SI{300}{\kilo\Hz}$, and $\kappa_m=\kappa_\mathrm{b,int}=0$ and individual qubit $T_2^*$ are infinite. The pump tones are perfectly tuned, and the input pulse is resonant with the buffer mode ($\delta=0$). All modes are truncated to two level systems. (b-g) We track the mode populations as a function of time, and define the efficiency of the detection mechanism by $\langle\hat{q}_2^\dagger\hat{q}_2\rangle$ at the end of the simulation window. (b-d) Linear scale. (e-g) Log10 scale. (b,e) $C=0.1$. (c,f) $C=1$. (d,g) $C=10$.}
    \label{fig: coop demo}
\end{figure*}

The quantum simulation result in three different cooperativity regimes (0.1, 1, and 10) are illustrated on \cref{fig: coop demo}(b-d). We observe a  small detector efficiency in the $C\ll 1 $ and $C\gg 1$, although for different reasons. In the $C\ll 1 $ case, the input pulse is reflected due to the poor rate matching. In the $C\gg 1$ case, we observe oscillations on the buffer occupancy (orange curve). This is the sign of reflections happening internally to the detector: the first 4WM mixing is faster than the second one, hence the population of the memory is partially retro-converted into the buffer mode, suffering from the buffer coupling rate to the environment and the absence of destructive interference to cancel those losses. The case $C=1$ shows that we can reach $\eta=1$ with our cSMPD scheme.

\section{Resonance conditions and qubit self-Kerr correction to the three mode model}
\label{sec: Kerr correction to 3 cav semiclassical model}
 
\subsection{Resonance conditions of the cascaded process}
\label{Appendix_spuriousparametricprocess}
 
\begin{figure}[h]
    \centering
    \includegraphics[width=\linewidth]{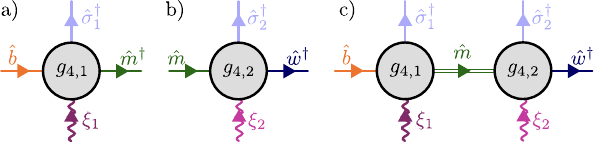}
    \caption{\textbf{Diagrams of the conversion processes}~\cite{xiao_diagrammatic_2023}. Straight arrows: excitations, labeled by their mode. Wavy arrows: pump photons. Double arrow: virtual excitation. (a) First 4WM process (buffer, $q_1$, memory). (b) Second 4WM process (memory, $q_2$, waste). (c) Second-order process with a virtual memory excitation.}
    \label{fig:appL_diagrams}
\end{figure}

\Cref{figMAIN: cSMPD1 pump maps} shows the probability to find the qubits excited as a function of the two pump detunings. We draw the processes behind its lines with the diagrams of Ref.~\cite{xiao_diagrammatic_2023} [\cref{fig:appL_diagrams}]. A process is resonant when the sum of the incoming frequencies equals the sum of the outgoing frequencies. All frequencies include the shifts defined after \cref{eq: freqmatchcond CPD N=2}.
 
The first 4WM process [\cref{fig:appL_diagrams}(a)] converts a buffer photon and a pump-1 photon into a $q_1$ excitation and a memory photon. Its resonance condition is the first line of \cref{eq: freqmatchcond CPD N=2}. This condition does not depend on $\omega_{p_2}$, so it gives the horizontal line in \cref{figMAIN: cSMPD1 pump maps}(b). The memory frequency $\tilde{\omega}_m$ contains the term $-\chi_{q_1m}$ because the memory photon appears while $q_1$ is in $\ket{e}$. The second 4WM process [\cref{fig:appL_diagrams}(b)] needs a memory photon. It occurs only where the first process is also resonant, which gives the bright spot in \cref{figMAIN: cSMPD1 pump maps}(d,f).
 
The oblique line [black dashed line in \cref{figMAIN: cSMPD1 pump maps}(d)] comes from the second-order process of \cref{fig:appL_diagrams}(c). A buffer photon, a pump-1 photon and a pump-2 photon become one excitation of each qubit and one waste photon. The memory excitation is virtual. The resonance condition is the sum of the two lines of \cref{eq: freqmatchcond CPD N=2}:
\begin{equation}\label{eq: virtual resonance}
    \tilde{\omega}_b + \omega_{p_1} + \omega_{p_2} = \tilde{\omega}_{q_1} + \tilde{\omega}_{q_2} + \tilde{\omega}_w .
\end{equation}
This condition fixes only the sum $\omega_{p_1}+\omega_{p_2}$. If the dressed frequencies do not depend on the pump frequencies, the line has a slope of $-1$ [see \cref{figMAIN: fig2}(a)]. Along this line, the first conversion would emit a memory photon at a frequency detuned by $\Delta_m$ from $\tilde{\omega}_m$. For $\abs{\Delta_m}\gg\kappa_m$, the conversion amplitude scales as $g_{4,1}g_{4,2}/\Delta_m$. The three-coupled-mode model of \cref{eq: general Swb expression} contains this process through $\Delta_m$ in $\lambda_m$. The next subsection explains why the measured slope differs from $-1$.
 
\subsection{Qubit self-Kerr correction}
\label{appendix: self-Kerr correction}

Even though we do not need the Kerr terms to derive a model to estimate the bandwidth and 4WM efficiency, we can go further and include the most important Kerr correction to that model, i.e the qubit self-Kerrs. The energy conservation relations reads $\omega_b + \left(\omega_{p_1}-\Delta_{p_1}\right) = \omega_{q_1} -2\chi_{q_1q_1}\abs{\xi_1(\omega_{p_1})}^2 + \left(\omega_{m}-\Delta_{m}\right)$ and $\left(\omega_m - \Delta_{p_1}\right) + \left(\omega_{p_2}-\Delta_{p_2}\right) = \omega_{q_2} -2\chi_{q_2q_2}\abs{\xi_2(\omega_{p_2})}^2 + \left(\omega_{w}-\Delta_{w}\right)$.

We neglect the cross-Kerr terms as in practice $\abs{\xi_1}\sim\abs{\xi_2}$, and $2\chi_{q_1q_1}\gg\chi_{q_1b}, \chi_{q_1m}, \chi_{q_2m}$, $2\chi_{q_2q_2}\gg\chi_{q_2m}, \chi_{q_1m}, \chi_{q_2w}$. We introduce the generalized pump frequency detunings $\Delta_{p_k}'=\Delta_{p_k}-2\chi_{q_kq_k}\abs{\xi_k(\omega_{p_k})}^2,\ k\in\{1,2\}$. The previous 4WM resonance conditions transform into
    \begin{align}
        \omega_b + \left(\omega_{p_1}-\Delta_{p_1}'\right) &= \omega_{q_1} + \left(\omega_{m}-\Delta_{m}\right)\label{eq: 2.10}\\
        \left(\omega_m - \Delta_{p_1}'\right) + \left(\omega_{p_2}-\Delta_{p_2}'\right) &= \omega_{q_2} + \left(\omega_{w}-\Delta_{w}\right).\label{eq: 2.11}
    \end{align}
    \noindent 
4WM terms are time-independent as long as $\Delta_m=\Delta_{p_1}'$ and $\Delta_w = \Delta_{p_1}' + \Delta_{p_2}'$.  To illustrate the dependency of the qubit self-Kerr with respect to the pump frequencies, we need to know an experimental set of pump settings and qubit frequencies. We base the following quantitative analysis on the values provided in \cref{table: CPD-0002-02-A-and-W025-CPD-0007-03-01-C1-circuit-params} and \cref{table: circuit-params-part2-merged}. The algorithm to compute the transmission coefficient with qubit self-Kerr corrections follows:
\begin{enumerate}
    \item Choose $\Delta_{p_1}$ and $\Delta_{p_2}$ frequency detuning arrays.
    \item Compute the swept pump frequency arrays as $\omega_{p_k,\mathrm{sweep}} = \omega_{p_k}-\Delta_{p_k},\ k\in\{1,2\}$. $\omega_{p_k}$ are the center settings, also called optimal setting, i.e the settings at zero pump detuning.
    \item Compute $\Delta_{p_1}'$ and $\Delta_{p_2}'$.
    \item Compute $\abs{g_{4,1,\mathrm{sweep}}} = \abs{\dfrac{\epsilon_1}{\omega_{q_1} - \omega_{p_1,\mathrm{sweep}}}}\sqrt{\chi_{q_1b}\chi_{q_1m}}$ and $\abs{g_{4,2,\mathrm{sweep}}} = \abs{\dfrac{\epsilon_2}{\omega_{q_2} - \omega_{p_2,\mathrm{sweep}}}}\sqrt{\chi_{q_2m}\chi_{q_2w}}$. The pump amplitudes $\epsilon_k$ were not measured directly with the sample CPD-0002-02-A, but we use the fitted $g_{4,k}$ values combined with the cross-Kerr coefficients to infer them.
    \item Compute $\abs{S_{wb}(\delta,\Delta_1',\Delta_2', \abs{\epsilon_1}, \abs{\epsilon_2})}^2$
    \item Eventually center the 2D map onto its maximum.
\end{enumerate}

\begin{figure}[h]
    \centering
    \includegraphics[width=1\linewidth]{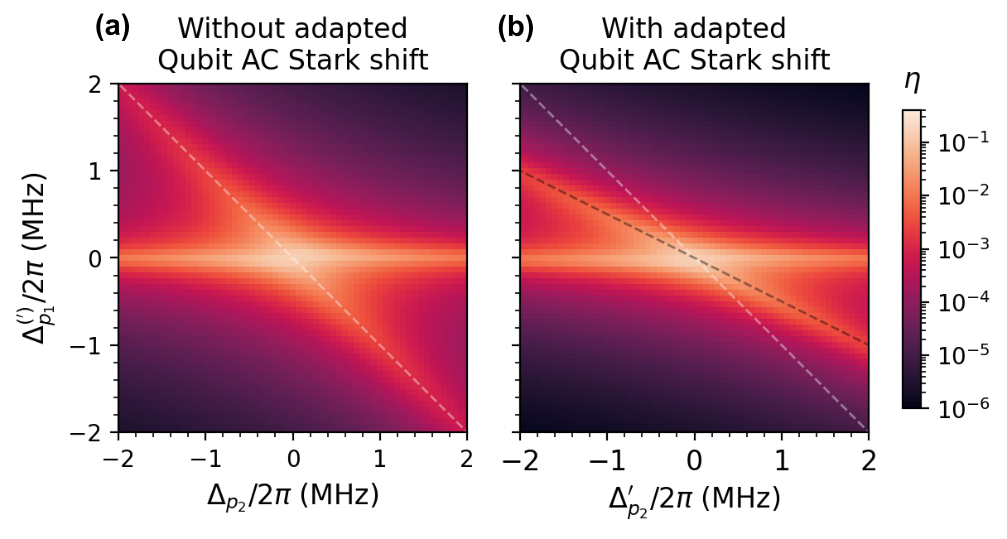}
    \caption{\textbf{Operational efficiency with/without qubits self-Kerr contribution}. Simulation based on cSMPD1 parameters. (a) Model without including the qubit AC Stark shifts. Virtual process on the memory has a slope -1 (white dashed line). (b) Model including the qubit AC Stark shifts. Virtual process on the memory has a slope $\approx -0.5$ (black dashed line).}
    \label{fig: selfKerr correction 3cav model}
\end{figure}

In \cref{fig: selfKerr correction 3cav model} we sweep the pump frequencies while the probe tone is always resonant ($\delta=0$). We scale the transmission coefficient by the factor $\eta_\mathrm{cycle}\ \eta_{q_1}\ \eta_{q_2}\ \mathcal{F}_{RO_1}\ \mathcal{F}_{RO_2}$ to directly plot the probability to find both qubits in their excited states at the end of a detection window. We observe that the slope of the virtual process on the memory mode is lower than one and is almost matching quantitatively the experimental data. The remaining slope discrepancy is attributed to the neglected cross-Kerr terms.

\section{Dark counts}
\label{Appendix_noise}

The noise coming from the buffer input line is modeled as the thermal
noise of a matched load at an effective temperature $T$. In the quantum
regime $hf\gg k_BT$, the available noise power in a bandwidth $\Delta f$
is $P_{\mathrm{th}}=hf\,n_\mathrm{th}(T,f)\,\Delta f$, with
$n_\mathrm{th}(T,f)=1/(e^{hf/k_BT}-1)$ the Bose--Einstein occupation
(it reduces to the classical $k_BT\Delta f$ for $hf\ll k_BT$), so that
the thermal photon flux is $P_{\mathrm{th}}/\hbar\omega=n_\mathrm{th}\Delta f$.
Here $T$ is the effective temperature of the microwave field at the
detector input, in general higher than the cryostat base temperature.

We consider the operational limit where the detection bandwidth is
narrow compared with the frequency scale over which $n_\mathrm{th}$
varies, and where the detector response is Lorentzian, centered on
$f_b$ with a full width at half maximum $\kappa_d/2\pi$, which holds for
$\abs{g_{4,1}}\ll\kappa_b$ and $\abs{g_{4,2}}\ll\kappa_w$.
The expected dark count rate due to thermal photons at $f_b$ then
follows:

\begin{equation}
    \alpha_{\mathrm{th}}=\eta\ n_\mathrm{th}(T, f_b)\int_{\mathbb{R}}\dfrac{1}{1+\left(\dfrac{f-f_b}{\kappa_d/(4\pi)}\right)^2}\mathrm{d}f
\end{equation}

\noindent
The latter integral is an example of  \cref{alpha_th}, and leads to the Lorentzian limit expression given in the main text..

\section{cSMPD2 extended dataset under pumps}
\label{appendix: cSMPD2 extended dataset}
This appendix presents detailed measurements results related to the \cref{figMAIN: fig3} in the main text. We present the $T_1$ data of the two qubits and the memory mode, the coherence time of both qubits, the method we used to calibrate the 4WM strengths, the cryostat heating and a dark count budget while varying the pump amplitudes $\xi_\mathrm{p}$.

\subsection{Memory and qubits $T_1$ under pumps}
\label{appendix: T1s under pump}

\begin{figure}[h]
    \centering
    \includegraphics[width=\linewidth]{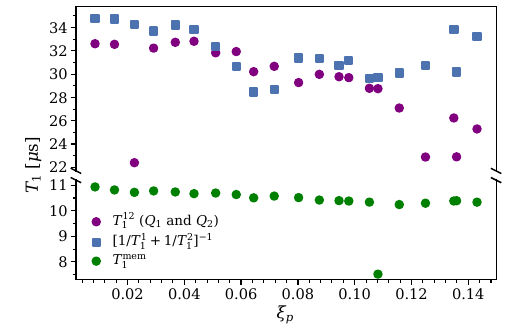}
    \caption{\textbf{Relaxation times under pumping.} Relaxation times measured as a function of the mean pump amplitude $\xi_p$, at a fixed pump ratio. Purple: relaxation time $T_{e_1e_2}$ of the joint excitation $P(e_1e_2)$. Blue: joint relaxation time expected from the individually measured qubit relaxation times. Green: memory relaxation time.}
    \label{fig: T1 under pumps}
\end{figure}

The cSMPD encodes the arrival of an itinerant photon in the joint excitation of both qubits, $P(e_1e_2)$, so that the lifetime $T_{ee}$ of this state directly limits the detector efficiency [\cref{eq: qubit decay kernel}]. We measured it by preparing both qubits in their excited state and fitting the decay of $P(e_1e_2)$ with an exponential. This was repeated for mean pump amplitudes $\xi_p$ from 0.01 to 0.14 at a fixed pump ratio $\beta_\xi \approx 0.62$, with the pumps detuned by \SI{20}{\mega\hertz} from their working point to suppress resonant 4WM conversion. The joint relaxation time decreases from \SI{33}{\micro\second} to \SI{24}{\micro\second} with increasing $\xi_p$, with a sharp dip down to \SI{22}{\micro\second} at $\xi_p \approx 0.02$ [\cref{fig: T1 under pumps}, purple]. It differs from the value expected from the individually measured qubit relaxation times, $\left[1/T_1^{(1)} + 1/T_1^{(2)}\right]^{-1}$ (blue), for reasons that remain unclear. Finally, using the protocol of \cref{Appendix_Memoryloss calibration}, we measured the memory relaxation time for the same pump settings (green). It decreases slightly, from \SI{11}{\micro\second} to \SI{10.5}{\micro\second}, apart from a sharp dip to \SI{7.5}{\micro\second} around $\xi_p \approx 0.11$.

\subsection{Calibration of 4WM mixing strengths}
\label{appendix: 4wmcal under pump}

\begin{figure}[h]
    \centering
    \includegraphics[width=\linewidth]{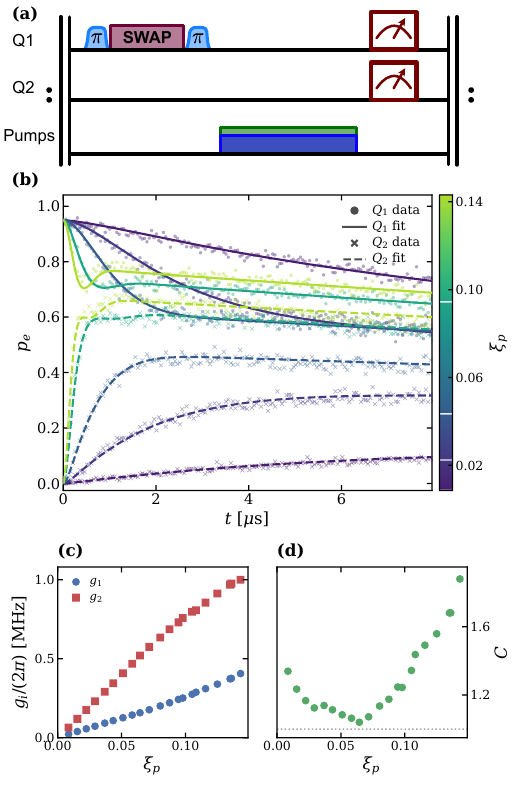}
    \caption{\textbf{Measurement of the 4WM strengths.} The cascaded system is prepared in an intermediate state, with qubit 1 excited and a single photon in the memory, and is then left to evolve under the two competing pumps. (a) Pulse sequence. A $\pi$ pulse on qubit 1 is followed by a parametric SWAP gate that transfers its excitation into the memory, leaving the memory in $\ket{1}$ and the qubit in its ground state. A second $\pi$ pulse brings qubit 1 back into its excited state. The pumps are then applied for a variable duration before both qubits are read out. (b) Averaged populations of qubit 1 (circles) and qubit 2 (crosses). Colors encode the mean pump amplitude $\xi_p$, with white ticks on the color bar marking the plotted values. Solid (dashed) lines are fits of the qubit 1 (qubit 2) population to the QuTiP model described in the text. (c) Extracted 4WM strengths $g_{4,1}$ (blue circles) and $g_{4,2}$ (red squares) as a function of $\xi_p$. (d) Cooperativity evaluated from \cref{eq: coop n=2} at zero detuning, $\delta = 0$.}
    \label{fig: Protocol to fit the strength of the 4WM strengths}
\end{figure}

We extracted the 4WM strengths directly using the sequence shown in \cref{fig: Protocol to fit the strength of the 4WM strengths}(a). We first prepare the intermediate state of the cSMPD detection process, with qubit 1 excited and the memory in the Fock state $\ket{1}$, which we denote $\ket{0,e,1,g,0}$, where the entries refer to the buffer, qubit 1, memory, qubit 2, and waste, respectively. To do so, we excite qubit 1, swap its excitation into the memory with the parametric SWAP gate described in \cref{Appendix_Memoryloss calibration}, and apply a second $\pi$ pulse to bring qubit 1 back into its excited state. The pumps, set to the cSMPD 4WM resonance condition, are then applied for a variable duration, after which both qubits are read out and the results are averaged over many cycles. \Cref{fig: Protocol to fit the strength of the 4WM strengths}(b) shows the measured populations of qubit 1 (circles) and qubit 2 (crosses). The pumps couple $\ket{0,e,1,g,0}$ to $\ket{1,g,0,g,0}$ and $\ket{0,e,0,e,1}$ with strengths $g_{4,1}$ and $g_{4,2}$, respectively. The memory photon is therefore either converted back into the buffer and lost through it, leaving both qubits in their ground state, or converted into the waste, leaving both qubits excited. The ratio of the 4WM strengths determines which process dominates, while their absolute values set the dynamics, so that both strengths can be determined uniquely.

Given the independently measured circuit parameters, namely the total linewidths of the buffer and waste and the relaxation times of the qubits and the memory, we model these dynamics with the Hamiltonian of \cref{H4wm cascaded N=2} and the Lindblad operators $\sqrt{\kappa_b}\,\hat{b}$, $\sqrt{1/T_1^{(1)}}\,\hat{q}_1$, $\sqrt{1/T_1^{(\mathrm{mem})}}\,\hat{m}$, $\sqrt{1/T_1^{(2)}}\,\hat{q}_2$, and $\sqrt{\kappa_w}\,\hat{w}$. We implemented this model in QuTiP~\cite{Qutip} and fitted the time evolution of the qubit populations with the 4WM strengths as free parameters. The resulting strengths are shown in \cref{fig: Protocol to fit the strength of the 4WM strengths}(c).

From the extracted strengths, we then estimate the cooperativity at each pump amplitude [\cref{fig: Protocol to fit the strength of the 4WM strengths}(d)]. For most pump amplitudes it exceeds unity, which is a consequence of how the sweep was performed. We calibrated the cooperativity to be close to 1 at $\xi_p \approx 0.08$, and swept the mean pump amplitude by scaling the output voltages of our control electronics by a common factor, in order to keep the ratio of the 4WM strengths fixed. Because of nonlinearities in the pump lines, however, the 4WM strengths did not scale linearly with this factor, and the cooperativity drifted away from unity. According to \cref{eq:14}, the resulting resonant 4WM efficiency ranges from 0.91, at the highest cooperativity, to 1, at the cooperativity closest to unity.

\subsection{Qubit $T_2^*$ measurements under 4WM pumps}

\begin{figure}[h]
    \centering
    \includegraphics[width=1\linewidth]{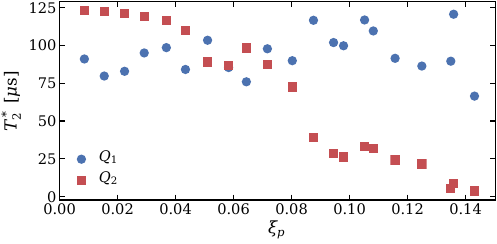}
    \caption{\textbf{$T_2^*$ under pumps.} Evolution of the dephasing time for the two qubits with varying pump amplitudes $\xi_p$. The pumps are detuned by \SI{20}{\mega\hertz} from the 4WM conditions. Qubit 1, blue circles, show no change with the increasing pumps, and is found to be around \SI{100}{\micro\second}. On the other hand, qubit 2 experiences a significant decrease with increasing pumps, dropping from \SI{125}{\micro\second} at low pump amplitudes down to \SI{4}{\micro\second} for the highest pump amplitudes.}
    \label{fig: T2Star under pumps}
\end{figure}

Dephasing times were recorded for different mean pump amplitudes $\xi_p$, while being detuned by \SI{20}{\mega\hertz} from the 4WM conditions. The result is shown in \cref{fig: T2Star under pumps}. Qubit 1 dephasing times are shown with blue circles, and no noticeable change can be observed over the swept interval, being constant around \SI{100}{\micro\second}. On the other side, qubit 2 dephasing times, shown with red squares, suffer a significant decrease with the increasing pumps, decreasing from \SI{125}{\micro\second} to \SI{4}{\micro\second}.

\subsection{Fridge heating under 4WM pumps}
\label{appendix: fridge heating under strong pumps}
\begin{figure}[H]
    \centering
    \includegraphics[width=1\linewidth]{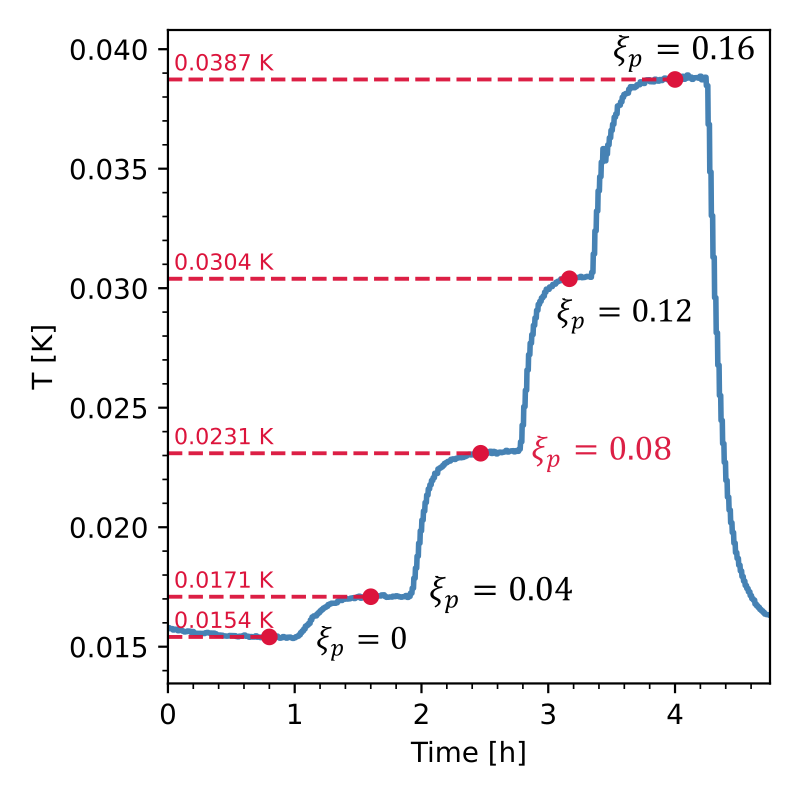}
    \caption{\textbf{Mixing chamber temperature under pump.} Mixing chamber
    temperature measured with a RuO$_2$ thermometer while the cSPD2 pump tone is
    applied at increasing amplitudes $\abs{\xi_p}$. Each setpoint is held for
    $\sim\SI{1}{\hour}$, long enough for the mixing chamber to thermalize; the
    red markers indicate the stabilized temperature at the end of each plateau.
    The base temperature is \SI{15.4}{\milli\kelvin} with the pump off and
    rises monotonically with pump power, reaching \SI{23.1}{\milli\kelvin} at
    the operating point $\abs{\xi_p}=0.08$ and \SI{38.7}{\milli\kelvin} at
    $\abs{\xi_p}=0.16$. The pump is switched off at $t\approx\SI{4.3}{\hour}$,
    after which the fridge cools back down.}
    \label{fig: Mixing chamber temperature under pumps}
\end{figure}

To quantify the heat load associated with the strong pump tone, we monitor the
mixing chamber temperature with a Ruthenium Oxide thermometer while stepping the
pump amplitude $\abs{\xi_p}$ (Fig.~\ref{fig: Mixing chamber temperature under
pumps}). Five amplitudes are applied successively over $\sim\SI{5}{\hour}$, each
held for one hour so that the fridge reaches a steady state before the next
step. With the pump off, the base temperature is \SI{15.4}{\milli\kelvin}. At
the operating point of cSMPD2, $\abs{\xi_p}=0.08$, the mixing chamber stabilizes
at \SI{23.1}{\milli\kelvin}, i.e.\ a \SI{7.7}{\milli\kelvin} increase, and the
temperature keeps rising with pump amplitude up to \SI{38.7}{\milli\kelvin} at $\abs{\xi_p}=0.16$. This pump-induced heating of the sample stage accounts for the higher thermal occupancy $n_\mathrm{th}$ measured
at large pump amplitudes.

\subsection{Dark count budget as a function of pump amplitudes}
\label{appendix: dck vs pump amplitudes}

\begin{figure}[h]
    \centering
    \includegraphics[width=\linewidth]{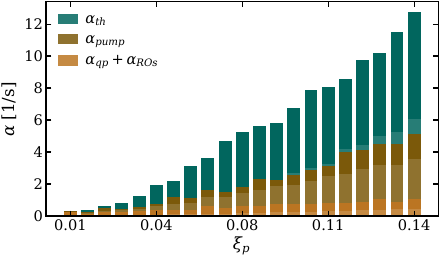}
    \caption{\textbf{Breakdown of the dark counts for different pump amplitudes.} For each mean pump amplitude $\xi_p$, the total bar height represents the total dark count rate. The contributions of thermal photons (green), the pumps (brown), and quasiparticles and readout (orange) are stacked, and the shaded areas show the contributions expected from the uncorrelated model.}
    \label{fig: DC breakdown}
\end{figure}

In \cref{subsection: Dark counts} of the main text, we described the protocol used to separate the contributions to the total dark count rate and applied it at the working point $\xi_p = 0.08$ (\cref{figMAIN: DC breakdown}). \Cref{fig: DC breakdown} extends this breakdown to mean pump amplitudes from $0.01$ to $0.14$. The thermal photon contribution increases with $\xi_p$ mainly because the detection bandwidth broadens, so that thermal photons are collected over a wider frequency range. The rise of the pump-induced contribution is instead caused mostly by heating of the mixing chamber by the pumps (\cref{fig: Mixing chamber temperature under pumps}). Since each quasiparticle and readout measurement was performed immediately after the corresponding pumped measurement, these contributions also increase slightly with the pump amplitude, even though the pumps were off during their acquisition.

\section{Sample holder material investigation}
\label{Appendix_sh}

We investigated the noise characteristics of the cSMPD over three different cooldowns, focusing on the influence of the sample holder lid material. The bottom part of the sample holder is made of bare OFHC copper, and the chip is mechanically secured at the four corners by CuBr clamps without any adhesive. We tested three lid configurations: (1) Aluminum 2024 (without specific cleaning), (2) bare OFHC copper (cleaned a few days prior with citric acid and rinsed with IPA), and (3) the same OFHC copper with a thin ~200 nm aluminum layer evaporated on the surface facing the chip.

Contrary to expectations, the results showed that the bare OFHC copper, presumed to provide the best thermalization, actually contributed the most to detector noise (three times more compared to Aluminum 2024 alone). The OFHC copper with the aluminum flash layer reduced the noise slightly, but it remained twice that in the full Aluminum 2024 configuration.

The noise could originate from the surface or bulk properties of the OFHC copper. Two potential sources of this noise are natural radioactivity (bulk) and reflectivity at the gap frequency. We hypothesize that applying an aluminum layer on the ceiling of the sample holder, which would naturally form a thin aluminum oxide layer, might absorb some of the infrared radiation, thereby reducing absorption at the Josephson junction level. This hypothesis will be tested in further studies.

\begin{table}[h]
    \centering
    \begin{tabular}{|c|c|c|c|}
        \toprule
        \textbf{Cooldown} & \textbf{Lid material} & $\mathbf{\alpha}$ (\SI{}{\per\second}) & $\mathbf{\eta}$ \\
        \midrule
        \midrule
         1 & Aluminum 2024            & $6.4\pm 0.7$  & $0.25\pm0.02$ \\
         2 & OFHC copper            & $15\pm 1$ & $0.19\pm0.02$ \\
         3 & OFHC copper + Al flash & $7.9\pm 0.3$ & $0.17\pm0.02$\\
        \bottomrule
    \end{tabular}
    \caption{Noise of the cSMPD for different sample holder lid materials.}
\end{table}

\section{$N=3$ and generalization to a $N$ qubit cascaded SMPD}
\label{Appendix: N qubit CPD}

In this section, we first examine the case of $N=3$,  followed by the general case of arbitrary $N$, where the behavior can be described using a simple binomial law.

A qubit chain with an odd number of qubits provides the flexibility to choose between different decoding schemes. One option is the "all-or-nothing" approach, where only the state $\ket{1\dotso1}$ is considered a positive outcome, with all other states treated as errors or negative outcomes. This method prioritizes a very low dark count rate at the cost of reduced operational efficiency. Alternatively, the "majority vote" scheme records a positive outcome if at least half of the qubits plus one are measured in their excited state.

We present a simplified scenario in which all qubit efficiencies, readout fidelities, and the average duty cycle approach unity. Each linear resonator is assumed to have minimal internal losses, and every subsystem of the transmission line is characterized by an individual conversion efficiency, denoted $\eta$, which depends solely on the subsystem's cooperativity. We define $\bar{\eta}\equiv 1 - \eta$, and consider the case of a nearly perfectly tuned system, i.e., $\eta\rightarrow 1$. All transmon qubits are assumed to be identical, producing the same conversion efficiency and exhibiting intrinsic noise dominated by their equilibrium excited state population, $\pth$. Let $p$ represent the probability of a positive outcome and $\bar{p}$ the probability of a negative one. Both scenarios are summarized in \cref{table: cSMPD N=3 ideal truth table}.

\begin{table}[h]
    \centering
    \begin{tabular}{|c|c|c|c|c|}
    \toprule
    Bitstring & $p$ & $\bar{p}$ & All-or-nothing & Majority \\
    \midrule
    \midrule
    000 & $(1-\eta)^3$      & $(1-\pth)^3$     & \multirow{7}{*}{False} & \multirow{4}{*}{False}\\
    001 & $\eta(1-\eta)^2$  & $\pth(1-\pth)^2$ &                        &                       \\
    010 & $\eta(1-\eta)^2$  & $\pth(1-\pth)^2$ &                        &                       \\
    100 & $\eta(1-\eta)^2$  & $\pth(1-\pth)^2$ &                        &                       \\
    \cmidrule{5-5}
    101 & $\eta^2(1-\eta)$  & $\pth^2(1-\pth)$ &                        & \multirow{7}{*}{True}\\
    110 & $\eta^2(1-\eta)$  & $\pth^2(1-\pth)$ &                        &                      \\
    011 & $\eta^2(1-\eta)$  & $\pth^2(1-\pth)$ &                        &                      \\
    \cmidrule{4-4}
    111 & $\eta^3$          & $\pth^3$         & True                  &                      \\
    \bottomrule
    \end{tabular}
    \caption{Truth table, cSMPD $N=3$ in an idealistic scenario.}
    \label{table: cSMPD N=3 ideal truth table}
\end{table}

In practice, we are interested by the assigment probabilities $P(\mathrm{True}\vert p)$ (the detection efficiency) and $P(\mathrm{True}\vert \bar{p})$ (a dark count). For the majority vote scheme, the detection efficiency is given by $P(\mathrm{True}\vert p)=3\eta^2(1-\eta)+\eta^3\xrightarrow{\eta\rightarrow 1}1-3\bar{\eta}^2$, and the dark count rate is $P(\mathrm{True}\vert \bar{p})=3\pth^2(1-\pth)+\pth^3 \xrightarrow{\pth\rightarrow 0} 3\pth^2$. In the all-or-nothing scheme, the detection efficiency is $P(\mathrm{True}\vert p)=\eta^3 \xrightarrow{\eta\rightarrow 1} 1-3\bar{\eta}$, and the dark count rate is $P(\mathrm{True}\vert \bar{p})=\pth^3$. 
Considering experimentally relevant equilibrium populations for transmons with transition frequencies $\omega^{ge}/2\pi$ on the order of a few gigahertz, typically in the range $\interval{10^{-4}}{10^{-2}}$, we conclude that the all-or-nothing decoding scheme heavily biases towards achieving a low intrinsic dark count rate. In contrast, the majority vote scheme is sufficient to maintain an intrinsic noise level of less than 1 count/s, with the noise primarily dominated by the detection of spurious microwave photons propagating through the setup. 

For a device with $N$ qubits ($N$ odd), in the majority vote scenario,
the leading-order terms for the effective equilibrium population and
the effective operational efficiency are given by
$\binom{N}{(N+1)/2}\,p_\mathrm{eq}^{(N+1)/2}$ and
$1-\binom{N}{(N+1)/2}\,\bar{\eta}^{(N+1)/2}$, respectively. If the transmon qubits are not fully equivalent in terms of their $T_1$ times, it is possible to design an optimized decoder that provides a non-uniform, weighted response to readout outcomes, improving overall performance.

% !TEX root = main.tex

\section{Pump-induced lifetime limitations}
\label{Appendix_high_order_RWA}

The pumps are on during the whole detection window, so the efficiency depends on the lifetimes under pump. In cSMPD1, $T_1$
of $q_1$ stays at $30\,\mu$s, $T_1$ of $q_2$ drops from 30--50 to $15\,\mu$s, and $\kappa_m$ grows from $6.8\times10^4$ to
$3.7\times10^5\,\mathrm{s^{-1}}$ (\cref{table: circuit-params-part2-merged,table: CPD-0002-02-A-and-W025-CPD-0007-03-01-C1-circuit-params}). Pumps can shorten lifetimes by dressing the decay of the modes into the
lines, by activating conversions, by multi-photon transitions and by heating~\cite{dai_characterization_2026}. The two devices differ in their filters and
in their frequency arrangements. We model the pumped cSMPD1 circuit to identify the processes it contains.

\emph{Circuit.} The model is the finite-element (FEM) simulation of cSMPD1 (\cref{tab:Q:fem}): linear modes,
zero-point phase fluctuations $\varphi_i^{(j)}$ of each mode across the junction of $q_j$, Josephson energies, and decay rates
$\kappa_x$ of the buffer, readout and waste modes into their lines $x=b,r_1,w$. Once dressed by the junctions, the simulated
modes lie $0.1$--$0.2$\,GHz from the measured ones and the simulated $\kappa_x$ are 3 to 5 times smaller. The analysis is
therefore qualitative.

\begin{table}
  \caption{FEM parameters of cSMPD1. $\omega_i$: linear mode; $\tilde\omega_i$: dressed frequency without pump;
  $\kappa_x$: decay rate into the line the mode serves; $\varphi_i^{(j)}$: zero-point phase fluctuation across the
  junction of $q_j$.}
  \label{tab:Q:fem}
  \small
  \begin{tabular}{lccccc}
    \hline\hline
    mode & $\omega_i/2\pi$ & $\tilde\omega_i/2\pi$ & $\kappa_x$ & $\varphi_i^{(1)}$ & $\varphi_i^{(2)}$\\
         & (GHz) & (GHz) & ($10^6\,\mathrm{s^{-1}}$) & & \\
    \hline
    $b$   & 8.994 & 8.993 & 1.15 & 0.027 & $9\times10^{-5}$\\
    $q_1$ & 6.767 & 6.519 & --   & 0.373 & 0.009\\
    $r_1$ & 7.764 & 7.763 & 1.20 & 0.031 & 0.002\\
    $m$   & 8.197 & 8.193 & --   & 0.034 & 0.032\\
    $q_2$ & 6.404 & 6.150 & --   & 0.006 & 0.387\\
    $w$   & 7.571 & 7.569 & 1.15 & 0.002 & 0.034\\
    \hline
    \multicolumn{4}{r}{$E_{J}/h$ (GHz)} & 24.27 & 21.57\\
    \hline\hline
  \end{tabular}
\end{table}

\emph{Hamiltonian.} The transmons are treated as Kerr-nonlinear bosonic modes $\hat q_1$, $\hat q_2$. With
$\hat a_i\in\{\hat b,\hat q_1,\hat r_1,\hat m,\hat q_2,\hat w\}$,
\begin{align}
  \frac{\hat H}{\hbar}&=\sum_i\omega_i\hat a_i^\dagger\hat a_i
  -\sum_{j=1,2}\frac{E_{J_j}}{\hbar}\Big[\cos\hat\varphi_j-1+\frac{\hat\varphi_j^2}{2}\Big],
  \label{eq:Q:H}\\
  \hat\varphi_j&=\sum_i\varphi_i^{(j)}\big(\hat a_i+\hat a_i^\dagger\big)+\varphi_{p_j}\cos\omega_{p_j}t .\nonumber
\end{align}
The quadratic part of the cosine is already contained in the linear modes. Pump $p_j$ enters as a classical phase
$\varphi_{p_j}\propto\xi_j$ across the junction of $q_j$. We set $\varphi_{p_1}$ and $\varphi_{p_2}$ so that the working point,
$|\xi_p|=0.07$ and $\beta_\xi=1.02$, reproduces $g_{4,1}$ and $g_{4,2}$ of~\cref{table: circuit-params-part2-merged}, and the pump frequencies satisfy Eq.~(3).
The model then gives ac Stark shifts of $1.6$ and $1.2$\,MHz for $q_1$ and $q_2$, close to the $1.2$ and $1.1$\,MHz given by
$2|\chi_{q_kq_k}||\xi_k|^2$.

\emph{Master equation.} Each mode $x=b,r_1,w$ couples to its line through its quadrature $\hat x+\hat x^\dagger$. A
time-dependent Schrieffer--Wolff transformation~\cite{venkatraman_static_2022,carde_flux-pump_2025} removes the rotating terms of
\cref{eq:Q:H}, leaves a static Hamiltonian $\hat H_{\rm eff}$, and turns each quadrature into
$\sum_j\hat C_x(\omega_j)e^{-i\omega_jt}+\mathrm{h.c.}$ In the Born--Markov approximation,
\begin{equation}
  \dot\rho=-i\big[\hat H_{\rm eff},\rho\big]+\sum_x\sum_j S_x(\omega_j)\,\mathcal D\big[\hat C_x(\omega_j)\big]\rho ,
  \label{eq:Q:me}
\end{equation}
where $S_x(\omega)$ is the spectral density of line $x$, with $S_x(\omega_x)=\kappa_x$. $\hat H_{\rm eff}$ contains the Lamb and ac
Stark shifts, the Kerr and cross-Kerr terms, and the two engineered conversions of Eq.~(2). The collapse operators
$\hat C_x(\omega_j)=\sum c\,\hat O$ are sums of products $\hat O$ of mode operators. We expand the cosine to fifth order in
operators and pump amplitude, and the transformation to second order.

\emph{Pump-induced Purcell decay.} For a flat bath, $S_x=\kappa_x$, the simulated Purcell decay rates of the qubits and of the
memory into the three lines are below $5\times10^2\,\mathrm{s^{-1}}$, and the pumps increase them by about $1\,\%$ at the working
point. These values change by less than $2\,\%$ at third order of the transformation. The measured relaxation rates are above
$2\times10^4\,\mathrm{s^{-1}}$, and the pumps increase them by a factor 2 for $q_2$ and 5.4 for the memory. The pump-induced
correction to the Purcell effect thus does not explain the measured degradation.

\begin{figure*}
  \includegraphics[width=\textwidth]{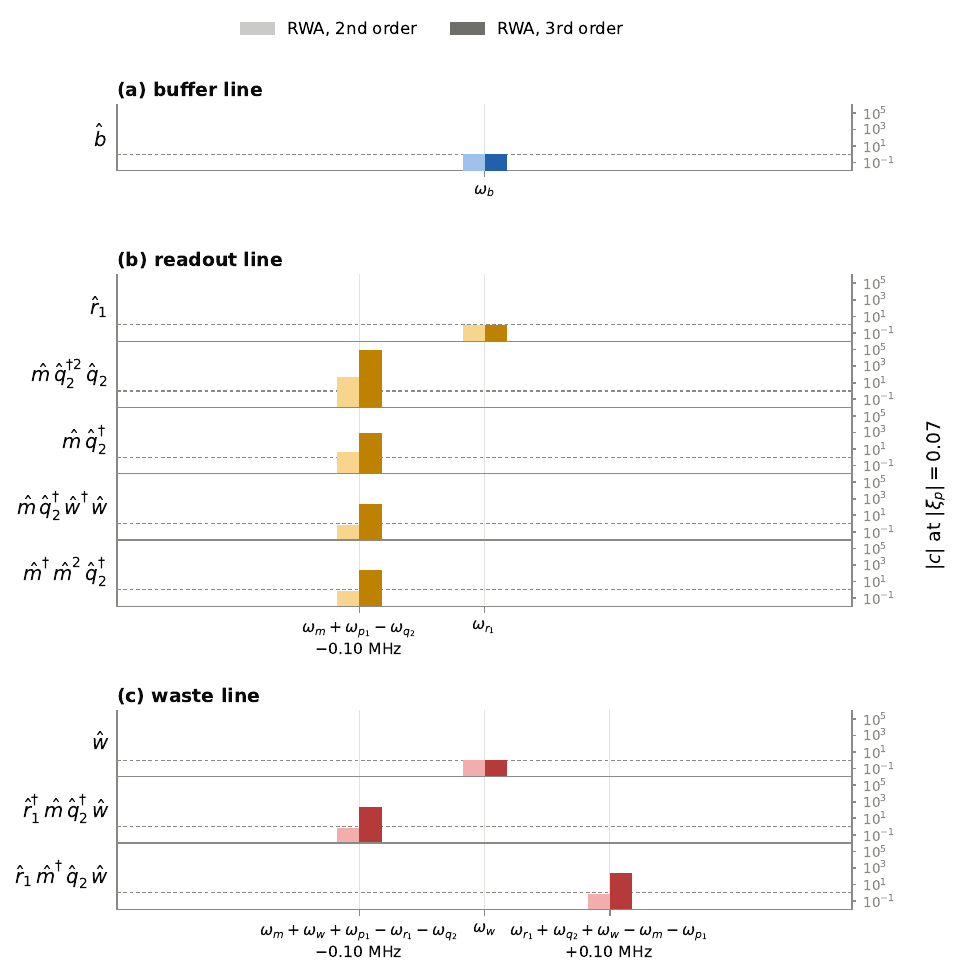}
  \caption{Failure of the expansion for near-resonant processes, cSMPD1 FEM circuit at the working point. One block per line:
  (a)~buffer, (b)~readout of $q_1$, (c)~waste. One row per collapse operator with $|c|\ge0.5$ at second order of the
  transformation. Light bars: second order; dark bars: third order; dashed line: $|c|=1$. The horizontal axis is linear, $\pm0.3$\,MHz around the mode of each line;
  ticks give the frequency at which each operator exchanges energy with its line, and its offset from the mode.}
  \label{fig:Q:divergence}
\end{figure*}

\emph{Near-resonant processes.} The transformation is a multi-frequency extension of the high-frequency expansion of \cite{venkatraman_static_2022}, in which a single drive frequency sets the perturbative parameter. Here, each rotating term is
eliminated with its own frequency $\delta$, a combination of mode and pump frequencies, and an accidental near-resonance makes
$\delta$ small and invalidates the perturbative assumption. This happens when $\delta$ is comparable to the couplings or to the
Kerr and anharmonic shifts, which the expansion treats as perturbations of the mode frequencies: a few MHz for the cross-Kerr
terms and 255--265\,MHz for the anharmonicities. \Cref{tab:Q:near} lists the processes of cSMPD1 in this situation. In the
FEM circuit, $m+p_1\to r_1+q_2(g\to e)$ is detuned by only $-0.09$\,MHz. We computed the collapse operators at third order of
the transformation: those of this process grow by a factor $2\times10^2$ to $2\times10^3$, whereas the decay of each mode into its
own line does not change (\cref{fig:Q:divergence}). The three other processes bring a transmon to $|f\rangle$: their detuning
differs by an anharmonicity from the one the bosonic expansion assigns to them. The analysis thus finds a near-resonance
problem, but the limits of the perturbative procedure prevent a quantitative estimate of the corresponding rates.

\begin{widetext}
\begin{table}
  \caption{Near-resonant processes of cSMPD1: processes with
  $g/2\pi\ge1$\,kHz and $|\delta|/2\pi<50$\,MHz for the FEM or the measured frequencies, the engineered conversions and their
  reverse excluded. Term of $\hat H$: the product of mode operators of Eq.~\eqref{eq:Q:H} that dominates the coupling (pump
  implicit). $g$: coupling between the initial and final states, from the FEM Hamiltonian at the working point. $\delta$:
  detuning from level energies that include the anharmonicities and the dispersive shifts, for the FEM model and for the
  measured frequencies of~\cref{table: CPD-0002-02-A-and-W025-CPD-0007-03-01-C1-circuit-params}.}
  \label{tab:Q:near}
  \small
  \begin{tabular}{lllcrr}
    \hline\hline
    process & term of $\hat H$ & effect & $g/2\pi$ (kHz) & \multicolumn{2}{c}{$\delta/2\pi$ (MHz)}\\
            &                  &        & FEM            & FEM & \cref{table: CPD-0002-02-A-and-W025-CPD-0007-03-01-C1-circuit-params}\\
    \hline
    $p_2+q_1(e\to g)\to q_2(g\to f)$ & $\hat q_1\hat q_2^{\dagger2}$                  & $q_1$ relaxes, $q_2$ to $|f\rangle$ & 245 & 8.9    & $-80$\\
    $m+p_1\to r_1+q_1(e\to f)$       & $\hat r_1^\dagger\hat m\hat q_1^\dagger$       & photon lost, $q_1$ to $|f\rangle$   & 183 & $-112$ & $-34$\\
    $w+q_1(e\to g)\to m+q_2(e\to f)$ & $\hat m^\dagger\hat w\hat q_1\hat q_2^\dagger$ & $q_1$ relaxes, $q_2$ to $|f\rangle$ & 101 & 11     & $-61$\\
    $m+p_1\to r_1+q_2(g\to e)$       & $\hat r_1^\dagger\hat m\hat q_2^\dagger$       & photon still counted                & 2.2 & $-0.09$ & 56\\
    \hline\hline
  \end{tabular}
\end{table}
\end{widetext}

\emph{Discussion.} The detunings of \cref{tab:Q:near} change by tens of MHz between the FEM and the measured
frequencies, more than the detunings themselves, and the measured frequencies reproduce the measured pump frequencies through \cref{eq: freqmatchcond CPD N=2} only to about 15\,MHz. The model therefore cannot tell which of these processes, if any, is near-resonant in the
device. Only $m+p_1\to r_1+q_1(e\to f)$ loses the memory photon, with a coupling of 0.18\,MHz, and none relaxes $q_2$. In
cSMPD1, several four-wave processes have detunings of only a few tens of MHz, a gap that the fabrication precision, also a few
tens of MHz (\cref{appendix: cSMPD1}), could bridge. A process that emits close to the mode of its line lies inside any passband: filters
cannot suppress it, only the frequency arrangement can. This frequency crowding, rather than the pump-induced correction to the
Purcell effect, is the plausible cause of the pump-induced degradation of cSMPD1. The model includes neither the pump lines
nor thermal baths, and its fifth-order expansion misses higher-order multi-photon transitions~\cite{dai_characterization_2026}.

\bibliography{references}

@article{distante_detecting_2021,
	title = {Detecting an {Itinerant} {Optical} {Photon} {Twice} without {Destroying} {It}},
	volume = {126},
	issn = {0031-9007, 1079-7114},
	url = {http://arxiv.org/abs/2105.14354},
	doi = {10.1103/PhysRevLett.126.253603},

	number = {25},
	urldate = {2024-09-26},
	journal = {Physical Review Letters},
	author = {Distante, Emanuele and Daiss, Severin and Langenfeld, Stefan and Hartung, Lukas and Thomas, Philip and Morin, Olivier and Rempe, Gerhard and Welte, Stephan},
	month = jun,
	year = {2021},
	pages = {253603},
}

@article{
travesedo_all-microwave_2024,
author = {Jaime Travesedo  and James O'Sullivan  and Louis Pallegoix  and Zhiyuan W. Huang  and Patrick Hogan  and Philippe Goldner  and Thierry Chaneliere  and Sylvain Bertaina  and Daniel Estève  and Patrick Abgrall  and Denis Vion  and Emmanuel Flurin  and Patrice Bertet },
title = {All-microwave spectroscopy and polarization of individual nuclear spins in a solid},
journal = {Science Advances},
volume = {11},
number = {10},
pages = {eadu0581},
year = {2025},
doi = {10.1126/sciadv.adu0581},
URL = {https://www.science.org/doi/abs/10.1126/sciadv.adu0581},
eprint = {https://www.science.org/doi/pdf/10.1126/sciadv.adu0581}}

@article{opremcak_high-fidelity_2021,
	title = {High-{Fidelity} {Measurement} of a {Superconducting} {Qubit} {Using} an {On}-{Chip} {Microwave} {Photon} {Counter}},
	volume = {11},
	url = {https://link.aps.org/doi/10.1103/PhysRevX.11.011027},
	doi = {10.1103/PhysRevX.11.011027},
	number = {1},
	urldate = {2025-02-04},
	journal = {Physical Review X},
	author = {Opremcak, A. and Liu, C. H. and Wilen, C. and Okubo, K. and Christensen, B. G. and Sank, D. and White, T. C. and Vainsencher, A. and Giustina, M. and Megrant, A. and Burkett, B. and Plourde, B. L. T. and McDermott, R.},
	month = feb,
	year = {2021},
	pages = {011027},

}

@article{chen_microwave_2011,
	title = {Microwave {Photon} {Counter} {Based} on {Josephson} {Junctions}},
	volume = {107},
	url = {https://link.aps.org/doi/10.1103/PhysRevLett.107.217401},
	doi = {10.1103/PhysRevLett.107.217401},
	number = {21},
	urldate = {2025-02-04},
	journal = {Physical Review Letters},
	author = {Chen, Y.-F. and Hover, D. and Sendelbach, S. and Maurer, L. and Merkel, S. T. and Pritchett, E. J. and Wilhelm, F. K. and McDermott, R.},
	month = nov,
	year = {2011},
	pages = {217401},
}

@article{hutin_monitoring_2024,
	title = {Monitoring the {Energy} of a {Cavity} by {Observing} the {Emission} of a {Repeatedly} {Excited} {Qubit}},
	volume = {133},
	url = {https://link.aps.org/doi/10.1103/PhysRevLett.133.153602},
	doi = {10.1103/PhysRevLett.133.153602},
	number = {15},
	urldate = {2025-02-03},
	journal = {Physical Review Letters},
	author = {Hutin, H. and Essig, A. and Assouly, R. and Rouchon, P. and Bienfait, A. and Huard, B.},
	month = oct,
	year = {2024},
	pages = {153602},
}

@article{essig_multiplexed_2021,
	title = {Multiplexed {Photon} {Number} {Measurement}},
	volume = {11},
	url = {https://link.aps.org/doi/10.1103/PhysRevX.11.031045},
	doi = {10.1103/PhysRevX.11.031045},
	number = {3},
	urldate = {2025-02-03},
	journal = {Physical Review X},
	author = {Essig, Antoine and Ficheux, Quentin and Peronnin, Théau and Cottet, Nathanaël and Lescanne, Raphaël and Sarlette, Alain and Rouchon, Pierre and Leghtas, Zaki and Huard, Benjamin},
	month = aug,
	year = {2021},
	pages = {031045},
}

@article{dassonneville_number-resolved_2020,
	title = {Number-{Resolved} {Photocounter} for {Propagating} {Microwave} {Mode}},
	volume = {14},
	url = {https://link.aps.org/doi/10.1103/PhysRevApplied.14.044022},
	doi = {10.1103/PhysRevApplied.14.044022},
	number = {4},
	urldate = {2025-02-03},
	journal = {Physical Review Applied},
	author = {Dassonneville, R. and Assouly, R. and Peronnin, T. and Rouchon, P. and Huard, B.},
	month = oct,
	year = {2020},
	pages = {044022},
}

@article{gleyzes_quantum_2007,
	title = {Quantum jumps of light recording the birth and death of a photon in a cavity},
	volume = {446},
	copyright = {2007 Springer Nature Limited},
	issn = {1476-4687},
	url = {https://www.nature.com/articles/nature05589},
	doi = {10.1038/nature05589},
	number = {7133},
	urldate = {2025-02-03},
	journal = {Nature},
	author = {Gleyzes, Sébastien and Kuhr, Stefan and Guerlin, Christine and Bernu, Julien and Deléglise, Samuel and Busk Hoff, Ulrich and Brune, Michel and Raimond, Jean-Michel and Haroche, Serge},
	month = mar,
	year = {2007},
	pages = {297--300},
}

@article{schuster_2007,
	title = {Resolving photon number states in a superconducting circuit},
	volume = {445},
	copyright = {2006 Springer Nature Limited},
	issn = {1476-4687},
	url = {https://www.nature.com/articles/nature05461},
	doi = {10.1038/nature05461},
	number = {7127},
	urldate = {2025-02-03},
	journal = {Nature},
	author = {Schuster, D. I. and Houck, A. A. and Schreier, J. A. and Wallraff, A. and Gambetta, J. M. and Blais, A. and Frunzio, L. and Majer, J. and Johnson, B. and Devoret, M. H. and Girvin, S. M. and Schoelkopf, R. J.},
	month = feb,
	year = {2007},
	pages = {515--518},
}

@article{stanisavljevic_efficient_2024,
	title = {Efficient {Microwave} {Photon}-to-{Electron} {Conversion} in a {High}-{Impedance} {Quantum} {Circuit}},
	volume = {133},
	url = {https://link.aps.org/doi/10.1103/PhysRevLett.133.076302},
	doi = {10.1103/PhysRevLett.133.076302},
	number = {7},
	urldate = {2025-02-03},
	journal = {Physical Review Letters},
	author = {Stanisavljević, O. and Philippe, J.-C. and Gabelli, J. and Aprili, M. and Estève, J. and Basset, J.},
	month = aug,
	year = {2024},
	pages = {076302},
}

@article{leppakangas_multiplying_2018,
	title = {Multiplying and detecting propagating microwave photons using inelastic {Cooper}-pair tunneling},
	volume = {97},
	url = {https://link.aps.org/doi/10.1103/PhysRevA.97.013855},
	doi = {10.1103/PhysRevA.97.013855},
	number = {1},
	urldate = {2025-01-12},
	journal = {Physical Review A},
	author = {Leppäkangas, Juha and Marthaler, Michael and Hazra, Dibyendu and Jebari, Salha and Albert, Romain and Blanchet, Florian and Johansson, Göran and Hofheinz, Max},
	month = jan,
	year = {2018},
	pages = {013855},
}

@article{albert_microwave_2024,
	title = {Microwave {Photon}-{Number} {Amplification}},
	volume = {14},
	url = {https://link.aps.org/doi/10.1103/PhysRevX.14.011011},
	doi = {10.1103/PhysRevX.14.011011},
	number = {1},
	urldate = {2025-01-12},
	journal = {Physical Review X},
	author = {Albert, R. and Griesmar, J. and Blanchet, F. and Martel, U. and Bourlet, N. and Hofheinz, M.},
	month = feb,
	year = {2024},
	pages = {011011},
}

@article{braggio_quantum-enhanced_2024,
  title = {Quantum-Enhanced Sensing of Axion Dark Matter with a Transmon-Based Single Microwave Photon Counter},
  author = {Braggio, C. and Balembois, L. and Di Vora, R. and Wang, Z. and Travesedo, J. and Pallegoix, L. and Carugno, G. and Ortolan, A. and Ruoso, G. and Gambardella, U. and D'Agostino, D. and Bertet, P. and Flurin, E.},
  journal = {Phys. Rev. X},
  volume = {15},
  issue = {2},
  pages = {021031},
  numpages = {17},
  year = {2025},
  month = {Apr},
  publisher = {American Physical Society},
  doi = {10.1103/PhysRevX.15.021031},
  url = {https://link.aps.org/doi/10.1103/PhysRevX.15.021031}
}

@article{osullivan_nuclearspinregister2024,
  author  = {O'Sullivan, J. and Travesedo, J. and Pallegoix, L. and others},
  title   = {Individual solid-state nuclear spin qubits with coherence exceeding seconds},
  journal = {Nature Physics},
  volume  = {21},
  pages   = {1794--1800},
  year    = {2025},
  doi     = {10.1038/s41567-025-03049-7},
  url     = {https://doi.org/10.1038/s41567-025-03049-7}
}

@article{pallegoix_smpd2024,
  title = {Enhancing the sensitivity of single-microwave-photon detection with bandwidth tunability},
  author = {Pallegoix, Louis and Travesedo, Jaime and May, Alexandre S. and Balembois, L{\'e}o and Vion, Denis and Bertet, Patrice and Flurin, Emmanuel},
  journal = {Phys. Rev. Appl.},
  volume = {25},
  issue = {4},
  pages = {044064},
  numpages = {13},
  year = {2026},
  month = {Apr},
  publisher = {American Physical Society},
  doi = {10.1103/dbqj-qld5},
  url = {https://link.aps.org/doi/10.1103/dbqj-qld5}
}

@article{dixit_searching_2021,
	title = {Searching for {Dark} {Matter} with a {Superconducting} {Qubit}},
	volume = {126},
	url = {https://link.aps.org/doi/10.1103/PhysRevLett.126.141302},
	doi = {10.1103/PhysRevLett.126.141302},
	number = {14},
	urldate = {2024-10-08},
	journal = {Physical Review Letters},
	author = {Dixit, Akash V. and Chakram, Srivatsan and He, Kevin and Agrawal, Ankur and Naik, Ravi K. and Schuster, David I. and Chou, Aaron},
	month = apr,
	year = {2021},
	pages = {141302},
}

@article{lee_graphene-based_2020,
	title = {Graphene-based {Josephson} junction microwave bolometer},
	volume = {586},
	copyright = {2020 The Author(s), under exclusive licence to Springer Nature Limited},
	issn = {1476-4687},
	url = {https://www.nature.com/articles/s41586-020-2752-4},
	doi = {10.1038/s41586-020-2752-4},
	number = {7827},
	urldate = {2024-02-25},
	journal = {Nature},
	author = {Lee, Gil-Ho and Efetov, Dmitri K. and Jung, Woochan and Ranzani, Leonardo and Walsh, Evan D. and Ohki, Thomas A. and Taniguchi, Takashi and Watanabe, Kenji and Kim, Philip and Englund, Dirk and Fong, Kin Chung},
	month = oct,
	year = {2020},
	pages = {42--46},
}

@article{kokkoniemi_nanobolometer_2019,
	title = {Nanobolometer with ultralow noise equivalent power},
	volume = {2},
	copyright = {2019 The Author(s)},
	issn = {2399-3650},
	url = {https://www.nature.com/articles/s42005-019-0225-6},
	doi = {10.1038/s42005-019-0225-6},

	number = {1},
	urldate = {2024-02-25},
	journal = {Communications Physics},
	author = {Kokkoniemi, Roope and Govenius, Joonas and Vesterinen, Visa and Lake, Russell E. and Gunyhó, András M. and Tan, Kuan Y. and Simbierowicz, Slawomir and Grönberg, Leif and Lehtinen, Janne and Prunnila, Mika and Hassel, Juha and Lamminen, Antti and Saira, Olli-Pentti and Möttönen, Mikko},
	month = oct,
	year = {2019},
	pages = {1--8},
}

@article{richards_bolometers_1994,
	title = {Bolometers for infrared and millimeter waves},
	volume = {76},
	issn = {0021-8979, 1089-7550},
	url = {https://pubs.aip.org/jap/article/76/1/1/497857/Bolometers-for-infrared-and-millimeter-waves},
	doi = {10.1063/1.357128},
	number = {1},
	urldate = {2024-07-26},
	journal = {Journal of Applied Physics},
	author = {Richards, P. L.},
	month = jul,
	year = {1994},
	pages = {1--24},
}

@article{bal_systematic_2024,
	title = {Systematic Improvements in Transmon Qubit Coherence Enabled by Niobium Surface Encapsulation},
	author = {Bal, Mustafa and Murthy, Akshay A. and Zhu, Shaojiang and Crisa, Francesco and You, Xinyuan and Huang, Ziwen and Roy, Tanay and Lee, Jaeyel and van Zanten, David and Pilipenko, Roman and Nekrashevich, Ivan and Lunin, Andrei and Bafia, Daniel and Krasnikova, Yulia and Kopas, Cameron J. and Lachman, Ella O. and Miller, Duncan and Mutus, Josh Y. and Reagor, Matthew J. and Cansizoglu, Hilal and Marshall, Jayss and Pappas, David P. and Vu, Kim and Yadavalli, Kameshwar and Oh, Jin-Su and Zhou, Lin and Kramer, Matthew J. and Lecocq, Florent Q. and Goronzy, Dominic P. and Torres-Castanedo, Carlos G. and Pritchard, Graham and Dravid, Vinayak P. and Rondinelli, James M. and Bedzyk, Michael J. and Hersam, Mark C. and Zasadzinski, John and Koch, Jens and Sauls, James A. and Romanenko, Alexander and Grassellino, Anna},
	journal = {npj Quantum Information},
	volume = {10},
	pages = {43},
	year = {2024},
	doi = {10.1038/s41534-024-00840-x},
}

@article{besse_single-shot_2018,
	title = {Single-{Shot} {Quantum} {Nondemolition} {Detection} of {Individual} {Itinerant} {Microwave} {Photons}},
	volume = {8},
	url = {https://link.aps.org/doi/10.1103/PhysRevX.8.021003},
	doi = {10.1103/PhysRevX.8.021003},
	number = {2},
	urldate = {2024-10-04},
	journal = {Physical Review X},
	author = {Besse, Jean-Claude and Gasparinetti, Simone and Collodo, Michele C. and Walter, Theo and Kurpiers, Philipp and Pechal, Marek and Eichler, Christopher and Wallraff, Andreas},
	month = apr,
	year = {2018},
	pages = {021003},
}

@article{briegel_measurement-based_2009,
	title = {Measurement-based quantum computation},
	volume = {5},
	copyright = {2009 Springer Nature Limited},
	issn = {1745-2481},
	url = {https://www.nature.com/articles/nphys1157},
	doi = {10.1038/nphys1157},
	number = {1},
	urldate = {2024-10-04},
	journal = {Nature Physics},
	author = {Briegel, H. J. and Browne, D. E. and Dür, W. and Raussendorf, R. and Van den Nest, M.},
	month = jan,
	year = {2009},
	pages = {19--26},

}

@article{bartolucci_fusion-based_2023,
	title = {Fusion-based quantum computation},
	volume = {14},
	copyright = {2023 The Author(s)},
	issn = {2041-1723},
	url = {https://www.nature.com/articles/s41467-023-36493-1},
	doi = {10.1038/s41467-023-36493-1},
	number = {1},
	urldate = {2025-02-12},
	journal = {Nature Communications},
	author = {Bartolucci, Sara and Birchall, Patrick and Bombín, Hector and Cable, Hugo and Dawson, Chris and Gimeno-Segovia, Mercedes and Johnston, Eric and Kieling, Konrad and Nickerson, Naomi and Pant, Mihir and Pastawski, Fernando and Rudolph, Terry and Sparrow, Chris},
	month = feb,
	year = {2023},
	pages = {912},
}

@article{raussendorf_measurement-based_2003,
	title = {Measurement-based quantum computation on cluster states},
	volume = {68},
	url = {https://link.aps.org/doi/10.1103/PhysRevA.68.022312},
	doi = {10.1103/PhysRevA.68.022312},
	number = {2},
	urldate = {2024-10-04},
	journal = {Physical Review A},
	author = {Raussendorf, Robert and Browne, Daniel E. and Briegel, Hans J.},
	month = aug,
	year = {2003},
	pages = {022312},
}

@article{narla_robust_2016,
	title = {Robust {Concurrent} {Remote} {Entanglement} {Between} {Two} {Superconducting} {Qubits}},
	volume = {6},
	url = {https://link.aps.org/doi/10.1103/PhysRevX.6.031036},
	doi = {10.1103/PhysRevX.6.031036},
	number = {3},
	urldate = {2024-10-04},
	journal = {Physical Review X},
	author = {Narla, A. and Shankar, S. and Hatridge, M. and Leghtas, Z. and Sliwa, K.M. and Zalys-Geller, E. and Mundhada, S.O. and Pfaff, W. and Frunzio, L. and Schoelkopf, R.J. and Devoret, M.H.},
	month = sep,
	year = {2016},
	pages = {031036},
}

@article{assouly_quantum_2023,
	title = {Quantum advantage in microwave quantum radar},
	volume = {19},
	copyright = {2023 The Author(s), under exclusive licence to Springer Nature Limited},
	issn = {1745-2481},
	url = {https://www.nature.com/articles/s41567-023-02113-4},
	doi = {10.1038/s41567-023-02113-4},
	number = {10},
	urldate = {2024-10-04},
	journal = {Nature Physics},
	author = {Assouly, R. and Dassonneville, R. and Peronnin, T. and Bienfait, A. and Huard, B.},
	month = oct,
	year = {2023},
	pages = {1418--1422},
}

@article{lamoreaux_analysis_2013,
	title = {Analysis of single-photon and linear amplifier detectors for microwave cavity dark matter axion searches},
	volume = {88},
	url = {https://link.aps.org/doi/10.1103/PhysRevD.88.035020},
	doi = {10.1103/PhysRevD.88.035020},
	number = {3},
	urldate = {2024-10-04},
	journal = {Physical Review D},
	author = {Lamoreaux, S. K. and van Bibber, K. A. and Lehnert, K. W. and Carosi, G.},
	month = aug,
	year = {2013},
	pages = {035020},
}

@article{orrit_single_1990,
	title = {Single pentacene molecules detected by fluorescence excitation in a p-terphenyl crystal},
	volume = {65},
	url = {https://link.aps.org/doi/10.1103/PhysRevLett.65.2716},
	doi = {10.1103/PhysRevLett.65.2716},
	number = {21},
	urldate = {2024-10-04},
	journal = {Physical Review Letters},
	author = {Orrit, M. and Bernard, J.},
	month = nov,
	year = {1990},
	pages = {2716--2719},
}

@article{klar_fluorescence_2000,
	title = {Fluorescence microscopy with diffraction resolution barrier broken by stimulated emission},
	volume = {97},
	issn = {0027-8424},
	doi = {10.1073/pnas.97.15.8206},
	number = {15},
	journal = {Proceedings of the National Academy of Sciences of the United States of America},
	author = {Klar, T. A. and Jakobs, S. and Dyba, M. and Egner, A. and Hell, S. W.},
	month = jul,
	year = {2000},
	pmid = {10899992},
	pmcid = {PMC26924},
	pages = {8206--8210},
}

@article{betzig_imaging_2006,
	title = {Imaging intracellular fluorescent proteins at nanometer resolution},
	volume = {313},
	issn = {1095-9203},
	doi = {10.1126/science.1127344},
	number = {5793},
	journal = {Science (New York, N.Y.)},
	author = {Betzig, Eric and Patterson, George H. and Sougrat, Rachid and Lindwasser, O. Wolf and Olenych, Scott and Bonifacino, Juan S. and Davidson, Michael W. and Lippincott-Schwartz, Jennifer and Hess, Harald F.},
	month = sep,
	year = {2006},
	pmid = {16902090},
	pages = {1642--1645},
}

@article{bruschini_single-photon_2019,
	title = {Single-photon avalanche diode imagers in biophotonics: review and outlook},
	volume = {8},
	issn = {2047-7538},
	shorttitle = {Single-photon avalanche diode imagers in biophotonics},
	doi = {10.1038/s41377-019-0191-5},
	journal = {Light, Science \& Applications},
	author = {Bruschini, Claudio and Homulle, Harald and Antolovic, Ivan Michel and Burri, Samuel and Charbon, Edoardo},
	year = {2019},
	pmid = {31645931},
	pmcid = {PMC6804596},
	pages = {87},
}

@article{hadfield_single-photon_2009,
	title = {Single-photon detectors for optical quantum information applications},
	volume = {3},
	issn = {1749-4885},
	doi = {10.1038/nphoton.2009.230},
	number = {12},
	journal = {Nature Photonics},
	author = {Hadfield, Robert H.},
	month = dec,
	year = {2009},
	pages = {696--705},
}

@article{kimble_strong_1998,
	title = {Strong {Interactions} of {Single} {Atoms} and {Photons} in {CavityQED}},
	volume = {T76},
	issn = {0031-8949},
	url = {https://iopscience.iop.org/article/10.1238/Physica.Topical.076a00127},
	doi = {10.1238/Physica.Topical.076a00127},

	number = {1},
	urldate = {2024-09-15},
	journal = {Physica Scripta},
	author = {Kimble, H. J.},
	year = {1998},
	pages = {127},
}

@article{sunada_fast_2022,
	title = {Fast {Readout} and {Reset} of a {Superconducting} {Qubit} {Coupled} to a {Resonator} with an {Intrinsic} {Purcell} {Filter}},
	volume = {17},
	url = {https://link.aps.org/doi/10.1103/PhysRevApplied.17.044016},
	doi = {10.1103/PhysRevApplied.17.044016},
	number = {4},
	urldate = {2023-07-21},
	journal = {Physical Review Applied},
	author = {Sunada, Y. and Kono, S. and Ilves, J. and Tamate, S. and Sugiyama, T. and Tabuchi, Y. and Nakamura, Y.},
	month = apr,
	year = {2022},
	pages = {044016},
}

@article{gambetta_qubit-photon_2006,
	title = {Qubit-photon interactions in a cavity: {Measurement}-induced dephasing and number splitting},
	volume = {74},
	shorttitle = {Qubit-photon interactions in a cavity},
	url = {https://link.aps.org/doi/10.1103/PhysRevA.74.042318},
	doi = {10.1103/PhysRevA.74.042318},
	number = {4},
	urldate = {2023-08-18},
	journal = {Physical Review A},
	author = {Gambetta, Jay and Blais, Alexandre and Schuster, D. I. and Wallraff, A. and Frunzio, L. and Majer, J. and Devoret, M. H. and Girvin, S. M. and Schoelkopf, R. J.},
	month = oct,
	year = {2006},
	pages = {042318},
}

@article{clerk_hybrid_2020,
	title = {Hybrid quantum systems with circuit quantum electrodynamics},
	volume = {16},
	copyright = {2020 Springer Nature Limited},
	issn = {1745-2481},
	url = {https://www.nature.com/articles/s41567-020-0797-9},
	doi = {10.1038/s41567-020-0797-9},
	number = {3},
	urldate = {2024-09-24},
	journal = {Nature Physics},
	author = {Clerk, A. A. and Lehnert, K. W. and Bertet, P. and Petta, J. R. and Nakamura, Y.},
	month = mar,
	year = {2020},
	note = {Publisher: Nature Publishing Group},
	pages = {257--267},
}

@article{amay_thesis,
  author  = "Alexandre MAY",
    title   = {Noise Mitigation in Single Microwave Photon Counting},
   journal = {PhD thesis},
  school  = "Paris Saclay University",
  year    = "2025"
}

@article{yeh_microwave_2017,
	title = {Microwave attenuators for use with quantum devices below 100 {mK}},
	volume = {121},
	issn = {0021-8979},
	url = {https://doi.org/10.1063/1.4984894},
	doi = {10.1063/1.4984894},
	pages = {224501},
	number = {22},
	journal = {Journal of Applied Physics},
	author = {Yeh, Jen-Hao and {LeFebvre}, Jay and Premaratne, Shavindra and Wellstood, F. C. and Palmer, B. S.},
	date = {2017-06-08},
}

@article{wang_cavity_2019,
	title = {Cavity Attenuators for Superconducting Qubits},
	volume = {11},
	url = {https://link.aps.org/doi/10.1103/PhysRevApplied.11.014031},
	doi = {10.1103/PhysRevApplied.11.014031},
	pages = {014031},
	number = {1},
	journal = {Physical Review Applied},
	publisher = {American Physical Society},
	author = {Wang, Z. and Shankar, S. and Minev, Z.K. and Campagne-Ibarcq, P. and Narla, A. and Devoret, M.H.},
	date = {2019-01-16},
}

@article{yan_broadband_2023,
	title = {Broadband {Bandpass} {Purcell} {Filter} for {Circuit} {Quantum} {Electrodynamics}},
	volume = {123},
	issn = {0003-6951, 1077-3118},
	url = {http://arxiv.org/abs/2306.06258},
	doi = {10.1063/5.0161893},
	number = {13},
	urldate = {2025-05-03},
	journal = {Applied Physics Letters},
	author = {Yan, Haoxiong and Wu, Xuntao and Lingenfelter, Andrew and Joshi, Yash J. and Andersson, Gustav and Conner, Christopher R. and Chou, Ming-Han and Grebel, Joel and Miller, Jacob M. and Povey, Rhys G. and Qiao, Hong and Clerk, Aashish A. and Cleland, Andrew N.},
	month = sep,
	year = {2023},
	note = {arXiv:2306.06258 [quant-ph]},
	pages = {134001},
}

@article{Qutip,
  title = {QuTiP: An open-source Python framework for the dynamics of open quantum systems.},
  author = {J. R. Johansson and P. D. Nation and F. Nori},
  journal = {Comp. Phys. Comm.},
  volume = {183},
  pages = {1760–1772},
  year = {2012},
  doi = {10.1016/j.cpc.2012.02.021},
  url = {https://www.sciencedirect.com/science/article/abs/pii/S0010465512000835?via%3Dihub}
}

@article{Qutip2,
  title = {QuTiP 2: A Python framework for the dynamics of open quantum systems.},
  author = {J. R. Johansson and P. D. Nation and F. Nori},
  journal = {Comp. Phys. Comm.},
  volume = {184},
  pages = {1234},
  year = {2013},
  doi = {10.1016/j.cpc.2012.11.019},
  url = {https://www.sciencedirect.com/science/article/abs/pii/S0010465512003955?via%3Dihub}
}

@article{lescanne_irreversible_2020,
	title = {Irreversible {Qubit}-{Photon} {Coupling} for the {Detection} of {Itinerant} {Microwave} {Photons}},
	volume = {10},
	issn = {2160-3308},
	url = {https://link.aps.org/doi/10.1103/PhysRevX.10.021038},
	doi = {10.1103/PhysRevX.10.021038},

	number = {2},
	urldate = {2023-02-23},
	journal = {Physical Review X},
	author = {Lescanne, Raphaël and Deléglise, Samuel and Albertinale, Emanuele and Réglade, Ulysse and Capelle, Thibault and Ivanov, Edouard and Jacqmin, Thibaut and Leghtas, Zaki and Flurin, Emmanuel},
	month = may,
	year = {2020},
	pages = {021038},
}

@article{wang_single-electron_2023,
	title = {Single-electron spin resonance detection by microwave photon counting},
	volume = {619},
	copyright = {2023 The Author(s), under exclusive licence to Springer Nature Limited},
	issn = {1476-4687},
	url = {https://www.nature.com/articles/s41586-023-06097-2},
	doi = {10.1038/s41586-023-06097-2},

	number = {7969},
	urldate = {2024-02-15},
	journal = {Nature},
	author = {Wang, Z. and Balembois, L. and Rančić, M. and Billaud, E. and Le Dantec, M. and Ferrier, A. and Goldner, P. and Bertaina, S. and Chanelière, T. and Esteve, D. and Vion, D. and Bertet, P. and Flurin, E.},
	month = jul,
	year = {2023},
	pages = {276--281},
}

@article{balembois_cyclically2024,
  title = {Cyclically Operated Microwave Single-Photon Counter with Sensitivity of ${10}^{\ensuremath{-}22}\phantom{\rule{0.2em}{0ex}}\mathrm{W}/\sqrt{\mathrm{Hz}}$},
  author = {Balembois, L. and Travesedo, J. and Pallegoix, L. and May, A. and Billaud, E. and Villiers, M. and Est\`eve, D. and Vion, D. and Bertet, P. and Flurin, E.},
  journal = {Phys. Rev. Appl.},
  volume = {21},
  issue = {1},
  pages = {014043},
  numpages = {12},
  year = {2024},
  month = {Jan},
  publisher = {American Physical Society},
  doi = {10.1103/PhysRevApplied.21.014043},
  url = {https://link.aps.org/doi/10.1103/PhysRevApplied.21.014043}
}

@article{albertinale_detecting_2021,
	title = {Detecting spins by their fluorescence with a microwave photon counter},
	volume = {600},
	copyright = {2021 The Author(s), under exclusive licence to Springer Nature Limited},
	issn = {1476-4687},
	url = {https://www.nature.com/articles/s41586-021-04076-z},
	doi = {10.1038/s41586-021-04076-z},

	number = {7889},
	urldate = {2024-09-15},
	journal = {Nature},
	author = {Albertinale, Emanuele and Balembois, Léo and Billaud, Eric and Ranjan, Vishal and Flanigan, Daniel and Schenkel, Thomas and Estève, Daniel and Vion, Denis and Bertet, Patrice and Flurin, Emmanuel},
	month = dec,
	year = {2021},
	pages = {434--438},
}

@article{kono_quantum_2018,
	title = {Quantum non-demolition detection of an itinerant microwave photon},
	volume = {14},
	issn = {1745-2473, 1745-2481},
	doi = {10.1038/s41567-018-0066-3},
	number = {6},
	urldate = {2024-02-25},
	journal = {Nature Physics},
	author = {Kono, S. and Koshino, K. and Tabuchi, Y. and Noguchi, A. and Nakamura, Y.},
	month = jun,
	year = {2018},
	pages = {546--549},
}

@article{inomata_single_2016,
	title = {Single microwave-photon detector using an artificial \${\textbackslash}{Lambda}\$-type three-level system},
	volume = {7},
	issn = {2041-1723},
	doi = {10.1038/ncomms12303},
	number = {1},
	urldate = {2024-02-25},
	journal = {Nature Communications},
	author = {Inomata, Kunihiro and Lin, Zhirong and Koshino, Kazuki and Oliver, William D. and Tsai, Jaw-Shen and Yamamoto, Tsuyoshi and Nakamura, Yasunobu},
	month = jul,
	year = {2016},
	pages = {12303},
}

@article{pankratov_observation_2024,
  author  = {Pankratov, A. L. and Gordeeva, A. V. and Chiginev, A. V. and others},
  title   = {Detection of single-mode thermal microwave photons using an underdamped Josephson junction},
  journal = {Nature Communications},
  volume  = {16},
  pages   = {3457},
  year    = {2025},
  doi     = {10.1038/s41467-025-56040-4},
  url     = {https://doi.org/10.1038/s41467-025-56040-4}
}

@article{khezri_measurement-induced_2022,
  title = {Measurement-induced state transitions in a superconducting qubit: Within the rotating-wave approximation},
  author = {Khezri, Mostafa and Opremcak, Alex and Chen, Zijun and Miao, Kevin C. and McEwen, Matt and Bengtsson, Andreas and White, Theodore and Naaman, Ofer and Sank, Daniel and Korotkov, Alexander N. and Chen, Yu and Smelyanskiy, Vadim},
  journal = {Phys. Rev. Appl.},
  volume = {20},
  issue = {5},
  pages = {054008},
  numpages = {12},
  year = {2023},
  month = {Nov},
  publisher = {American Physical Society},
  doi = {10.1103/PhysRevApplied.20.054008},
  url = {https://link.aps.org/doi/10.1103/PhysRevApplied.20.054008}
}

@article{shillito_dynamics_2022,
	title = {Dynamics of {Transmon} {Ionization}},
	volume = {18},
	issn = {2331-7019},
	url = {http://arxiv.org/abs/2203.11235},
	doi = {10.1103/PhysRevApplied.18.034031},
	number = {3},
	urldate = {2023-07-28},
	journal = {Physical Review Applied},
	author = {Shillito, Ross and Petrescu, Alexandru and Cohen, Joachim and Beall, Jackson and Hauru, Markus and Ganahl, Martin and Lewis, Adam G. M. and Vidal, Guifre and Blais, Alexandre},
	month = sep,
	year = {2022},
	pages = {034031},
}

@article{cohen_reminiscence_2022,
  title = {Reminiscence of Classical Chaos in Driven Transmons},
  author = {Cohen, Joachim and Petrescu, Alexandru and Shillito, Ross and Blais, Alexandre},
  journal = {PRX Quantum},
  volume = {4},
  issue = {2},
  pages = {020312},
  numpages = {27},
  year = {2023},
  month = {Apr},
  publisher = {American Physical Society},
  doi = {10.1103/PRXQuantum.4.020312},
  url = {https://link.aps.org/doi/10.1103/PRXQuantum.4.020312}
}

@article{lescanne_escape_2019,
	title = {Escape of a {Driven} {Quantum} {Josephson} {Circuit} into {Unconfined} {States}},
	volume = {11},
	url = {https://link.aps.org/doi/10.1103/PhysRevApplied.11.014030},
	doi = {10.1103/PhysRevApplied.11.014030},
	number = {1},
	urldate = {2023-07-28},
	journal = {Physical Review Applied},
	author = {Lescanne, Raphaël and Verney, Lucas and Ficheux, Quentin and Devoret, Michel H. and Huard, Benjamin and Mirrahimi, Mazyar and Leghtas, Zaki},
	month = jan,
	year = {2019},
	pages = {014030},
}

@article{sank_measurement-induced_2016,
	title = {Measurement-{Induced} {State} {Transitions} in a {Superconducting} {Qubit}: {Beyond} the {Rotating} {Wave} {Approximation}},
	volume = {117},
	shorttitle = {Measurement-{Induced} {State} {Transitions} in a {Superconducting} {Qubit}},
	url = {https://link.aps.org/doi/10.1103/PhysRevLett.117.190503},
	doi = {10.1103/PhysRevLett.117.190503},
	number = {19},
	urldate = {2023-07-28},
	journal = {Physical Review Letters},
	author = {Sank, Daniel and Chen, Zijun and Khezri, Mostafa and Kelly, J. and Barends, R. and Campbell, B. and Chen, Y. and Chiaro, B. and Dunsworth, A. and Fowler, A. and Jeffrey, E. and Lucero, E. and Megrant, A. and Mutus, J. and Neeley, M. and Neill, C. and O’Malley, P.J.J. and Quintana, C. and Roushan, P. and Vainsencher, A. and White, T. and Wenner, J. and Korotkov, Alexander N. and Martinis, John M.},
	month = nov,
	year = {2016},
	pages = {190503},
}

@article{dumas_unified_2024,
  title = {Measurement-Induced Transmon Ionization},
  author = {Dumas, Marie Fr\'ed\'erique and Groleau-Par\'e, Benjamin and McDonald, Alexander and Mu\~noz-Arias, Manuel H. and Lled\'o, Crist\'obal and D'Anjou, Benjamin and Blais, Alexandre},
  journal = {Phys. Rev. X},
  volume = {14},
  issue = {4},
  pages = {041023},
  numpages = {31},
  year = {2024},
  month = {Oct},
  publisher = {American Physical Society},
  doi = {10.1103/PhysRevX.14.041023},
  url = {https://link.aps.org/doi/10.1103/PhysRevX.14.041023}
}

@article{venkatraman_static_2022,
  title = {Static Effective Hamiltonian of a Rapidly Driven Nonlinear System},
  author = {Venkatraman, Jayameenakshi and Xiao, Xu and Corti\~nas, Rodrigo G. and Eickbusch, Alec and Devoret, Michel H.},
  journal = {Phys. Rev. Lett.},
  volume = {129},
  issue = {10},
  pages = {100601},
  numpages = {6},
  year = {2022},
  month = {Aug},
  publisher = {American Physical Society},
  doi = {10.1103/PhysRevLett.129.100601},
  url = {https://link.aps.org/doi/10.1103/PhysRevLett.129.100601}
}

@article{carde_flux-pump_2025,
	title = {Flux-Pump-Induced Degradation of $T_1$ for Dissipative Cat Qubits},
	author = {Carde, Léon and Rouchon, Pierre and Cohen, Joachim and Petrescu, Alexandru},
	journal = {Physical Review Applied},
	volume = {23},
	pages = {024073},
	year = {2025},
	doi = {10.1103/PhysRevApplied.23.024073},
}

@article{xiao_diagrammatic_2023,
	title = {A Diagrammatic Method to Compute the Effective Hamiltonian of Driven Nonlinear Oscillators},
	author = {Xiao, Xu and Venkatraman, Jayameenakshi and Cortiñas, Rodrigo G. and Chowdhury, Shoumik and Devoret, Michel H.},
	journal = {Physical Review Applied},
	volume = {24},
	pages = {044021},
	year = {2025},
}

@article{bertet_dephasing_2005,
	title = {Dephasing of a Superconducting Qubit Induced by Photon Noise},
	volume = {95},
	url = {https://link.aps.org/doi/10.1103/PhysRevLett.95.257002},
	pages = {257002},
	number = {25},
	journal = {Physical Review Letters},
	publisher = {American Physical Society},
	author = {Bertet, P. and Chiorescu, I. and Burkard, G. and Semba, K. and Harmans, C. J. P. M. and {DiVincenzo}, D. P. and Mooij, J. E.},
	date = {2005-12-13},
}

@article{yan_flux_2016,
	title = {The flux qubit revisited to enhance coherence and reproducibility},
	volume = {7},
	issn = {2041-1723},
	url = {https://www.nature.com/articles/ncomms12964},
	doi = {10.1038/ncomms12964},
	pages = {12964},
	number = {1},
	journal = {Nature Communications},
	publisher = {Nature Publishing Group},
	author = {Yan, Fei and Gustavsson, Simon and Kamal, Archana and Birenbaum, Jeffrey and Sears, Adam P. and Hover, David and Gudmundsen, Ted J. and Rosenberg, Danna and Samach, Gabriel and Weber, S. and Yoder, Jonilyn L. and Orlando, Terry P. and Clarke, John and Kerman, Andrew J. and Oliver, William D.},
	date = {2016-11-03},
}

@article{scigliuzzo_primary_2020,
	title = {Primary Thermometry of Propagating Microwaves in the Quantum Regime},
	volume = {10},
	url = {https://link.aps.org/doi/10.1103/PhysRevX.10.041054},
	doi = {10.1103/PhysRevX.10.041054},
	pages = {041054},
	number = {4},
	journal = {Physical Review X},
	publisher = {American Physical Society},
	author = {Scigliuzzo, Marco and Bengtsson, Andreas and Besse, Jean-Claude and Wallraff, Andreas and Delsing, Per and Gasparinetti, Simone},
	date = {2020-12-17},
}

@article{andrews_bidirectional_2014,
	title = {Bidirectional and efficient conversion between microwave and optical light},
	volume = {10},
	issn = {1745-2481},
	url = {https://doi.org/10.1038/nphys2911},
	doi = {10.1038/nphys2911},
	pages = {321--326},
	number = {4},
	journal = {Nature Physics},
	author = {Andrews, R. W. and Peterson, R. W. and Purdy, T. P. and Cicak, K. and Simmonds, R. W. and Regal, C. A. and Lehnert, K. W.},
	date = {2014-04-01},
}

@article{Brubaker2022,
  title = {Optomechanical Ground-State Cooling in a Continuous and Efficient Electro-Optic Transducer},
  author = {Brubaker, B. M. and Kindem, J. M. and Urmey, M. D. and Mittal, S. and Delaney, R. D. and Burns, P. S. and Vissers, M. R. and Lehnert, K. W. and Regal, C. A.},
  journal = {Phys. Rev. X},
  volume = {12},
  issue = {2},
  pages = {021062},
  numpages = {25},
  year = {2022},
  month = {Jun},
  publisher = {American Physical Society},
  doi = {10.1103/PhysRevX.12.021062},
  url = {https://link.aps.org/doi/10.1103/PhysRevX.12.021062}
}

@article{bultink_active_2016,
	title = {Active Resonator Reset in the Nonlinear Dispersive Regime of Circuit {QED}},
	volume = {6},
	issn = {2331-7019},
	url = {https://link.aps.org/doi/10.1103/PhysRevApplied.6.034008},
	doi = {10.1103/PhysRevApplied.6.034008},
	pages = {034008},
	number = {3},
	journal = {Physical Review Applied},
	author = {Bultink, C.C. and Rol, M.A. and O’Brien, T.E. and Fu, X. and Dikken, B.C.S. and Dickel, C. and Vermeulen, R.F.L. and De Sterke, J.C. and Bruno, A. and Schouten, R.N. and {DiCarlo}, L.},
	urldate = {2023-04-23},
	date = {2016-09-13},
	langid = {english},
}

@article{mori_suppression_2025,
	title = {Suppression of Measurement-Induced State Transitions in $\cos\varphi$-Coupling Transmon Readout},
	author = {Mori, Cyril and D'Esposito, Francesca and Petrescu, Alexandru and Ruela, Lucas and Kumar, Shelender and Suresh, Vishnu Narayanan and Ardati, Wael and Nicolas, Dorian and Cappelli, Giulio and Ranadive, Arpit and Le Gal, Gwenael and Esposito, Martina and Ficheux, Quentin and Roch, Nicolas and Buisson, Olivier},
	journal = {PRX Quantum},
	volume = {7},
	pages = {020369},
	year = {2026},
	doi = {10.1103/ylsw-t92p},
}

@article{dixit_millimeter_2026,
	title = {Millimeter Wave Readout of a Superconducting Qubit},
	author = {Dixit, Akash V. and Parrott, Zachary L. and Chunikhin, Dennis and Larson, Trevyn F. Q. and Hauer, Bradley D. and Teufel, John D.},
	journal = {Physical Review Letters},
	year = {2026},
	doi = {10.1103/c84y-ggqs},
	note = {Accepted},
}

@article{hazra_benchmarking_2025,
	title = {Benchmarking the Readout of a Superconducting Qubit for Repeated Measurements},
	author = {Hazra, S. and Dai, W. and Connolly, T. and Kurilovich, P. D. and Wang, Z. and Frunzio, L. and Devoret, M. H.},
	journal = {Physical Review Letters},
	volume = {134},
	pages = {100601},
	year = {2025},
	doi = {10.1103/PhysRevLett.134.100601},
}

@article{kono_mechanically_2024,
	title = {Mechanically induced correlated errors on superconducting qubits with relaxation times exceeding 0.4 ms},
	volume = {15},
	rights = {2024 The Author(s)},
	issn = {2041-1723},
	url = {https://www.nature.com/articles/s41467-024-48230-3},
	doi = {10.1038/s41467-024-48230-3},
	pages = {3950},
	number = {1},
	journal = {Nature Communications},
	publisher = {Nature Publishing Group},
	author = {Kono, Shingo and Pan, Jiahe and Chegnizadeh, Mahdi and Wang, Xuxin and Youssefi, Amir and Scigliuzzo, Marco and Kippenberg, Tobias J.},
	urldate = {2025-01-31},
	date = {2024-05-10},
	langid = {english},
}

@article{tuokkola_methods_2024,
	title = {Methods to Achieve Near-Millisecond Energy Relaxation and Dephasing Times for a Superconducting Transmon Qubit},
	author = {Tuokkola, Mikko and Sunada, Yoshiki and Kivijärvi, Heidi and Albanese, Jonatan and Grönberg, Leif and Kaikkonen, Jukka-Pekka and Vesterinen, Visa and Govenius, Joonas and Möttönen, Mikko},
	journal = {Nature Communications},
	volume = {16},
	pages = {5421},
	year = {2025},
	doi = {10.1038/s41467-025-61126-0},
}

@article{swiadek_enhancing_2024,
	title = {Enhancing Dispersive Readout of Superconducting Qubits through Dynamic Control of the Dispersive Shift: Experiment and Theory},
	volume = {5},
	url = {https://link.aps.org/doi/10.1103/PRXQuantum.5.040326},
	doi = {10.1103/PRXQuantum.5.040326},
	shorttitle = {Enhancing Dispersive Readout of Superconducting Qubits through Dynamic Control of the Dispersive Shift},
	pages = {040326},
	number = {4},
	journal = {{PRX} Quantum},
	publisher = {American Physical Society},
	author = {Swiadek, François and Shillito, Ross and Magnard, Paul and Remm, Ants and Hellings, Christoph and Lacroix, Nathan and Ficheux, Quentin and Zanuz, Dante Colao and Norris, Graham J. and Blais, Alexandre and Krinner, Sebastian and Wallraff, Andreas},
	date = {2024-11-20},
}

@article{spring_fast_2025,
	title = {Fast Multiplexed Superconducting-Qubit Readout with Intrinsic Purcell Filtering Using a Multiconductor Transmission Line},
	volume = {6},
	url = {https://link.aps.org/doi/10.1103/PRXQuantum.6.020345},
	doi = {10.1103/PRXQuantum.6.020345},
	pages = {020345},
	number = {2},
	journal = {{PRX} Quantum},
	publisher = {American Physical Society},
	author = {Spring, Peter A. and Milanovic, Luka and Sunada, Yoshiki and Wang, Shiyu and van Loo, Arjan F. and Tamate, Shuhei and Nakamura, Yasunobu},
	date = {2025-06-05},
}

@article{bland_2d_2025,
	title = {Millisecond Lifetimes and Coherence Times in {2D} Transmon Qubits},
	author = {Bland, Matthew P. and Bahrami, Faranak and Martinez, Jeronimo G. C. and Prestegaard, Paal H. and Smitham, Basil M. and Joshi, Atharv and Hedrick, Elizabeth and Kumar, Shashwat and Yang, Ambrose and Pakpour-Tabrizi, Alexander C. and Jindal, Apoorv and Chang, Ray D. and Cheng, Guangming and Yao, Nan and Cava, Robert J. and de Leon, Nathalie P. and Houck, Andrew A.},
	journal = {Nature},
	volume = {647},
	pages = {343--348},
	year = {2025},
}

@article{serniak_hot_2018,
	title = {Hot Nonequilibrium Quasiparticles in Transmon Qubits},
	volume = {121},
	url = {https://link.aps.org/doi/10.1103/PhysRevLett.121.157701},
	doi = {10.1103/PhysRevLett.121.157701},
	pages = {157701},
	number = {15},
	journal = {Physical Review Letters},
	publisher = {American Physical Society},
	author = {Serniak, K. and Hays, M. and de Lange, G. and Diamond, S. and Shankar, S. and Burkhart, L.D. and Frunzio, L. and Houzet, M. and Devoret, M.H.},
	date = {2018-10-10},
}

@article{connolly_coexistence_2023,
	title = {Coexistence of Nonequilibrium Density and Equilibrium Energy Distribution of Quasiparticles in a Superconducting Qubit},
	author = {Connolly, Thomas and Kurilovich, Pavel D. and Diamond, Spencer and Nho, Heekun and Bøttcher, Charlotte G. L. and Glazman, Leonid I. and Fatemi, Valla and Devoret, Michel H.},
	journal = {Physical Review Letters},
	shortjournal = {Phys. Rev. Lett.},
	volume = {132},
	number = {21},
	pages = {217001},
	date = {2024},
	doi = {10.1103/PhysRevLett.132.217001},
	url = {https://link.aps.org/doi/10.1103/PhysRevLett.132.217001}
}

@article{kerschbaum_assessing_2026,
	title = {Assessing the Sensitivity of Niobium- and Tantalum-Based Superconducting Qubits to Infrared Radiation},
	author = {Kerschbaum, Michael and Wagner, Felix and Ognjanović, Uroš and Vio, Giovanni and Knapp, Kuno and Zanuz, Dante Colao and Flasby, Alexander and Panah, Mohsen Bahrami and Wallraff, Andreas and Besse, Jean-Claude},
	journal = {arXiv:2602.05806},
	year = {2026},
	doi = {10.48550/arXiv.2602.05806},
}

@article{vepsalainen_impact_2020,
	title = {Impact of ionizing radiation on superconducting qubit coherence},
	volume = {584},
	issn = {0028-0836, 1476-4687},
	url = {https://www.nature.com/articles/s41586-020-2619-8},
	doi = {10.1038/s41586-020-2619-8},
	pages = {551--556},
	number = {7822},
	journal = {Nature},
	author = {Vepsäläinen, Antti P. and Karamlou, Amir H. and Orrell, John L. and Dogra, Akshunna S. and Loer, Ben and Vasconcelos, Francisca and Kim, David K. and Melville, Alexander J. and Niedzielski, Bethany M. and Yoder, Jonilyn L. and Gustavsson, Simon and Formaggio, Joseph A. and {VanDevender}, Brent A. and Oliver, William D.},
	urldate = {2023-01-27},
	date = {2020-08-27},
	langid = {english},
}

@article{mcewen_resisting_2024,
	title = {Resisting High-Energy Impact Events through Gap Engineering in Superconducting Qubit Arrays},
	author = {{McEwen}, Matt and Miao, Kevin C. and Atalaya, Juan and Bilmes, Alex and Crook, Alex and Bovaird, Jenna and Kreikebaum, John Mark and Zobrist, Nicholas and Jeffrey, Evan and Ying, Bicheng and Bengtsson, Andreas and Chang, Hung-Shen and Dunsworth, Andrew and Kelly, Julian and Zhang, Yaxing and Forati, Ebrahim and Acharya, Rajeev and Iveland, Justin and Liu, Wayne and Kim, Seon and Burkett, Brian and Megrant, Anthony and Chen, Yu and Neill, Charles and Sank, Daniel and Devoret, Michel and Opremcak, Alex},
	journal = {Physical Review Letters},
	volume = {133},
	pages = {240601},
	year = {2024},
	doi = {10.1103/PhysRevLett.133.240601},
}

@article{mcewen_resolving_2022,
	title = {Resolving catastrophic error bursts from cosmic rays in large arrays of superconducting qubits},
	volume = {18},
	rights = {2021 The Author(s), under exclusive licence to Springer Nature Limited},
	issn = {1745-2481},
	url = {https://www.nature.com/articles/s41567-021-01432-8},
	doi = {10.1038/s41567-021-01432-8},
	pages = {107--111},
	number = {1},
	journal = {Nature Physics},
	publisher = {Nature Publishing Group},
	author = {{McEwen}, Matt and Faoro, Lara and Arya, Kunal and Dunsworth, Andrew and Huang, Trent and Kim, Seon and Burkett, Brian and Fowler, Austin and Arute, Frank and Bardin, Joseph C. and Bengtsson, Andreas and Bilmes, Alexander and Buckley, Bob B. and Bushnell, Nicholas and Chen, Zijun and Collins, Roberto and Demura, Sean and Derk, Alan R. and Erickson, Catherine and Giustina, Marissa and Harrington, Sean D. and Hong, Sabrina and Jeffrey, Evan and Kelly, Julian and Klimov, Paul V. and Kostritsa, Fedor and Laptev, Pavel and Locharla, Aditya and Mi, Xiao and Miao, Kevin C. and Montazeri, Shirin and Mutus, Josh and Naaman, Ofer and Neeley, Matthew and Neill, Charles and Opremcak, Alex and Quintana, Chris and Redd, Nicholas and Roushan, Pedram and Sank, Daniel and Satzinger, Kevin J. and Shvarts, Vladimir and White, Theodore and Yao, Z. Jamie and Yeh, Ping and Yoo, Juhwan and Chen, Yu and Smelyanskiy, Vadim and Martinis, John M. and Neven, Hartmut and Megrant, Anthony and Ioffe, Lev and Barends, Rami},
	urldate = {2024-04-16},
	date = {2022-01},
	langid = {english},
}

@article{motzoi_precise_2018,
	title = {Precise single-qubit control of the reflection phase of a photon mediated by a strongly-coupled ancilla–cavity system},
	volume = {20},
	issn = {1367-2630},
	url = {https://iopscience.iop.org/article/10.1088/1367-2630/aac0be},
	doi = {10.1088/1367-2630/aac0be},
	number = {5},
	urldate = {2025-04-09},
	journal = {New Journal of Physics},
	author = {Motzoi, F and Mølmer, K},
	month = may,
	year = {2018},
	pages = {053029},
}

@article{harrington_synchronous_2025,
	title = {Synchronous detection of cosmic rays and correlated errors in superconducting qubit arrays},
	volume = {16},
	rights = {2025 The Author(s)},
	issn = {2041-1723},
	url = {https://www.nature.com/articles/s41467-025-61385-x},
	doi = {10.1038/s41467-025-61385-x},
	pages = {6428},
	number = {1},
	journal = {Nature Communications},
	publisher = {Nature Publishing Group},
	author = {Harrington, Patrick M. and Li, Mingyu and Hays, Max and Van De Pontseele, Wouter and Mayer, Daniel and Pinckney, H. Douglas and Contipelli, Felipe and Gingras, Michael and Niedzielski, Bethany M. and Stickler, Hannah and Yoder, Jonilyn L. and Schwartz, Mollie E. and Grover, Jeffrey A. and Serniak, Kyle and Oliver, William D. and Formaggio, Joseph A.},
	urldate = {2026-06-18},
	date = {2025-07-11},
	langid = {english},
}

@misc{GardinerZoller_quantumNoise,
	title = {Quantum {Noise}: {A} {Handbook} of {Markovian} and {Non}-{Markovian} {Quantum} {Stochastic} {Methods} with {Applications} to {Quantum} {Optics} {\textbar} {SpringerLink}},
	author = { Crispin Gardiner , Peter Zoller},
	url = {https://link.springer.com/book/9783540223016},
	urldate = {2025-04-12},
}

@article{kiilerich_input-output_2019,
	title = {Input-{Output} {Theory} with {Quantum} {Pulses}},
	volume = {123},
	url = {https://link.aps.org/doi/10.1103/PhysRevLett.123.123604},
	doi = {10.1103/PhysRevLett.123.123604},
	number = {12},
	urldate = {2023-10-21},
	journal = {Physical Review Letters},
	author = {Kiilerich, Alexander Holm and Mølmer, Klaus},
	month = sep,
	year = {2019},
	note = {Publisher: American Physical Society},
	pages = {123604},
}

@article{kiilerich_quantum_2020,
	title = {Quantum interactions with pulses of radiation},
	volume = {102},
	url = {https://link.aps.org/doi/10.1103/PhysRevA.102.023717},
	doi = {10.1103/PhysRevA.102.023717},
	number = {2},
	urldate = {2023-10-21},
	journal = {Physical Review A},
	author = {Kiilerich, Alexander Holm and Mølmer, Klaus},
	month = aug,
	year = {2020},
	note = {Publisher: American Physical Society},
	pages = {023717},
}

@article{khanahmadi_multimode_2023,
	title = {Multimode character of quantum states released from a superconducting cavity},
	volume = {5},
	issn = {2643-1564},
	url = {https://link.aps.org/doi/10.1103/PhysRevResearch.5.043071},
	doi = {10.1103/PhysRevResearch.5.043071},
	number = {4},
	urldate = {2025-04-09},
	journal = {Physical Review Research},
	author = {Khanahmadi, Maryam and Lund, Mads Middelhede and Mølmer, Klaus and Johansson, Göran},
	month = oct,
	year = {2023},
	pages = {043071},
}

@article{larson_quasiparticle_2025,
	title = {Quasiparticle Poisoning of Superconducting Qubits with Active Gamma Irradiation},
	author = {Larson, C. P. and Yelton, E. and Dodge, K. and Okubo, K. and Batarekh, J. and Iaia, V. and Kurinsky, N. A. and Plourde, B. L. T.},
	journal = {PRX Quantum},
	volume = {6},
	pages = {030339},
	year = {2025},
	doi = {10.1103/2lyd-8swv},
}

@article{valenti_spatial_2025,
	title = {Seismometry of Radiation-Induced Quasiparticle Bursts in Superconducting Devices},
	author = {Valenti, Francesco and Murani, Anil and Paluch, Patrick and Gartmann, Robert and Scheller, Lukas and Weller, Richard and Kruk, Robert and Reisinger, Thomas and Ardila-Perez, Luis and Pop, Ioan M.},
	journal = {Physical Review B},
	volume = {113},
	pages = {174504},
	year = {2026},
	doi = {10.1103/xmyw-yc9h},
}

@article{kurilovich_correlated_2025,
	title = {Correlated Phase Error Bursts in a Gap-Engineered Superconducting Qubit Array},
	author = {Kurilovich, Vladislav D. and Roberts, Gabrielle and Martin, Leigh S. and {McEwen}, Matt and Eickbusch, Alec and Faoro, Lara and Ioffe, Lev B. and Atalaya, Juan and Bilmes, Alexander and Kreikebaum, John Mark and Bengtsson, Andreas and Klimov, Paul and Neeley, Matthew and Mruczkiewicz, Wojciech and Miao, Kevin and Aleiner, Igor L. and Kelly, Julian and Chen, Yu and Satzinger, Kevin and Opremcak, Alex},
	journal = {Physical Review X},
	volume = {16},
	pages = {021025},
	year = {2026},
	doi = {10.1103/1bl4-b2f7},
}

@article{yang_high-temporal-resolution_2026,
	title = {High-Temporal-Resolution Measurements of the Impacts of Ionizing Radiation on Superconducting Qubits},
	author = {Yang, Jihee and Carroll, Thomas J. and Mason, Philip and Schwartz, Robert and O'Hara, Kenneth M. and Lund, Jennifer and Gottschalk, Michael and Stephenson, Timothy and Friedman, Lawrence H. and Yumiceva, Francisco and Hackley, Justin and Graninger, Aurelius L. and Rotella, Chris and Warner, Pat and Cochran, Jonathan M. and Bruce, Adam V. and Wagner, Melody and Wenner, James and Steers, Stan and Moore, Christopher and Marakov, Alex and Christensen, Bradley G.},
	journal = {arXiv:2602.23544},
	year = {2026},
	doi = {10.48550/arXiv.2602.23544},
}

@article{lin_suppression_2026,
	title = {Suppression of Quasiparticle Poisoning to $10^{-11}$ Levels in Superconducting Qubits via Infrared Shielding},
	author = {Lin, Wei-En and Ma, Chen-Hsun and Yeh, Erh-Hsiang and Peng, Wei-Lun and Wei, Yu-Sen and Goan, Hsi-Sheng and Wu, Cen-Shawn and Ke, Chung-Ting and Chen, Yung-Fu and Chen, Chii-Dong},
	journal = {arXiv:2606.07339},
	year = {2026},
	doi = {10.48550/arXiv.2606.07339},
}

@article{celi_measuring_2026,
	title = {Measuring Quasiparticle Dynamics for Particle Impact Reconstruction in a Superconducting Qubit Chip},
	author = {Celi, E. and Linehan, R. and Harrington, P. M. and Li, M. and Pinckney, H. D. and Serniak, K. and Oliver, W. D. and Formaggio, J. A. and Figueroa-Feliciano, E. and Baxter, D.},
	journal = {arXiv:2604.13176},
	year = {2026},
	doi = {10.48550/arXiv.2604.13176},
}

@article{binney_distinguishing_2026,
	title = {Distinguishing Types of Correlated Errors in Superconducting Qubits},
	author = {Binney, Hannah P. and Pinckney, H. Douglas and Azar, Kate and Harrington, Patrick M. and Jha, Shantanu and Li, Mingyu and Yang, Jiatong and Contipelli, Felipe and {DePencier Piñero}, Renée and Gingras, Michael and Niedzielski, Bethany M. and Stickler, Hannah and Schwartz, Mollie E. and Grover, Jeffrey A. and Hays, Max and Serniak, Kyle and Formaggio, Joseph A. and Oliver, William D.},
	journal = {Physical Review Applied},
	volume = {26},
	pages = {024025},
	year = {2026},
	doi = {10.1103/j235-47t6},
}

@article{GoogleQuantumAI2021RepetitionCode,
  title = {Exponential suppression of bit or phase errors with cyclic error correction},
  author = {{Google Quantum AI}},
  journal = {Nature},
  volume = {595},
  number = {7867},
  pages = {383--387},
  year = {2021},
  doi = {10.1038/s41586-021-03588-y},
}

@article{xia_exceeding_2025,
	title = {Exceeding the Parametric Drive Strength Threshold in Nonlinear Circuits},
	author = {Xia, Mingkang and Lledó, Cristóbal and Capocci, Matthew and Repicky, Jacob and D'Anjou, Benjamin and Mondragon-Shem, Ian and Kaufman, Ryan and Koch, Jens and Blais, Alexandre and Hatridge, Michael},
	journal = {arXiv:2506.03456},
	year = {2025},
	doi = {10.48550/arXiv.2506.03456},
}

@article{connolly_full_2025,
	title = {Full Characterization of Measurement-Induced Transitions of a Superconducting Qubit},
	author = {Connolly, Thomas and Kurilovich, Pavel D. and Kurilovich, Vladislav D. and Bøttcher, Charlotte G. L. and Hazra, Sumeru and Dai, Wei and Ding, Andy Z. and Joshi, Vidul R. and Nho, Heekun and Diamond, Spencer and Weiss, Daniel K. and Fatemi, Valla and Frunzio, Luigi and Glazman, Leonid I. and Devoret, Michel H.},
	journal = {arXiv:2506.05306},
	year = {2025},
	doi = {10.48550/arXiv.2506.05306},
}

@article{dai_characterization_2026,
	title = {Characterization of Drive-Induced Unwanted State Transitions in Superconducting Circuits},
	volume = {16},
	issn = {2160-3308},
	url = {https://link.aps.org/doi/10.1103/zdpg-mhpc},
	doi = {10.1103/zdpg-mhpc},
	pages = {011011},
	number = {1},
	journal = {Physical Review X},
	author = {Dai, W. and Hazra, S. and Weiss, D. K. and Kurilovich, P. D. and Connolly, T. and Babla, H. K. and Singh, S. and Joshi, V. R. and Ding, A. Z. and Parakh, P. D. and Venkatraman, J. and Xiao, X. and Frunzio, L. and Devoret, M. H.},
	date = {2026-01-15},
	langid = {english},
}

\end{document}